\documentclass[12pt,preprint]{aastex}
\usepackage{emulateapj5}
\usepackage{natbib}
\usepackage{onecolfloat}
\usepackage{apjfonts}
\usepackage{subfigure}  
\usepackage{color}
\usepackage{appendix}

\def\aap{A\&A}

\def\apjl{ApJL}

\def\apjs{ApJS}

\newcommand{\hh}{\,h^{-1}}  

\newcommand{\chandra}{{\it Chandra}}

\newcommand{\be}{\begin{equation}}
\newcommand{\ee}{\end{equation}}
\newcommand{\ba}{\begin{eqnarray}}
\newcommand{\ea}{\end{eqnarray}}

\shorttitle{Lensing and X-ray mass estimates - Simulations}
\shortauthors{Rasia et al.}
\begin{document}   

\title{Lensing and X-ray Mass Estimates of Clusters (Simulations)}
\author{
E. Rasia\altaffilmark{1,2},
M. Meneghetti\altaffilmark{3},  
R. Martino\altaffilmark{4},
S. Borgani\altaffilmark{5,6,7},
A. Bonafede\altaffilmark{8},
K. Dolag\altaffilmark{9},
S. Ettori\altaffilmark{3,10}
D. Fabjan\altaffilmark{7,11,12},
C. Giocoli\altaffilmark{3},
P. Mazzotta\altaffilmark{4,13},
J. Merten\altaffilmark{14},
M. Radovich \altaffilmark{15},
L. Tornatore\altaffilmark{5,6,7}
}

\altaffiltext{1}{ Department of Astronomy, University of Michigan, 500 Church St., Ann Arbor, MI 48109-1120, USA, rasia@umich.edu}
\altaffiltext{2}{ Fellow of Michigan Society of Fellows }
\altaffiltext{3}{ INAF, Osservatorio Astronomico di Bologna, via Ranzani 1, I-40127, Bologna, Italy, meneghetti.massimo@oabo.inaf.it }
\altaffiltext{4}{ Dipartimento di Fisica, Universit\`a di Roma Tor Vergata,
via della Ricerca Scientifica 1, I-00133, Roma, Italy }
\altaffiltext{5}{Dipartimento di Fisica dellÕUniversit\`a di Trieste, Sezione di Astronomia, via Tiepolo 11, I-34131 Trieste, Italy}
\altaffiltext{6}{INAF -- Osservatorio Astronomico di Trieste, via Tiepolo 11, I-34131 Trieste, Italy}
\altaffiltext{7}{INFN -- Istituto Nazionale di Fisica Nucleare, Trieste, Italy}
\altaffiltext{8} {Jacobs University Bremen, Campus Ring 1, D-28759 Bremen, Germany}
\altaffiltext{9}{INAF -- Osservatorio Astronomico di Padova, vicolo dell'Osservatorio 5, I-35122, Padova, Italy}
\altaffiltext{10}{INFN, Sezione di Bologna, viale Berti Pichat 6/2, I-40127 Bologna, Italy}
\altaffiltext{11}{Center of Excellence SPACE-SI, A\v{s}ker\v{c}eva 12, 1000 Ljubljana, Slovenia}
\altaffiltext{12}{Faculty of Mathematics and Physics, University of Ljubljana, Jadranska 19, 1000 Ljubljana, Slovenia}
\altaffiltext{13}{ Harvard-Smithsonian Centre for Astrophysics, 60 Garden Street, Cambridge, MA 02138, USA}
\altaffiltext{14}{ITA, Zentrum f${\ddot u}$r Astronomie, Universit${\ddot a}$t Heidelberg, Germany}
\altaffiltext{15} {University Observatory M${\ddot u}$nchen, Scheinerstr. 1, 81679, M${\ddot u}$nchen, Germany}
 
\begin{abstract}
We present a comparison between weak-lensing and X-ray mass estimates of a sample of numerically simulated clusters. The sample consists on the $20$ most massive objects at redshift $z=0.25$ and $M_{\rm vir} > 5 \times 10^{14} M_{\odot} h^{-1}$. They were found in a cosmological simulation of volume $1 h^{-3}$ Gpc$^3$, evolved in the framework of a WMAP-7 normalized cosmology. Each cluster has been resimulated at higher resolution and with more complex gas physics. We  processed it thought {\tt Skylens}  and  {\tt X-MAS} to generate optical and X-ray mock observations along three orthogonal projections. The final sample consists on 60 cluster realizations. The optical simulations include lensing effects on background sources. Standard observational tools and methods of analysis are used to recover  the mass profiles of each cluster projection from the mock  catalogue. The resulting mass profiles from lensing and X-ray are individually compared to the input mass distributions. Given the size of our sample, we could also investigate the dependence of the results on cluster morphology, environment, temperature inhomogeneity, and mass.  We confirm previous results showing that lensing masses obtained from the fit of the cluster tangential shear profiles with NFW functionals are biased low by $\sim 5-10\%$ with a large scatter ($\sim 10-25\%$). We show that scatter could be reduced by optimally selecting clusters either having regular morphology or living in substructure-poor environment. The X-ray masses are biased low by a large amount ($\sim 25-35\%$), evidencing the presence of both non-thermal sources of pressure in the ICM and temperature inhomogeneity,  but they show a significantly lower scatter than weak-lensing-derived masses. The X-ray mass bias grows from the inner to the outer regions of the clusters.   We find that both biases are weakly correlated with the third-order power ratio, while a stronger correlation exists with the centroid shift. Finally, the X-ray bias is strongly connected with temperature inhomogeneities.
Comparison with previous analysis of simulations leads to the conclusion that the values of X-ray mass bias from simulations is still uncertain, showing dependences on the ICM physical treatment and, possibly, on the hydrodynamical scheme adopted. 
\end{abstract}



\section{INTRODUCTION   \label{sec:intro}}

Galaxy clusters are important test sites for cosmology and astrophysics. First, they are ideal laboratories for studying how the dark-matter behaves in dense environment and evolves in the non-linear regime.  
Second, their mass function is highly sensitive to cosmology, since its evolution traces the growth of the linear density perturbations with exponential magnification \citep{PR74.1,SH02.1,JE01.1,2006ApJ...646..881W}. Indeed, clusters are the most massive gravitationally bound structures in the universe and, in the framework of the hierarchical structure formation scenario, they are also the youngest systems formed  to date.
Therefore, clusters are a mine of cosmological information, a large fraction of which is contained in the  mass profile of these structures.
Several methods can be used to determine the matter distribution in galaxy clusters. Two widely used approaches are based on X-ray and lensing observations. \\
{\it X-ray observations} allow the cluster mass profiles to be derived by assuming that these systems are spherically symmetric and that the emitting gas is in hydrostatic equilibrium \citep[e.g.][]{henriksen&mushotzky,sarazin88,ettori.etal.02}. This method has the advantage that, since the X-ray emissivity is proportional to the square of the electron density, it is not very sensitive to projection effects of masses along the line of sight to the clusters. However, it is still not well established how safely the hydrostatic equilibrium approximation can be made \citep[][hereafter R06]{rtm,rasia.etal.06}.\\
As the highest mass concentrations in the universe, galaxy clusters are the most efficient {\it gravitational lenses} on the sky. Their matter distorts background-galaxy images with an intensity that increases from the outskirts to the inner regions. Strong distortions occur in the cores of some massive galaxy clusters, leading to the formation of ``gravitational arcs'' and/or to the formation of systems of multiple images of the same source. Weak distortions, which can be only measured statistically, are impressed on the shape of distant galaxies that lie on the sky at large angular distances from the cluster centers \citep[e.g.][]{bart&schne}.
Both these lensing regimes can be used to map the mass distribution in galaxy clusters. 
Gravitational lensing can directly probe the cluster {\it total mass} without any strong assumptions on the equilibrium state of the lens. Further, mass profiles can be measured over a wide range of scales, from $\lesssim 100$ kpc out to the virial radius. However, lensing measures the projected mass instead of the 3D mass and it is sensitive to projection effects, such as triaxiality and additional concentrations of mass along the line of sight. \\
{\it Given the pros and cons of each method, we can conclude that lensing and X-ray are complementary in many ways}. In particular, the comparison of these two mass estimates can greatly help in improving the accuracy of the measurements and understanding the systematic errors.

Numerical simulations provide a unique way to investigate the performance of the lensing and X-ray techniques for measuring the mass profiles of galaxy clusters. Several studies were performed in the past which made use of relatively simple descriptions of galaxy clusters and simulation set-ups, but still were able to assess some fundamental limits of these techniques and possibly suggest improvements \citep[see e.g.][]{Metzler:2001gg,Clowe:2004cb,becker&kravtsov,rtm,Piffaretti:2003kn}. In the last years, using the increasing number of observational constraints and profiting of the huge increment of computational efficiency, the simulations have become even more sophisticated and can now include a large number of realistic and important features. These improvements regard both the description of the physical processes determining the evolution of the cosmic structures \citep[see][for a review]{borgani&kravtsov} and the interface between simulations and observations. In particular, few pipelines have been developed which produce simulated observations  of the numerically modeled clusters at different wavelengths \citep{xmas,nagai.etal.07,rasia.etal.08,2008A&A...482..403M,heinz&brueggen}. These pipelines can simulate observations with a variety of existing and future instruments and include most observational noises which typically affect and limit real measurements. Thus, they are ideal for testing data reduction pipelines
\citep{rasia.etal.06,mazzotta.etal.04,nagai.etal.07}.

In \citet{2010A&A...514A..93M} (M10, hereafter) we combined our optical simulator, {\tt SkyLens} \citep{2008A&A...482..403M}, with our X-ray one, {\tt XMAS} \citep{rasia.etal.08, xmas}, to  study the systematic effects in mass measurements encountered following standard lensing and X-ray analysis. In that work, we used three simulated clusters and study them along three independent lines-of-sight. In this paper, we extend that work to a much larger sample. We consider 60 mock optical and X-ray images (20 independent clusters for three orthogonal lines-of-sight). Throughout the paper the quoted errors correspond to 1$\sigma$ level. 

The paper is structured as follow. Section 2 contains a short review of the results obtained in previous numerical studies, especially in M10.  Section 3 presents a description of the simulated clusters.  Section 4 and Section 5 describe the lensing and the X-ray simulation pipelines and the methods of analyses. We present the results in Section 6, where we first discuss the lensing and X-ray mass estimates individually. We show how the bias and the scatter of the mass measurements depend on the cluster morphology and environment in Section 7. Finally, we discuss our results in Section 8.

\section{Previous studies } \label{sec:study}

\subsection{Strong lensing}
In M10, we used the parametric code {\em Lenstool} to construct mass
models from the multiple image systems detected in synthetic
Hubble-Space-Telescope (HST) observations. In the region where strong
lensing constraints were found (within the Einstein radius),
the mass profiles recovered agree with the input mass distributions
with an accuracy of a few percent. Similar results were obtained by
\cite{2007NJPh....9..447J} testing the performances of {\tt Lenstool}
with lens models produced using semi-analytical methods. The
strong-lensing models are constructed by combining a main halo
component and additional massive clumps associated to star clumps (the
galaxies of the cluster). Fundamental, in this process, is the
  modeling of the central galaxy, BCG
  \citep{2011A&A...528A..73D,2009MNRAS.398..438D,
    2006ApJ...642...39C}.  M10 demonstrated that a wrong
  parametrization of the BCG leads to a severe under- or over-estimate
  of the strong-lensing masses extrapolated at large radii. Indeed,
  when the central galaxy was excluded during the creation and the
  analysis of the synthetic optical images, both the bias and scatter
  were largely reduced. Discrepancies were seen already at $R_{2500}$,
  a radius which is typically 2--3 times larger than the Einstein radius. The parametrization of the BCG is also important for a more realistic
estimate of the lensing cross-section: its presence  increases the strong lensing signal up to a twenty percent  in
cluster size haloes \citep{giocoli.etal.11} .
 
  The necessity of having an accurate model for the BCG in order to
  extrapolate the strong lensing mass at large radii makes the
  comparison between strong lensing and X-ray mass estimates highly
  uncertain. Indeed, X-ray emission from the central region
  ($\sim 70-100$ kpc) is often excluded from the X-ray analysis
  because more difficult to model \citep[see e.g.][]{vikh.etal.06}.

\subsection{Weak lensing}

The weak-lensing analysis is based on the measurement of the shape of galaxies in the background of the clusters, whose ellipticity can be used to estimate the shear produced by the lens. Details on the weak-lensing analysis can be found in Section~\ref{sec:lensing}. In M10, we found that fitting the reduced tangential shear profile with a Navarro-Frenk-White \citep[NFW,][]{nfw} functional or using the aperture mass densitometry produce quite similar results. The measured projected mass is accurate at level of $\sim 10\%$ for those clusters that do not show any massive substructures nearby. Two lens planes presented massive clumps just {\it outside} the virial radius of the cluster. This dilutes the shear tangential to the main cluster clump even at smaller radii. As a consequence, the mass profiles of these two lenses resulted to  be severely under-estimated. Such problem affects the methods where the shear is measured tangentially. 
Instead, we tested that techniques which combine strong- and weak-lensing , such as those by \cite{2006A&A...458..349C} and by \cite{2009A&A...500..681M} are not influenced.  To reconstruct the lensing potential the latter method uses an adaptive grid \citep[see also][]{Bradac:PbBNSt9K,DI07.1,2011arXiv1103.2772M} and naturally incorporates the effects of substructure. As result, the scatter in the projected mass measurement is  reduced and limited to $\lesssim 10\%$.

The main causes of substantial scatter in the deprojected masses are triaxiality and presence of substructure. Under the standard assumption of spherical symmetry, three-dimensional masses are over- or under-estimated, depending on the orientation of the cluster major axis with respect to the line-of-sight. For systems whose major axis points toward the observer, masses are typically over-estimated. The opposite occurs for clusters elongated in the plane of the sky. In M10, the resulting scatter of our sample is of order $\sim 17\%$. For clusters with substructure, the unknown location of the substructure along the line of sight also makes the three-dimensional mass estimate highly unsure.  

More recently, \cite{becker&kravtsov} used a large number of simulated halos extracted from a large cosmological box to discuss the accuracy of weak lensing masses measured by fitting the cluster shear profiles with NFW models. Their results are consistent with ours. 
Given the large size of their sample, they significantly probe that weak-lensing masses measured by fitting the tangential shear profiles are biased low, concluding that the NFW model is actually a poor description of the actual shear profiles of clusters at the radii used in the fitting. At the radius enclosing an overdensity of $500$ times the critical density of the universe, $R_{500}$, they found that the bias amounts to $\sim 10\%$ for both clusters at $z=0.25$ and $z=0.5$.  They varied the integration length to see the dependence on large-scale structure on the deflection field. The scatter found using our integration length (i.e. 20 $h^{-1}$ Mpc) is comparable to ours.

Within the integration depth we chose, the large-scale structure can be considered correlated. If the integration length is larger, we will include also uncorrelated structures. 
Their effects on the weak lensing mass estimates have been discussed in detail in several papers: uncorrelated structures introduce a noise in the mass estimates and their contribution  to the total error budget is comparable to the statistical errors \citep{1997ApJ...485...39C,1999ApJ...520L...9M,2001A&A...370..743H,2003MNRAS.339.1155H,2004APh....22...19W, 2011MNRAS.tmp...72H}. \cite{becker&kravtsov}, specifically, 
 showed that the scatter in the weak-lensing masses of low-mass halos increases more than for high-mass halos as a function of line-of-sight integration length because the high-mass halos generate more shear than the low-mass halos. The large scale structure has different impact depending on the redshift of the lenses and on the depth of the observations \citep{2003MNRAS.339.1155H}.  

Finally, \cite{becker&kravtsov} also discussed how the scatter and the bias changes under varying  number density of background sources, $n_g$. They show that as the number density increases, the shape noise contribution (due to the intrinsic ellipticity of the sources) to the scatter decreases and eventually becomes subdominant with respect to the intrinsic scatter in weak-lensing mass measurements. For  clusters at $z=0.25$, the total scatter on $M_{500}$ changes from $\sim 37\%$ for $n_g=10$ gals arcmin$^{-2}$ to $\sim 25\%$ for $n_g=40$ gals arcmin$^-2$. As for the bias, they found that fitting the NFW functional form within $R_{500}$ can reduce the bias by $\sim 5\%$.
 
\subsection{X-ray}  

Regarding the X-ray analysis, we tested two different approaches in M10 that we dubbed the {\it backward} and {\it forward} methods. \\
The {\it backward} procedure  assumed a priori a functional form for the mass (such as NFW), spherical symmetry and hydrostatic equilibrium (Eq. 26 in M10): 
\begin{equation}
-G \mu m_p n_{\rm gas} M_{\rm tot}(<r)/r^2= dP/dr=d(n_{\rm gas} \times T)/dr 
\label{eq:he}
\end{equation}
where $G$ is the gravitational constant, $\mu=0.59$ is mean molecular weight in a.m.u., $m_p$ is the proton mass, $k$ is the Boltzmann constant, and $n_{\rm gas}$ and T are the gas density and temperature profiles. These are estimated at once by geometrically de-projecting the measured X-ray surface brightness and temperature data. The 3D temperature is computed following the recipe by \cite{mazzotta.etal.04}. More details can be found in \citep[][R06, M10]{ettori.etal.02, 2007MNRAS.379..518M}. \\
The {\it forward} method, instead, uses a complex parametric formulae to fit the projected surface brightness and temperature profiles. Subsequently, the analytic 2D expressions are de-projected assuming sphericity and finally the total mass is computed through Eq.\ref{eq:he}.\\
The two methods are based on the same basic hypothesis: spherical symmetry and hydrostatic equilibrium. They differ for the quantity they analytically parametrized. The first method imposes a fixed mass profile (usually NFW) while the second uses parametric formulas for the surface brightness and the temperature distributions (see Sec.5.2). In this way, the forward approach has smoother radial profiles to be derived, but also more parameters. 
The two procedures consistently  reconstruct both the total and the gas masses, as we demonstrated. For this reason, here, we limit our X-ray analysis to the forward method (presented with more detail later on the paper). The X-ray masses were shown to systematically underestimate the true mass of the simulated clusters by 5-20\% with an average bias of 10\% between $R_{2500}$ and $R_{200}$ and a scatter of 6\%. The gas masses reconstructed were usually 5\% higher than the true ones within the region with sufficient signal. Thanks to the high exposure time used (500 ks) and the field of view of the images, we compared the mass profiles up to $R_{200}$. 

M10 results were similar to the findings of \cite{nagai.etal.07}. In the same fashion, the authors created mock X-ray images of 16 objects simulated with an adaptive mesh code. The exposure time was 1 Msec and the field of view selected extended well beyond $R_{200}$. Processing the images, they followed the forward method. In their whole sample, the average difference between the total mass and the X-ray derived mass was 16\% at $R_{500}$ with a scatter of 9\%, reducing to 13\% $\pm$ 10\% for regular systems.

\cite{rasia.etal.06} studied a smaller sample of five objects. We follow the backward method assuming different parametrization for the total mass: $\beta$ model either isothermal or with polytropic temperature profile, NFW and the model presented by \cite{rtm}. Under the condition of perfect background subtraction, we found an averaged bias of 23\% at $R_{500}$  and $20.6\%$ at $R_{2500}$. The causes of the bias were double: the neglect contribution of the gas bulk motions to the total energy budget and the temperature bias towards lower values of the X-ray temperatures. The contribution of the last factor was confirmed by \cite{piffaretti&valdarnini} who analyzed more than 150 SPH-simulated clusters. Both papers found a temperature  bias of 10-15\% \cite[see also][]{rasia.etal.05}. \cite{ameglio.etal.09} pointed out the direct correlation between this bias and the cluster mass (or temperature): the bias is higher in most massive systems because they have a larger spread in temperature.

\section{Simulations}
\label{sec:sim}

Our analysis is based on 20 simulated clusters identified at $z=0.25$,
all having virial mass $M_{\rm vir}>5\times 10^{14}h^{-1}M_\odot$ at
that redshift, and each observed along three orthogonal projection
directions.  These clusters belong to the set of radiative simulations
presented by \cite{fabjan.etal.11}, whose initial conditions have been
described in details by \cite{bonafede.etal.11}. The Lagrangian
regions around each of these clusters have been identified within a
low--resolution N-body cosmological simulation, that followed $1024^3$
DM particles within a box having a comoving side of $1\,
h^{-1}$Gpc. The cosmological model assumed is a flat $\Lambda$CDM one,
with $\Omega_m=0.24$ for the matter density parameter, $\Omega_{\rm
  bar}=0.04$ for the contribution of baryons, $H_0=72\,{\rm
  km\,s^{-1}Mpc^{-1}}$ for the present-day Hubble constant, $n_s=0.96$
for the primordial spectral index and $\sigma_8=0.8$ for the
normalization of the power spectrum, thus consistent with the CMB
WMAP7 constrains \citep{komatsu.etal.11}.  Within each Lagrangian
region mass resolution is increased following the Zoomed Initial
Condition (ZIC) technique \citep{tormen.etal.97}. Resolution is
progressively degraded outside such regions, so as to
save computational time, while preserving a correct description of the
large--scale tidal field. Within the high--resolution region, it is
$m_{\rm DM}= 8.47 \times 10^8 M_{\odot} h^{-1}$ and $m_{\rm DM}= 1.53
\times 10^8 M_{\odot} h^{-1}$ for the masses of the DM and gas
particles, respectively.

Simulations have been carried out using the TreePM/SPH GADGET-3 code,
a newer and more efficient version of the GADGET-2 code originally
presented by \cite{sp05.1}. A Plummer--equivalent softening length
for the computation of the gravitational force in the high--resolution
region was fixed to $\epsilon=5\,h^{-1}$kpc in physical units at
redshift $z<2$, while being kept fixed in comoving units at higher
redshift. As for the computation of hydrodynamic forces, we assume the
SPH smoothing length to reach a minimum allowed value of
$0.5\epsilon$. Our simulations include metal--dependent radiative
cooling and cooling/heating from a spatially uniform and evolving UV
background, according to the prescription presented by
\cite{wiersma.etal.09}. Following the star-formation model by
\cite{springel&hernquist03} gas particles whose density exceeds a
given threshold value are treated as multi-phase particles, where a
hot ionized phase coexists in pressure equilibrium with a cold phase,
which is the reservoir for star formation. We also include a detailed
description of metal enrichment from different stellar populations,
using the model originally described by \cite{tornatore.etal.07}. The
effect of SN feedback is included through the effect of galactic winds
having a velocity of 500 km s$^{-1}$.

The cluster significant radii ($R_{2500}$, $R_{1000}$, $R_{500}$,
$R_{200}$, $R_{\rm vir}$)\footnote{$R_{\Delta}$ and $M_{\Delta}$ are
  the radius and the mass of the sphere whose density is $\Delta$
  times the critical density at the cluster redshift.} and the
corresponding masses are listed in Table~\ref{tab:clusters}.  To
compute these quantities we center on the minimum of the potential
well as done in \cite{rasia.etal.11}. In the following, we will refer
to these numbers as the {\it true} or the {\it intrinsic} values.

\begin{table*}[htdp]
\caption{True radii in $\hh$ kpc and masses in $\hh 10^{14} M_{\odot}$ at different overdensity ($\Delta= 2500,1000, 500, 200, vir$).} 
\centering
\begin{tabular}{lcccccccccccc}
\hline
\hline
cluster & $R_{\rm 2500}$ &   $R_{\rm 1000}$ & $R_{\rm 500}$  &$R_{\rm 200}$ & $R_{\rm vir}$  & $M_{\rm 2500}$ & $M_{\rm 1000}$& $M_{\rm 500}$  & $M_{\rm 200}$ & $M_{\rm vir}$ \\
\hline
CL1  & 388 & 669 & 989 &1561& 1988 & 2.089 & 4.277 & 6.900 &10.852 &12.394  \\
CL2  & 491 & 823 &1161 &1731 & 2241 & 4.227 & 7.948 & 11.170&14.796 & 17.76 \\
CL3    & 341 & 558 & 790 &1181& 1515 & 1.410 & 2.484 & 3.510 & 4.702 & 5.489    \\
CL4 & 314 & 513 & 747 &1204& 1615 & 1.099 & 1.923 & 2.974 & 4.979 & 6.641    \\
CL5  & 415 & 654 & 925  &1495& 1962 & 2.557 & 3.985 & 5.637 & 9.518 &11.921   \\
CL6    & 437 & 719 & 1010&1557& 2048 & 2.966 & 5.296 & 7.342 &10.772 &13.543 \\
CL7  & 396 & 656 & 934 & 1476 & 1949& 2.218 & 4.021 & 5.807 & 9.169 &11.698  \\
CL8    & 404 & 655 & 921 & 1487 & 1951 & 2.357 & 4.003 & 5.563 & 9.367 &11.719  \\
CL9    & 372 & 615 & 857 & 1277& 1647 & 1.830 & 3.315 & 4.480 & 5.941 & 7.046   \\
CL10    & 393 & 708 &1052 &1637 & 2075 & 2.163 & 5.051 & 8.299 &12.514 &14.091 \\
CL11    & 458 & 739 &1019 &1528 & 1943 & 3.427 & 5.751 & 7.546 &10.187 &11.565 \\
CL12    & 317 & 568 & 836 & 1343 & 1763 & 1.131 & 2.617 & 4.171 & 6.902 & 8.640   \\
CL13  & 304 & 541 & 868 & 1405 & 1827 & 1.005 & 2.257 & 4.655 & 7.913 & 9.621   \\
CL14    & 452 & 723 & 998 & 1503 & 1930 & 3.289 & 5.381 & 7.079 & 9.686 &11.346  \\
CL15    & 373 & 608 & 902 & 1467 & 1965 & 1.847 & 3.200 & 5.238 & 9.008 &11.971  \\
CL16    & 400 & 653 & 911 & 1392 & 1822 & 2.278 & 3.970 & 5.392 & 7.691 & 9.547   \\
CL17    & 370 & 616 & 892 & 1459 & 1891 & 1.809 & 3.332 & 5.052 & 8.863 &10.662  \\
CL18    & 277 & 475 & 700 & 1147 & 1504 & 7.584 & 1.528 & 2.444 & 4.303 & 5.365   \\
CL19    & 289 & 513 & 780 & 1249 & 1585 & 0.858 & 1.922 & 3.380 & 5.551 & 6.279   \\
CL20    & 403 & 660 & 920 & 1410 & 1858 & 2.337 & 4.092 & 5.544 & 7.993 &10.121  \\
\hline
\hline
\end{tabular}
\label{tab:clusters}
\end{table*}%

\section{Weak Lensing{sec:lensing}}
\label{sec:lensing}


\subsection{{\tt SkyLens} simulations}
To simulate their lensing effects on a population of background sources, we process the halos using our well tested optical simulation pipeline {\tt SkyLens} \citep[e.g.][and M10]{2008A&A...482..403M,meneghetti.etal.11}. Here, we briefly summarize the basic steps toward the realization of the simulated images and refer the reader to those papers for further details.

We begin selecting particles falling into a cylinder centered on the cluster and having its width and depth set equal to $10$ and $20 \;h^{-1}$Mpc, respectively. This ensures to include in the simulation the effects of filaments apart from the cluster and of additional mass clumps that could produce addition shear signal. Since we are focusing on high resolution re-simulated clusters, we do not include the effects of un-correlated large-scale-structures.  
As matter of fact, the importance of matter along the line of sight is fairly small for rich clusters at intermediate redshifts, like those in our sample, provided that the bulk of the sources are at high redshift compared to the cluster (see Section~\ref{sec:intro}).

We project the mass distribution (i.e. the selected particles) on a {\em lens plane} at the redshift of the cluster, $z_{\rm L}=0.25$. For each cluster in the sample we derive three lens planes, corresponding to the projections (named 1, 2, and 3) along the three axes of the simulation box. The final number of lens planes used in this study is 60. This is a factor of $\sim 7$ larger than the sample previously investigated in M10.

The deflection field of each cluster is determined by tracing a bundle of $4096\times4096$ light-rays from the observer position through the lens plane (see M10 for the description of the tree-code). The final deflection matrix is used to further trace the light rays toward the background sources, allowing us to reconstruct their distorted images.  In short, the code uses a set of real galaxies decomposed into shapelets \citep{RE03.1} to model the source morphologies on a synthetic sky. In the current version of the simulator, the shapelet database contains $\sim 3000$ galaxies in the $z$-band from the GOODS/ACS  archive \citep{GIA04.1} and $\sim 10000$ galaxies in the $B,V,i,z$ bands from the {\it Hubble-Ultra-Deep-Field} (
HUDF)  archive \citep{BECK06.1}. Most galaxies have spectral classifications and photometric redshifts available \citep{2000ApJ...536..571B,2006AJ....132..926C}, which are used to generate a population of sources whose luminosity and redshift distributions resemble those of the HUDF.

{\tt SkyLens} allows us to mimic observations with a variety of telescopes, both from space and from the ground. For this work, we simulate wide field observations, on which we carry out a weak lensing analysis, using the SUBARU Suprime-Cam. All simulations include realistic background and instrumental noise. The galaxy colors are realistically reproduced by adopting $22$ SEDs to model the background galaxies, following the spectral classifications published by \cite{2006AJ....132..926C}.

Compared to M10, we use here a different setup. First, we assume an exposure time of $2000$s in the $I$-band, which is a factor of three shallower than in M10. This is aimed at testing the weak lensing analysis under more realistic conditions.  Second, we use real stars observed with SUBARU to model the PSF. The PSF model is characterized by a FWHM of $\sim 0.6^{\prime\prime}$. M10 used an isotropic gaussian PSF instead. For all lens planes, we produce wide-field images covering a region of $2400^{\prime\prime}\times2400^{\prime\prime}$ around the cluster center. This allows us to measure the shear signal up to a distance of $\sim 3.5 \;h^{-1}$Mpc at $z=z_{\rm L}$, well beyond the virial radius of any cluster in the sample.

\subsection{Weak-lensing analysis}
\label{sect:wl}
The weak lensing measurements are done using the standard Kaiser-Squires-Broadhurst (KSB) method, proposed by \cite{1995ApJ...449..460K} and subsequently extended by \cite{1997ApJ...475...20L} and by \cite{1998ApJ...504..636H}. The galaxy ellipticities are measured from the quadrupole moments of their surface brightness distributions, corrected for the PSF, and used to estimate the reduced shear under the assumption that the expectation value of the intrinsic source ellipticity vanishes.

In this study we use the publicly available pipeline {\tt KSBf90} by C. Heymans\footnote{\tt http://www.roe.ac.uk/$\sim$heymans/KSBf90/Home.html} to process our images and measure the shear fields. The final galaxy catalogs are constructed by selecting only the galaxies with signal-to-noise ratio $S/N>10$ (as provided by {\tt SExtractor}, \citealt{BE96.1}), half-light radius larger than 1.15 times the PSF size, and reduced shear $|g|<1$. Given the above mentioned exposure time and seeing conditions, the effective number density of galaxies in the final shear catalogs is $\sim 17$ arcmin$^{-2}$.  In observations exploiting a depth similar to our simulated images, the number of sources available for the lensing analysis may be smaller because the light emission from cluster galaxies (not included in these simulations) are potential contaminants. These non-lensed galaxies bias low the lensing signal if they are accidentally included in the shear catalogues. Color based techniques \citep[e.g.][] {medezinski.etal.07, medezinski.etal.10} allow to separate the foreground and the background galaxy populations efficiently, but, to be conservative, several sources which may have a dubious classification are usually excluded from the lensing catalogs. Among them several background galaxies.

In M10, we considered several methods to measure the total mass using the observed shear field. As discussed above, we found that the most precise mass measurements are obtained by combining weak and strong lensing non-parametrically \citep[see e.g.][]{2009A&A...500..681M}. The disadvantage of this approach is that it is very expensive both in terms of time needed to carry out the analysis and in terms of data requirements. The identification of the strong-lensing features, used to constrain the model in the inner region, usually requires deep and high-resolution Hubble-Space-Telescope imaging. Moreover, strong-lensing clusters are relatively rare and known to be affected by many biases \citep{meneghetti.etal.10,meneghetti.etal.11,Hennawi:2007fj}. Fitting the tangential shear profiles with functionals describing the cluster density profiles is a very common and easy alternative to measure the mass \cite[e.g.][]{CL02.1,Hoekstra:2000cq,2005ApJ...618...46J,2006ApJ...653..954D,PaulinHenriksson:2007ex,Bardeau:2007jj,2009ApJ...699.1038O,2007ApJ...671.1466K,2010PASJ...62..811O,2010A&A...514A..88R,2011ApJ...729..127U,Zitrin:2011ut}. Further, this method can be applied to clusters down to relatively small mass limits and in absence of strong lensing features.

Here, we assume that the density profiles of clusters are well described by the Navarro-Frenk-White profile \citep{nfw},
\begin{equation}
    \rho_{\rm NFW}(r)=\frac{\rho_s}{r/r_s(1+r/r_s)^2}
    \label{eq:nfw}       
\end{equation}
where $\rho_s$ and $r_s$ are the characteristic density and the scale radius, respectively.
The characteristic density is often written in terms of the concentration parameter, $c_{200}=r_{200}/r_{s}$, as
\begin{equation}
\rho_s=\frac{200}{3}\rho_{\rm cr}\frac{c_{200}^3}{[\ln(1+c_{200})-c_{200}/(1+c_{200})]} \ .
\label{equation:deltacpar}
\end{equation}
We derive the mass by fitting the one-dimensional reduced tangential shear profile with the corresponding NFW functional \citep{BA96.1,WR00.1,ME03.1}.

The tangential shear profile is derived from the data by radially binning the galaxies and averaging the tangential component of their ellipticity within each bin. The tangential and cross components are, respectively, defined as
\begin{equation}
    \epsilon_+=-{\rm Re}[\epsilon \;{\rm e}^{-2i\phi}] \; \; \mathrm{and} \;  \;   \epsilon_\times=-{\rm Im}[\epsilon \;{\rm e}^{-2i\phi}] \;.
\end{equation}
The angle $\phi$ specifies the direction from the galaxy centroid towards the center of the cluster, which we identify with most bound particle in the simulation. When averaging over many galaxies, the expectation value of the intrinsic source ellipticity vanishes, and the reduced tangential shear is given by $g_+=\langle\epsilon_+\rangle$.
On the contrary, in absence of systematics the averaged cross component of the ellipticity should be  zero.


\section{X-ray}
\label{sec:xray}

\subsection{\tt{X-MAS} simulations}

 Before producing the X-ray synthetic catalogue, we have applied the technique described in Appendix A to remove over-cooled particles.
%
Subsequently, our clusters are processed through {\tt X-MAS} to obtain \chandra\ mock images. The characteristics of this software package are described in detail in other works \citep{xmas,rasia.etal.08}.  
To create the photon event file, we assumed the Ancillary Response Function (ARF) and Redistribution Matrix Function (RMF) typical of the ACIS-S3 detector aimpoint.
We consider the redshift as that of the simulated time frame ($z=0.25$) and the metallicity constant and equal to 0.3 solar in respect to the tables by \cite{anders&grevesse}\footnote{The Helium abundance used in the plasma emission was modified from 9.77e-02 to 7.72e-2 to be consistent with the hydrogen mass fraction used as input in GADGET-3 code. }. The field of view of our images has a side of 16 arcmin. For our cosmology and redshift, this corresponds to 2561 $h^{-1}$ kpc. All the clusters have their $R_{500}$ regions within the field of view (see Table~\ref{tab:clusters}), even if some of them at that radius do not emit a sufficient number of photons to allow a precise spatial and spectral analysis (see more in the next session).  We account for the emission by all the particles within a depth of 10 Mpc along the line of sight direction and centered on the cluster. The exposure time chosen is 100 ksec. This setting differs from what adopted in M10: the reduction of the exposure time  allows a more realistic comparison with observed data. In the final event files, we add a contribution for the galactic absorption by a {\em WABS} model  with $N_H=5 \times 10^{20}$ cm$^{-2}$. As in M10, we do not include the influence of the background since R06 proved that its net effect is to enlarge the error on the mass estimates without introducing an extra bias. Furthermore, new background models are capable to predict the spatial variation of the Chandra background with an accuracy  better than 1\%  (Bartalucci et al. in preparation). 


  We note that tools as {\em X-MAS} are not suitable to address
  calibration problematics since the same response files are used both
  to create and to analyze the data. In this sense, in our analysis we
  assume a perfect knowledge of the instrument calibration and the
  results do not depend on the instrument reproduced. In the analysis
  of real observational data, systematic instrumental uncertainties
  are highly important, in particular in situation of high statistics
  (high number of counts). To treat them correctly one needs to
  include them in the analysis.  \cite{lee.etal.11} have recently
  provided a bayesian statistical method to tackle this problem.

\subsection{X-ray analysis}
\label{sec:x_analys}
Using {{\em CIAO}} tool \citep{fruscione.etal.06} we extract soft band  images in the [0.7 $-$ 2] keV band. We apply the wavelet algorithm of \cite{vikh.etal.98} to identify clumps. These and any major substructure have been masked and excluded from the following analysis.
The surface brightness profiles are centered in the X-ray centroid \citep{rasia.etal.11} and account 15-30 linearly spaced annuli with at least 100 counts. The innermost annulus is selected outside the central 10\% of $R_{500}$, the outermost one is always beyond $R_{1000}$ and reaches  $R_{500}$ in the majority of the cases (see Table~\ref{tab:class}).  
The radial coverage is comparable to recent observations, some of
  which extend beyond $R_{500}$ \citep{neumann05,
    ettori&balestra,leccardi&molendi,2009ApJ...692.1033V}.
The temperature profile is calculated in 6-10 annuli spanning over the same radial range of the surface brightness profile. The minimum number of photon per temperature annulus is 1000. The spectra are grouped and fitted by a single-temperature MEKAL model in the XSPEC  package\citep{arnaud96}. The statistics used is $\chi^2$ and the energy band considered is [0.8-7] keV. In the pipeline the values of galactic absorption, redshift, hydrogen column density and metallicity are fixed equal to the input ones.

To compute the total mass from the X-ray analysis, we follow the ``forward" method of M10 \cite[see also,][]{vikh.etal.06}. The surface brightness and the temperature profiles are fitted by the analytic formulae:
\begin{equation}
n_p n_e = n^2  \frac{(r/r_{c})^{-\alpha}}{[1+(r/r_{c})^2]^{3\beta-\alpha/2}}  \frac{1}{[1+(r/r_s)^{\gamma}]^{\epsilon/\gamma} }; \ \
T=T_0 \frac{(r/r_t)^{-a}}{[1+(r/r_t)^b]^{c/b}}.
\label{eq:xfit}
\end{equation}
Since we exclude the cluster central part from our analysis, we do not
model the cooling core region as done in \cite{vikh.etal.06}. 
  This excision is common in both simulations
  \citep[e.g. M10,][]{nagai.etal.07} and observations
  \citep[e.g.][]{2009ApJ...692.1033V,ettori.etal.10} to avoid, in the
  former case, the influence of the overcooled central region and, in
  the latter case, the presence of central active galaxy, cool-core
  regions, gas sloshing.  The 2D analytic formulae are deprojected
and the total mass is recovered by assuming hydrostatic equilibrium,
that from Eq. 1 it can be written as
\begin{equation}  
M_{\rm X}(<r) = -\frac{kT(r) r}{G \mu m_{\rm p}}   
\left( \frac{d \ln \rho}{d \ln r} + \frac{d \ln T} {d \ln r} \right) \ ,  
\label{eq:hydr}  
\end{equation}  
where $T$ and $\rho$ are the deprojected 3D analytic profiles.
Following this procedure, we obtain the X-ray mass that we compare
with the true mass of the simulated cluster.  The uncertainties in
  the estimate of this mass, $eM_{\rm X}$, were obtained through Monte
  Carlo simulations. In each Monte Carlo realization, surface
  brightness and temperature profiles were varied within their
  measured errors. Each time a new mass was then derived. The
  resulting uncertainty were defined as the standard deviation
  computed over 100 such realizations.


\section{Results } \label{sec:results}

\subsection{Weak-lensing mass estimates}

Weak-lensing allows us to measure the mass of the cluster projected on the plane of the sky. The NFW analytic formula of the integral along the line of sight of the mass contained in a cylinder is 
$M(R_{2D})=4 \rho_s r_s^3  h(x)$, where $x=R_{2D}/r_s$ and 
\begin{equation}
h(x) = \ln \frac{x}{2} +
\left\{
\begin{array}{r@{\quad \quad}l}
 \frac{2}{\sqrt{x^2-1}}\,
 \mbox{arctan}\sqrt{\frac{x-1}{x+1}} & (x>1) \\
 \frac{2}{\sqrt{1-x^2}}\,
 \mbox{arctanh}\sqrt{\frac{1-x}{1+x}} & (x<1) \\
 1 & (x=1)
\end{array} \right.
\;.
\nonumber
\end{equation}
The profile parameters $r_s$ and $\rho_s$ are obtained from fitting the tangential shear profile, as discussed above.

Unfortunately, in most cosmological applications the projected mass is not the interesting quantity. Rather, we need to measure the three-dimensional mass . To derive it 
by de-projecting the two-dimensional mass, one needs to make {\em strong} assumptions about the shape of the cluster, which is usually assumed spherical. 
In this case, the NFW model is given by
\begin{equation}
M(r) = 4\pi r_{s}^2\rho_s\left[\ln(1+y)-\frac{y}{1+y}\right]\;,
\end{equation}
where $y=r/r_s$.
 To both the 2D and the 3D masses we associate errors, $eM_{WL,2D}$ and $eM_{WL,3D}$, computed by propagating the errors on $r_s$ and $\rho_s$, as obtained from fitting the tangential shear profiles.

\begin{figure}
\begin{center}
\includegraphics[width=0.49\textwidth]{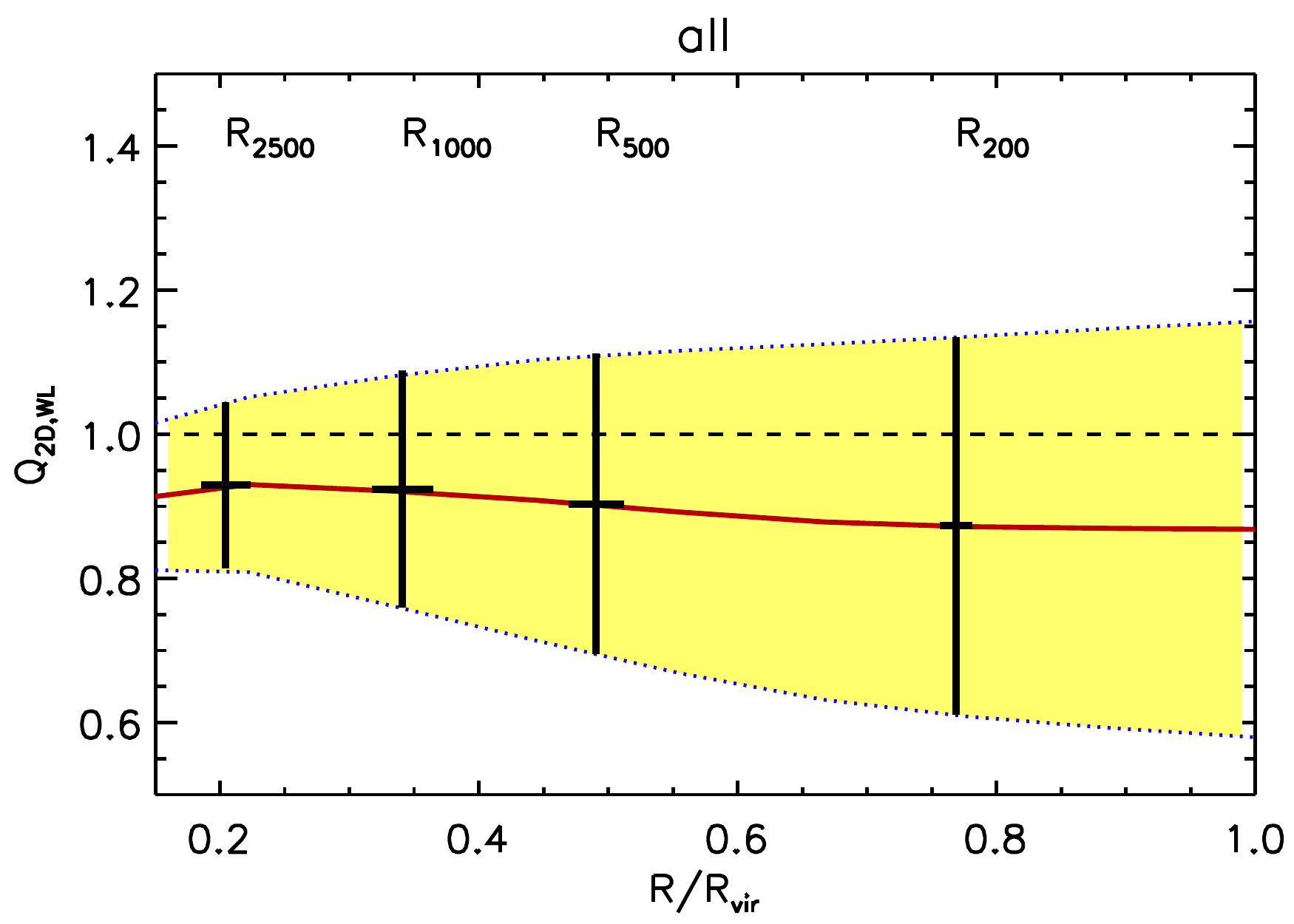}
\includegraphics[width=0.49\textwidth]{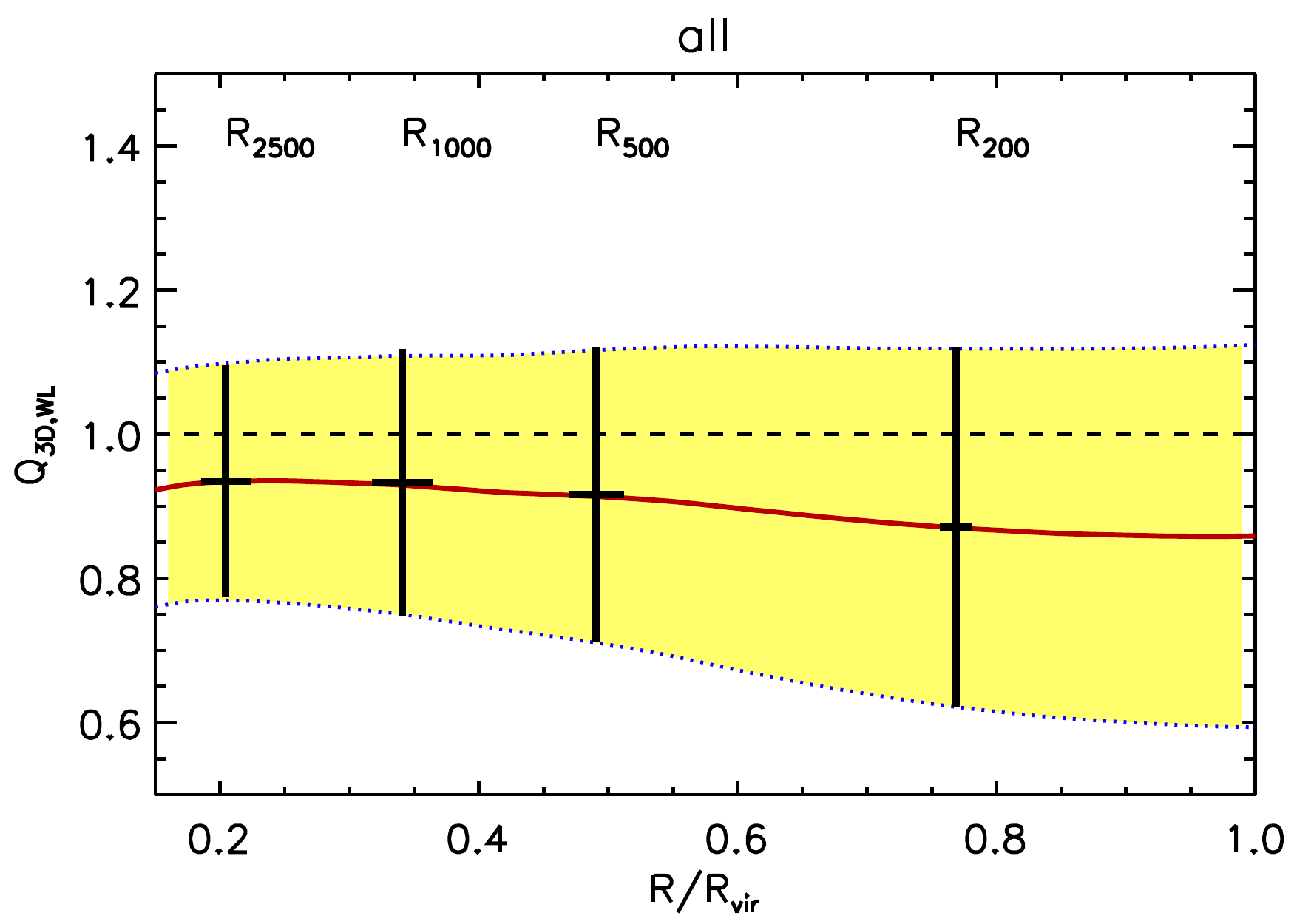}
\caption{Comparison between weak-lensing and true masses using all clusters in the sample (60 lens planes). The solid lines in the left and right panels show the average ratios between 2D- and the 3D-weak-lensing masses and the true masses, respectively. These are plotted as a function of the distance from the cluster center in units of the virial radius. The crosses in each panel mark the average locations of various over-density radii and their amplitude is the weighted average bias at each significant radius. The yellow-shaded region marks the standard deviation at each radius.}
\label{fig:wlall}
\end{center}  
\end{figure}

In the following, we show how well we measure projected and de-projected
mass profiles of the clusters in our sample. In both cases, the
quality of the mass measurement is assessed by means of the ratio
$Q_{\rm WL}$ between measured and true mass: $Q_{\rm WL}=M_{\rm
  WL}/M_{\rm true} $. The uncertainty on this ratio,  $eM_{WL}/M_{\rm true}$, accounts only for the errors in the weak-lensing mass since the true mass is perfectly known from
  simulations.   The weighted-mean of both the 2D and the 3D
  weak-lensing bias radial profiles, $<Q_{\rm WL}>$, is shown by the solid red line in
  Fig.~\ref{fig:wlall}. Its scatter is quantified by the standard
  deviation of its distribution, and is represented by the shaded yellow region. 
    In formulae:
    \begin{equation}
    <Q_{\rm WL}> = \frac{\Sigma_i Q_{\rm WL,i} (R/R_{\rm vir,i}) \times eM_{\rm WL,i}}{ \Sigma_i eM_{\rm WL,i}} \ {\rm and} \  \  {\rm scatter=}\left[\frac{\Sigma_i (Q_{\rm WL,i} (R/R_{\rm    vir,i})-<Q_{\rm WL}>)^2 \times eM_{\rm WL,i}}{\Sigma_i eM_{\rm WL,i}} \right]^{0.5}
    \end{equation}
    
     The
  profiles are plotted in units of $R_{vir}$. The over-imposed crosses
  refer to the weighted averaged $Q_{\rm WL}$ computed at the
  significant radius of each object, with average computed at a radius
  corresponding to a fixed overdensity $\Delta$, $(R_{\rm
    \Delta,i}/R_{\rm vir,i})$, and not over the whole radial profile
  ($R/R_{\rm vir,i}$). They are located at the respective averaged
over-density radii. The cross horizontal bars show the dispersion
around the radii in units of $R_{\rm vir}$.  The quantitive version of
Fig.~\ref{fig:wlall} is reported on Table~\ref{tab:bias} where we
resume all our results. Each value of $Q_{3D, \rm WL}$ and its
  corresponding uncertainty is listed in Table ~B~6 of Appendix B.

Two important conclusions emerge from this analysis. First, the mass measured fitting the reduced tangential shear profile with an NFW functional is biased low. The bias amounts to $\sim 7-10\%$ between $R_{2500}$ and $R_{500}$ and reaches $\sim13\%$ for larger distances from the cluster center. Second, the scatter in both two- and three-dimensional masses ranges between $\sim 10\%$ at small radii and $\sim 25\%$ at larger radii, being 20\% at $R_{500}$. 

These results agree with the findings of M10 and \cite{becker&kravtsov}, confirming that the weak-lensing analysis via the KSB pipeline does not introduce significant systematics

\begin{table*}[htdp]
\caption{Weighted average mass bias, $Q_{WL}$ and $Q_X$, and their standard deviation for the whole sample and the different sub-samples based on X-ray and environmental classification. The environmental classification is performed on the visual inspection of both intrinsic simulated maps, I. Poor, and on optical synthetic images, O. Poor (see Section~7 for details). }
\centering
\begin{tabular}{|c||c|c||c|c||c|c||c|c||c|c|}
\hline
\multicolumn{9}{|c|}{}\\
\multicolumn{9}{|c|}{$(1-Q_{2D,\rm WL})\times 100$} \\
\hline
radius &\multicolumn{2}{|c||}{all cluster}  & \multicolumn{2}{|c||}{regular}  & \multicolumn{2}{|c||}{I. Poor}   & \multicolumn{2}{|c||}{O. Poor}\\
&bias & rms &bias & rms & bias &rms & bias &rms \\
\hline
$R_{2500}$ &7.0& 11.5  &7.9 & 11.0       & 4.4  & 4.3   & 7.0& 9.5 \\
$R_{1000}$ & 7.6& 16.5 &8.7 & 15.8 &       1.7 &3.3 & 6.0  & 12.6\\
$R_{500}$   & 9.7& 20.8 & 10.1 & 19.5    & 0.0 & 5.0       & 4.9 & 16.4\\
$R_{200}$   & 12.7& 26.2&12.8 & 23.1    & -4.2 & 7.0     & 4.1 &  22.2\\

\hline
\multicolumn{9}{|c|}{}\\
\multicolumn{9}{|c|}{$(1-Q_{3D,\rm WL})\times 100$} \\
\hline
radius &\multicolumn{2}{|c||}{all cluster}  & \multicolumn{2}{|c||}{regular}  & \multicolumn{2}{|c||}{I. Poor}   & \multicolumn{2}{|c||}{O. Poor}\\
&bias & rms &bias & rms & bias &rms & bias &rms\\
\hline
$R_{2500}$ &  6.5 & 16.1 & 6.9 & 9.5 & 3.0& 13.3 & 6.2 & 15.2  \\
$R_{1000}$ &  6.7 & 18.5 & 8.2 & 12.8 & 4.5 & 12.2 & 5.2 & 16.4 \\
$R_{500}$   &  8.4 &  20.5 &8.9 & 16.9 &3.5& 11.3 & 4.8 & 16.7  \\
$R_{200}$   & 12.8 & 25.0 &13.3 & 22 &0.0 & 9.8 &      5.8 & 20.0  \\
\hline
\multicolumn{9}{|c|}{}\\
\multicolumn{9}{|c|}{$(1-Q_{X})\times 100$}\\
\hline
radius &\multicolumn{2}{|c||}{all cluster}  & \multicolumn{2}{|c||}{regular}  & \multicolumn{2}{|c||}{I. Poor}   & \multicolumn{2}{|c||}{O. Poor}\\
&bias & rms &bias & rms & bias &rms & bias &rms \\
\hline
$R_{2500}$ & 23.9 & 11.0& 19.0 & 7.6 & 21.9 & 4.9        & 20.8 & 8.2 \\
$R_{1000}$ & 27.5 & 7.9   & 25.6 & 7.8 & 22.5 & 2.2      & 26.4 & 6.2 \\
$R_{500}^*$    & 33.0 & 9.4   & 34.4 & 10.4 & 26.1 & 7.7 & 33.1 & 8.8\\

\hline

\multicolumn{9}{c}{$*$ The X-ray measures are extrapolated for some clusters.}\\
\end{tabular}
\label{tab:bias}
\end{table*}%

\subsection{X-ray mass estimates}

Contrarily to the optical mass measurements, the X-ray mass derivation gives directly the 3D mass profile. Therefore, we can straightly define the ratio between the X-ray mass and the true ones: $Q_X=M_{\rm X}/M_{\rm true}$.
\begin{figure}
\begin{center}
\includegraphics[width=0.49\textwidth]{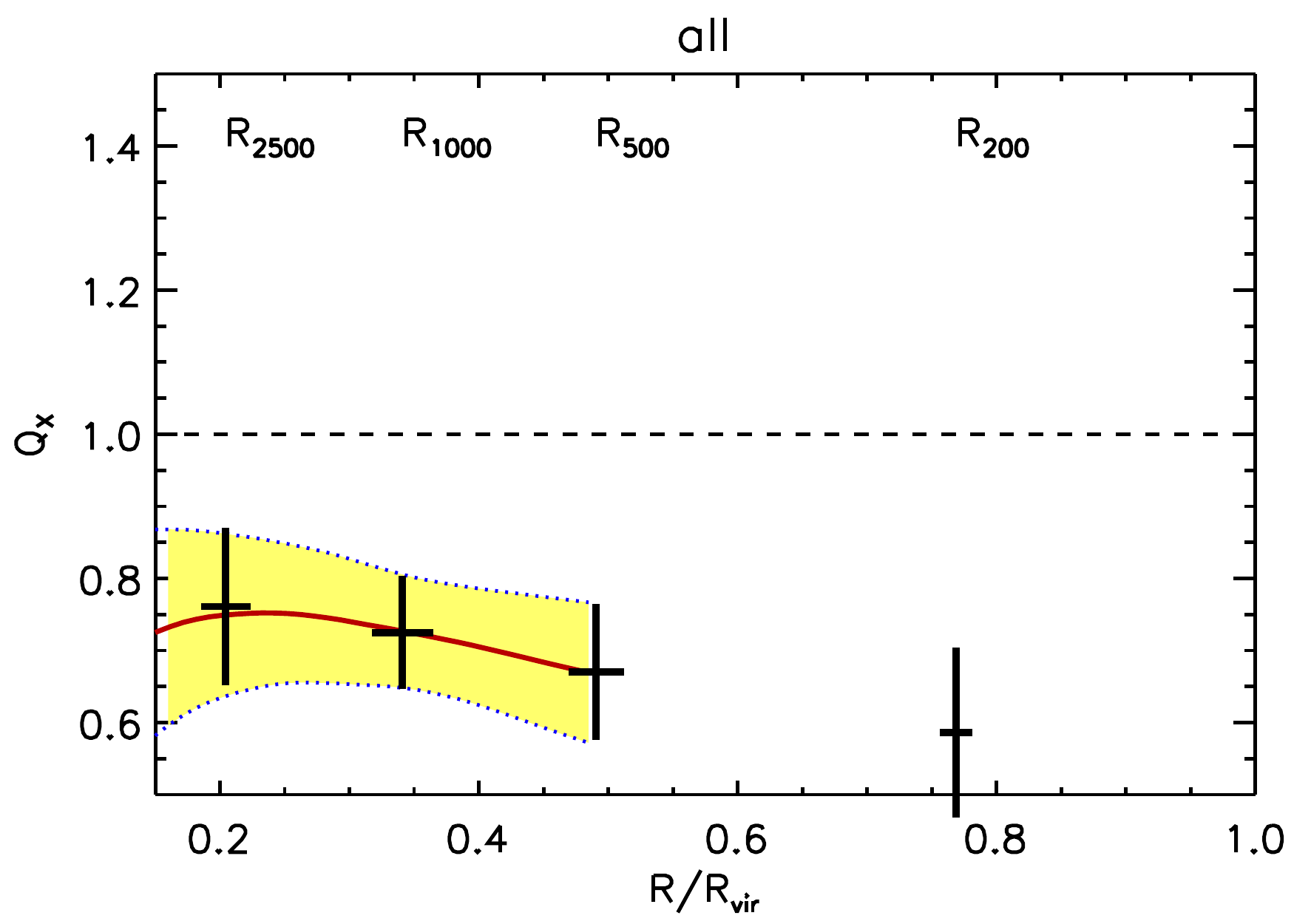}
\caption{Comparison between X-ray and true masses using the whole sample. The meaning of lines, crosses and shaded regions is the same of Fig.\ref{fig:wlall}.}
\label{fig:xall}
\end{center}
\end{figure}
Similarly to weak lensing analysis, we compute the weighted average of
$Q_X$ over the whole sample and the standard deviation of the
distribution (Eq.~9). In Fig.~\ref{fig:xall}, we plot the values only within
$R_{500}$, which is the radius reached by most of the surface
brightness and temperature profiles (Table~\ref{tab:class}).The cross
shown at $R_{200}$ is the result of the extrapolation of the analytic
formulae. On the third part of Table~\ref{tab:bias}, we report the
weighted-averaged $Q_X$ and its scatter. Each $Q_X$ value and its
  corresponding uncertainty is listed in Table~B~7 of Appendix B. 

Fig.~\ref{fig:xall} confirms some previous findings that  we will synthesize here postponing to Section~\ref{sec:conclu} a more profound discussion.\\
The average X-ray mass is consistently underestimating  the true mass. The X-ray bias is around 25\%  at the center and 30-35\% at $R_{500}$.
 The decline in the most external regions is expected since the cluster outskirts present a more dramatic lack of hydrostatic equilibrium \citep{lau.etal.09} and a stronger influence of gas clumpiness \citep{nagai.etal.11}. The presence of gas clumps affects the X-ray mass determination in two ways: it shallows the surface brightness profile and it cools X-ray temperature (see more on this in the Section 8.)
Massive systems, as the ones studied in this paper, are expected to be still accreating and therefore far for an equilibrium state. Moreover, the temperature bins in the external regions, where the temperature profile declines more steeply, are usually larger containing more temperature structure. Finally,
the large bias on the most external region has to be taken with caution since it is not the result of a measurement but of an extrapolation. The dispersion around the average is quite small.
The standard deviation is less than 10\% at all radii apart from $R_{2500}$ where it is 12\%. These numbers are twice or three times smaller than the gravitational lensing dispersion.

\section{Cluster classification} \label{sec:class}

We investigate in this Section the efficiency in reducing bias and scatter on both X-ray and gravitational lensing masses of two selecting criteria. We create different sub-samples determined by the morphology of the X-ray images or by the presence of substructures on their environment.

\subsection{Masses and X-ray morphology}

To limit the impact of the non-thermal processes on the X-ray mass estimates, clusters are often selected on the basis of their appearance.
The literature is rich of studies where clusters have been classified into {\em relaxed}, or {\em regular}, and {\em unrelaxed}, or {\em disturbed}, because of their X-ray morphology \citep[e.g.][]{Zhang.etal.08,2009ApJ...692.1033V}. Most of the time, the classification is done ``visually'', i.e. simply quantifying the regularity of an object from the X-ray image in the soft band.
More objective criteria, proposed in the past, are the {\em power ratios}, {\em centroid-shift}, {\em asymmetry and  fluctuation parameters}, and {\em hardness ratio}. We test all of them and present here our result.

\paragraph{Third order power ratio and centroid shift.}

\cite{buote&tsai} suggested to  decompose the surface brightness distribution in multipoles. The high order multipoles, usually normalized by the monopole and called  {\em power ratios},  are used to quantify the contribution of different scale components (asymmetries and substructures) to the surface-brightness power spectrum relative to the large-scale smooth cluster emission.  Most information in the power spectrum is contained in the first four multipoles.  $P_0$ is the monopole. The power ratio $P_1/P_0$ measures the dipole of the X-ray emission, which is zero if measured with respect to the X-ray centroid. The power ratio $P_2/P_0$ measures the ellipticity (quadrupole).  The third order power ratio $P_3/P_0$ can be used to quantify asymmetries and is the best indicator of clusters with multimodal distributions. Substructures on smaller scales contribute to higher order multipoles.

Another indicator of the dynamical state and of the asymmetry of the X-ray emission is the {\em centroid-shift},  i.e. the shift of the surface brightness centroid in apertures of increasing size. This parameter points out the dynamical state of the cluster as well as the asymmetry. Following \cite{poole.etal.06} and \cite{maughan.etal.08}, we define the centroid-shift  as
\begin{equation}
w=\frac{1}{R_{\rm max}} \times \sqrt{ \frac{\sum_i (\Delta_i-\langle \Delta \rangle)^2}{(N-1)}},
\end{equation}
where $R_{\rm max}$ is the radius of the largest aperture, and $\Delta_i=\vec{R}_{c,i}-\vec{R}_{c,{\rm max}}$
is the shift of the centroid in the $i-$th aperture with respect to the centroid in the largest aperture, $\vec{R}_{c,\rm max}$. $\langle \Delta \rangle$ is the mean value of the various $\Delta_i$ and the sum is done over all the $N$ apertures with radii up to $R_{\rm max}$.   In this work we assumed $N=17$ apertures with radii ranging between $R_{\rm min}=0.15 \times R_{500}$ and $R_{\rm max}=R_{500}$.

 The third--order power ratio and the centroid shift were shown to
  be effective in classifying clusters by two recent works by
  \cite{cassano.etal.10} and \cite{boehringer.etal.10}. Clusters are
  located in a rather well defined region in the $P_3/P_0-w$ plane:
  objects with small centroid shift and small $P_3/P_0$ are classified
  as ``regular''. The majority of them are cool core systems, not very
  dynamically active and showing absence or very little radio
  emission. For all these reasons, often, these objects are referred
  as ``relaxed''.

 In this work, we compute the power ratio, $P_3/P_0$, and the
 centroid-shift, from the signal of the region within $R_{500}$ of the
 masked images. With this attention, we aim to evaluate the
 ``irregularity'' of the actual portion of the image that we use to
 retrieve the mass.  Both the values and their uncertainties are
   derived from Monte Carlo simulations. We create 100 new images
   where the photons are re-distribuited accordingly to a Poisson
   statistics. We evaluate the estimators in each image. Finally, we
   extract the medians and the $16^{th}$ and 
$84^{th}$ 
percentile of the Monte Carlo distributions to represent the final
values of the morphological estimators and their uncertainties.
 
\begin{figure}
\begin{center}
\includegraphics[width=0.45\textwidth]{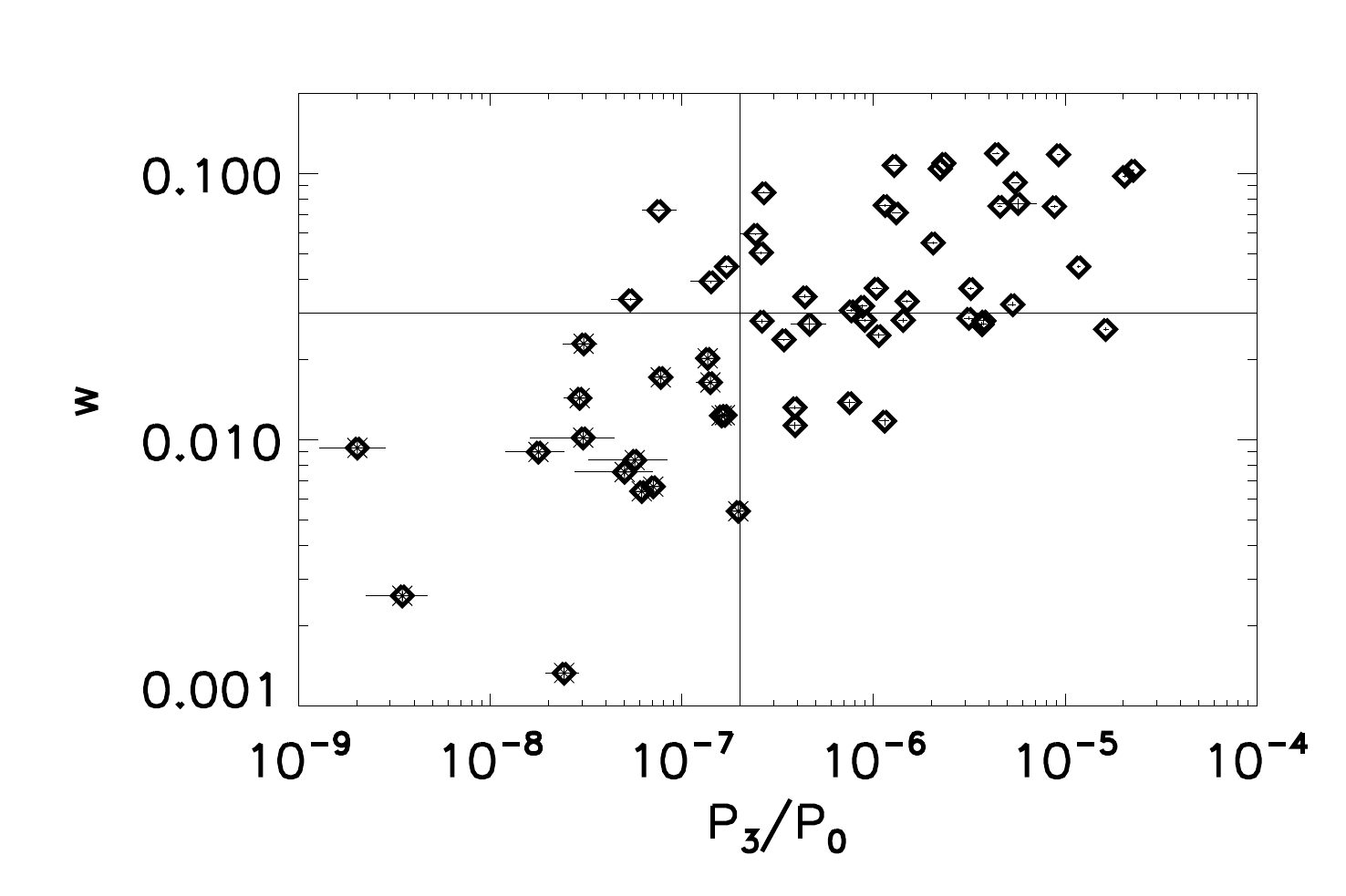}
\includegraphics[width=0.4\textwidth]{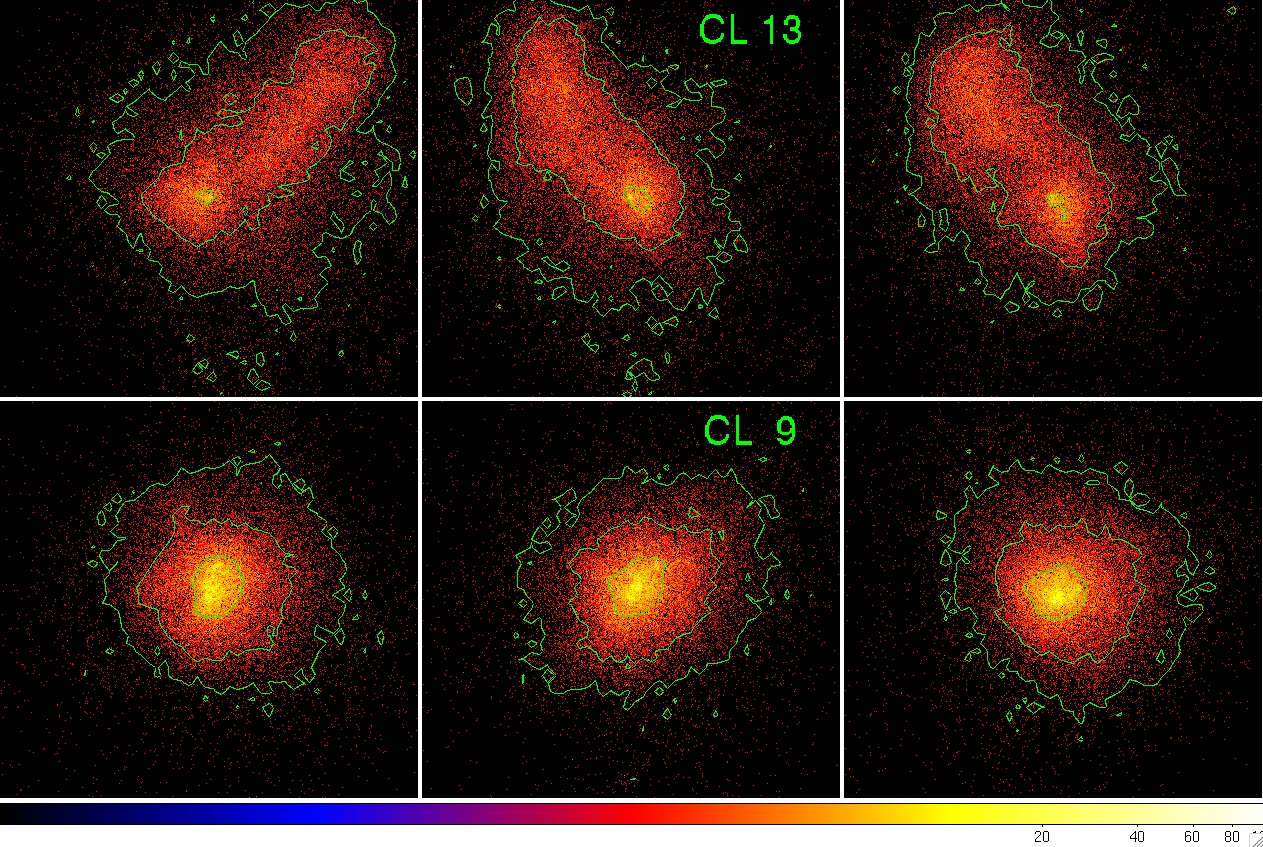}
\caption{On the left: the distribution of clusters in the $P_3/P_0-w$ plane.  The asterisks represent clusters classified as regular. On the right: soft X-ray images of a disturbed cluster and a regular one seen along the three projections: CL13 on the top and CL9 at the bottom. To emphasize the morphology we over-plotted the iso-flux contours in green. }
\label{fig:pw}
\end{center}  
\end{figure}  The third order power ratios and the centroid shifts of
  our sample with their uncertainties are shown in the left panel of
  Fig.~\ref{fig:pw}. We recognize the region with {\it regular}
  systems by slightly relaxing the criteria adopted by
  \cite{cassano.etal.10}. In our work, we define a cluster to be
  regular when $w<0.03$ and $P_3/P_0 < 2\times 10^{-7}$. Our choice is
  motivated by the fact that our aperture is equal to $R_{500}$, thus
  larger than the 500 kpc aperture radius analyzed by Cassano et
  al. (2010). Reducing this radius, they naturally measured lower
  values of the morphological estimators and, in particular, of the
  power ratios \citep{boehringer.etal.10}.

The 17 regular objects are denoted by asterisks in the figure.  In most cases, these are systems with small companions or some minor irregularity in the surface brightness map. The full classification is listed in Table~\ref{tab:class} below the column ``$P_3/P_0,w$''.
Two extreme examples are represented in the right panel of Fig.~\ref{fig:pw}. The X-ray images of the most disturbed system of our sample (CL 13) are shown on the top panels, while the bottom panels refer to a relaxed cluster (CL 9). 

 The uncertainties on our parameters are smaller than those of
  \cite{boehringer.etal.10} because of the better spatial resolution
  of Chandra with respect to XMM-Newton \citep[see][for a detailed
  discussion about the influence on spatial resolution or point-spread
  function]{boehringer.etal.10}. The comparison between their work and
  that by Cassano et al. (2010), based on Chandra data, confirms this
  statement. The large exposure time assumed in our mock observations
  ensures a high counts statistics and, therefore, a further reduction
  on the uncertainties. If future missions will reproduce the great
  spatial resolution of Chandra, both power-ratios and centroid shifts
  will be available with sufficient accuracy for a large number of
  objects, thus allowing highy detailed studies of cluster
  morphologies.

\begin{table*}[htdp]
\caption{ Per each cluster and projection, we checked those: whose X-ray data are available at $R_{500}$ (first column);  that are morphologically regular, $P_3/P_0 < 2.\times 10^{-7}$ and $w<0.03$ (second column); that {\it intrinsically} lie in an poor environment; that {\it observationally} are recognized as lying in a poor environment.}
\centering
\begin{tabular}{|l|c|c|c|c||c|c|c|c||c|c|c|c|}

\hline
&\multicolumn{4}{|c}{projection 1} &\multicolumn{4}{|c}{projection 2} &\multicolumn{4}{|c|}{projection 3} \\
cluster &  $R_{500,X}$  & $P_3/P_0,w$ & I.p. & O.p. & $R_{500,X}$ &  $P_3/P_0,w$ & I.p. & O.p. & $R_{500,X}$ & $P_3/P_0,w$ & I.p. & O.p. \\
\hline
%
%
CL1&    $\surd$- &         -        &    - &$\surd$     &$\surd$&   $\surd$    &  -&-         &$\surd$&   -     &   -&-          \\
CL2 & - & $\surd$           &   $\surd$&$\surd$ &$\surd$&    $\surd$    &$\surd$&$\surd$& $\surd$& - &$\surd$&$\surd$ \\
CL3  & $\surd$  &       -      &   -&-     &$\surd$&  $\surd$ &  -&-        & $\surd$&- & $\surd$&-\\
CL4 & $\surd$   & -         &    -&-  &$\surd$&   - &  -&$\surd$      &$\surd$&  $\surd$ & -&-       \\
CL5 & -& $\surd$           &    $\surd$ &- &-& $\surd$  &    -&-       &-&    -  &   -&-            \\
CL6    & -&    - &     -&-    &-&  $\surd$ &    -&-         &-&  -   &     -&$\surd$           \\
CL7     & $\surd$ & -              &    $\surd$&$\surd$&$\surd$&  $\surd$  & - &$\surd$           &$\surd$& - & $\surd$&$\surd$\\
CL8    &  -&    -           &   -&-   &-&  $\surd$  &     -&$\surd$       &$\surd$&$\surd$ &   -&$\surd$       \\
CL9   & - &    $\surd$        &   $\surd$&$\surd$ &- &  $\surd$ &   -&-       &-& $\surd$  &      -&-      \\
CL10    &-& -         &  - &$\surd$ &-&  -  &   -&-         &-& - & -&-     \\
CL11    & $\surd$ &  -         &   $\surd$&$\surd$ &$\surd$ & $\surd$  &-&$\surd$  &$\surd$&   - &$\surd$&-\\
CL12    &-& -    &  -&-  &-&- & -&$\surd$     &-& - &  -&-     \\
CL13  & $\surd$&  -     &     -&-   &$\surd$ & - &-&-     &$\surd$ & - &-&-     \\
CL14    &$\surd$ &  -           &  -&-     &$\surd$ & $\surd$   &-&-    &-& - &-&-      \\
CL15    &-& -  &   -&- &$\surd$& -&-&-      &$\surd$&-&-&$\surd$    \\
CL16    &- &  -          &   -&-   &-& -  &-&-     &-& - &-&-     \\
CL17    &-&-  &   -&-  &-& - &-&-   &- &-&-&-     \\
CL18    & -&-      &   -&- &-&- &-&-     &-& - &-&-      \\
CL19    & -& -         &    - &-& - &-&-     &-& -&-&-  &-   \\
CL20    & $\surd$ &   -&    -&-  &$\surd$&  $\surd$    &-&-     &$\surd$& -&-&-      \\
\hline
\hline
\end{tabular}
\label{tab:class}
\end{table*}%

\begin{figure}
\begin{center}
\includegraphics[width=0.49\textwidth]{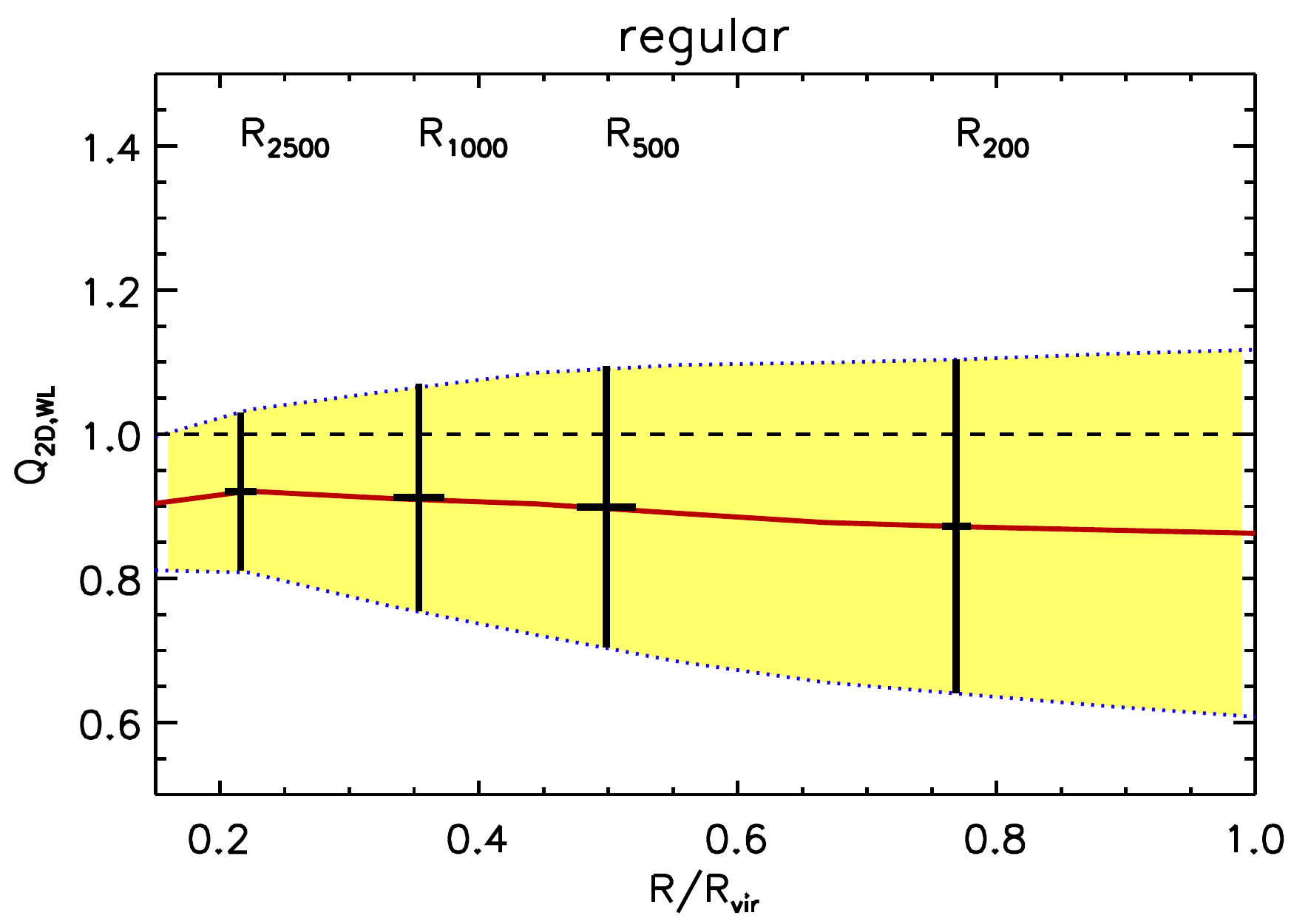}
\includegraphics[width=0.49\textwidth]{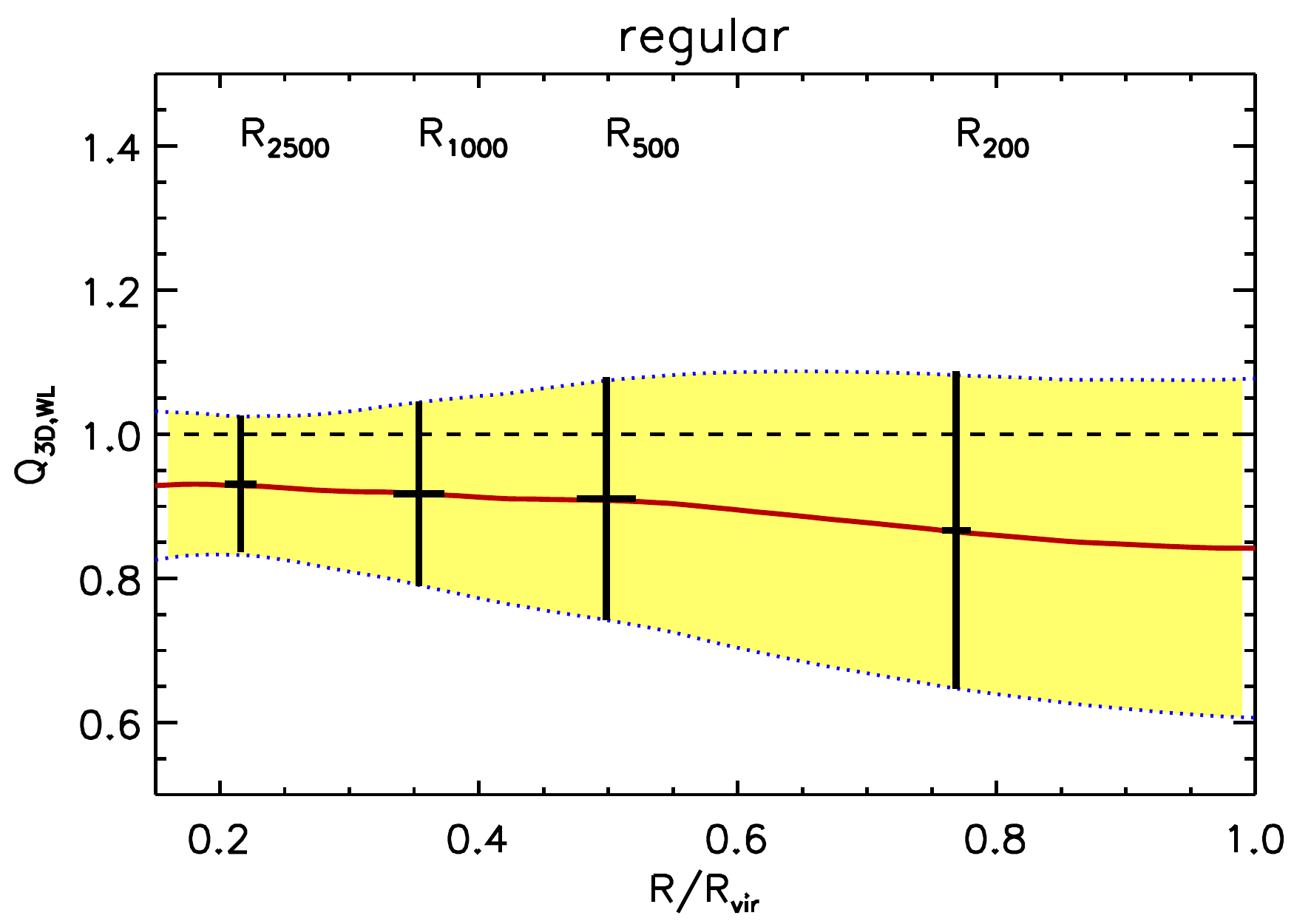}
\includegraphics[width=0.49\textwidth]{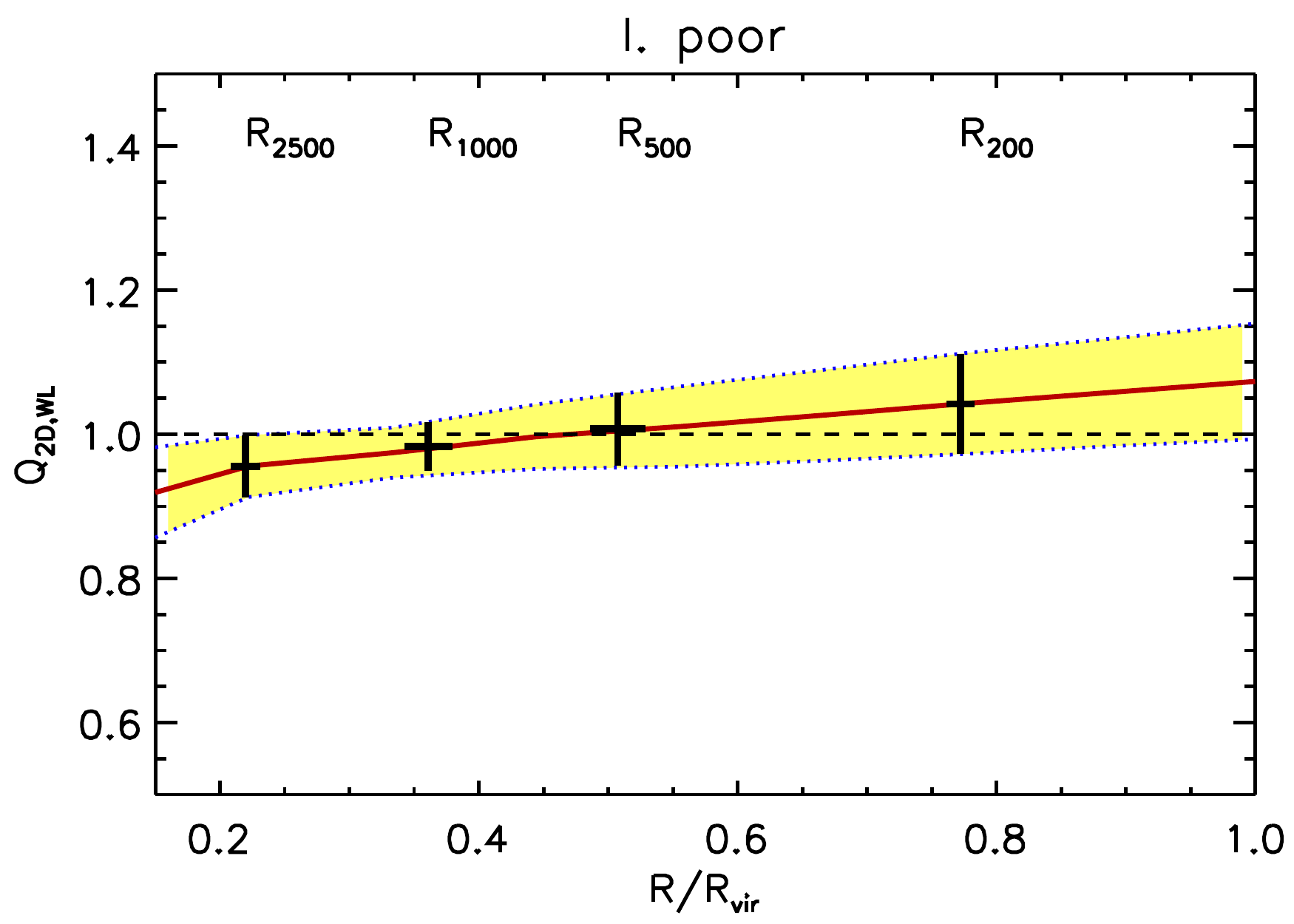}
\includegraphics[width=0.49\textwidth]{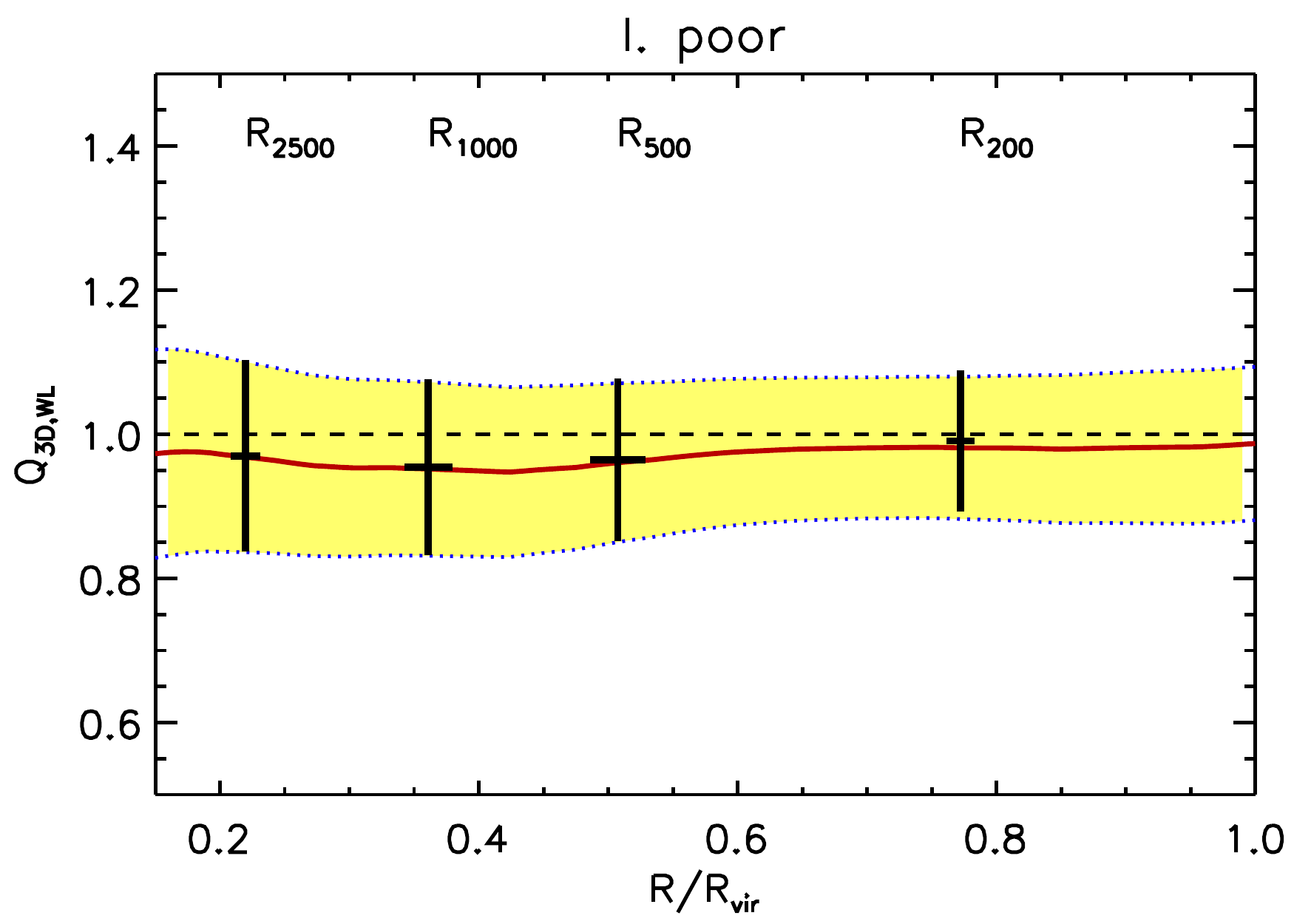}
\includegraphics[width=0.49\textwidth]{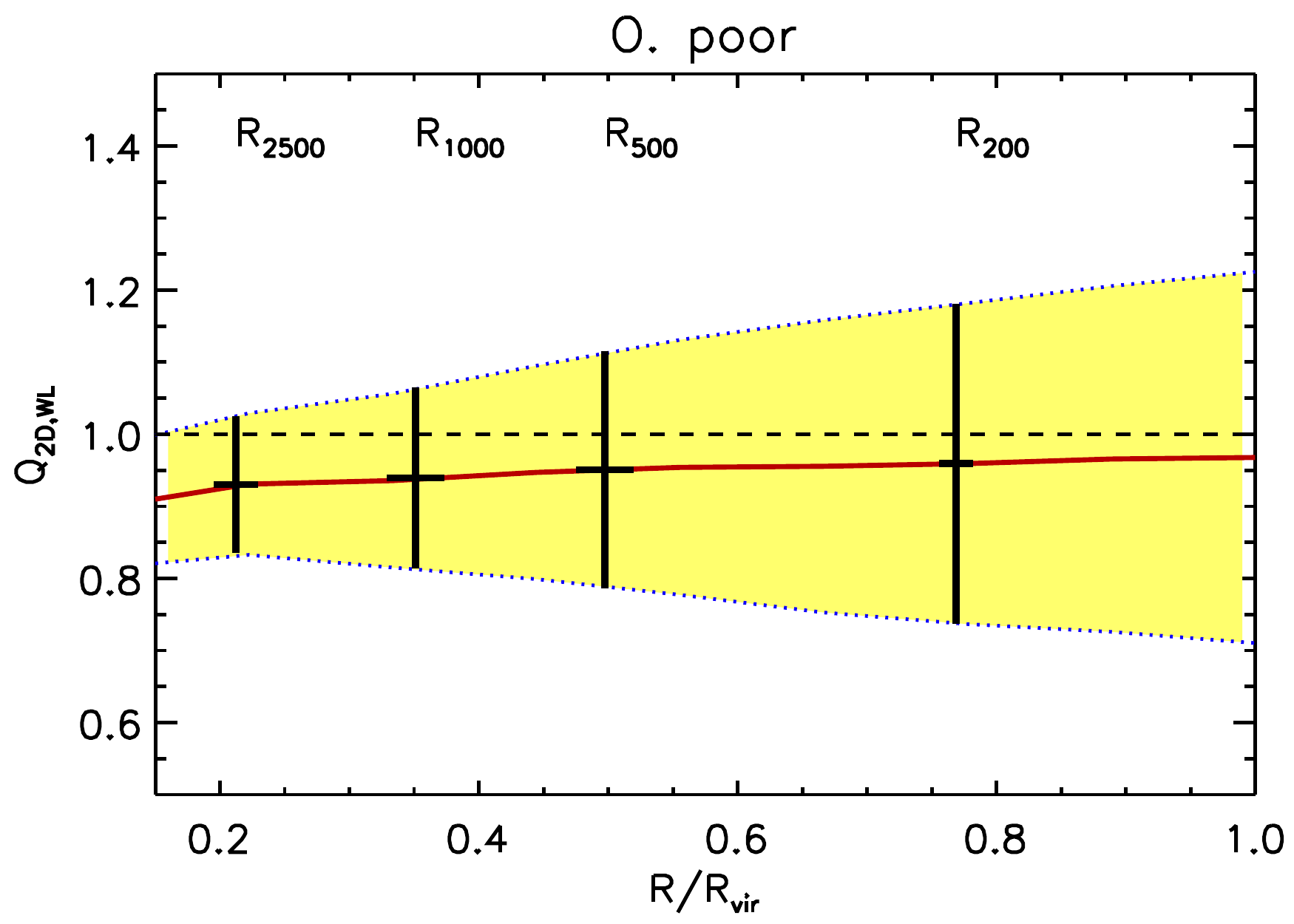}
\includegraphics[width=0.49\textwidth]{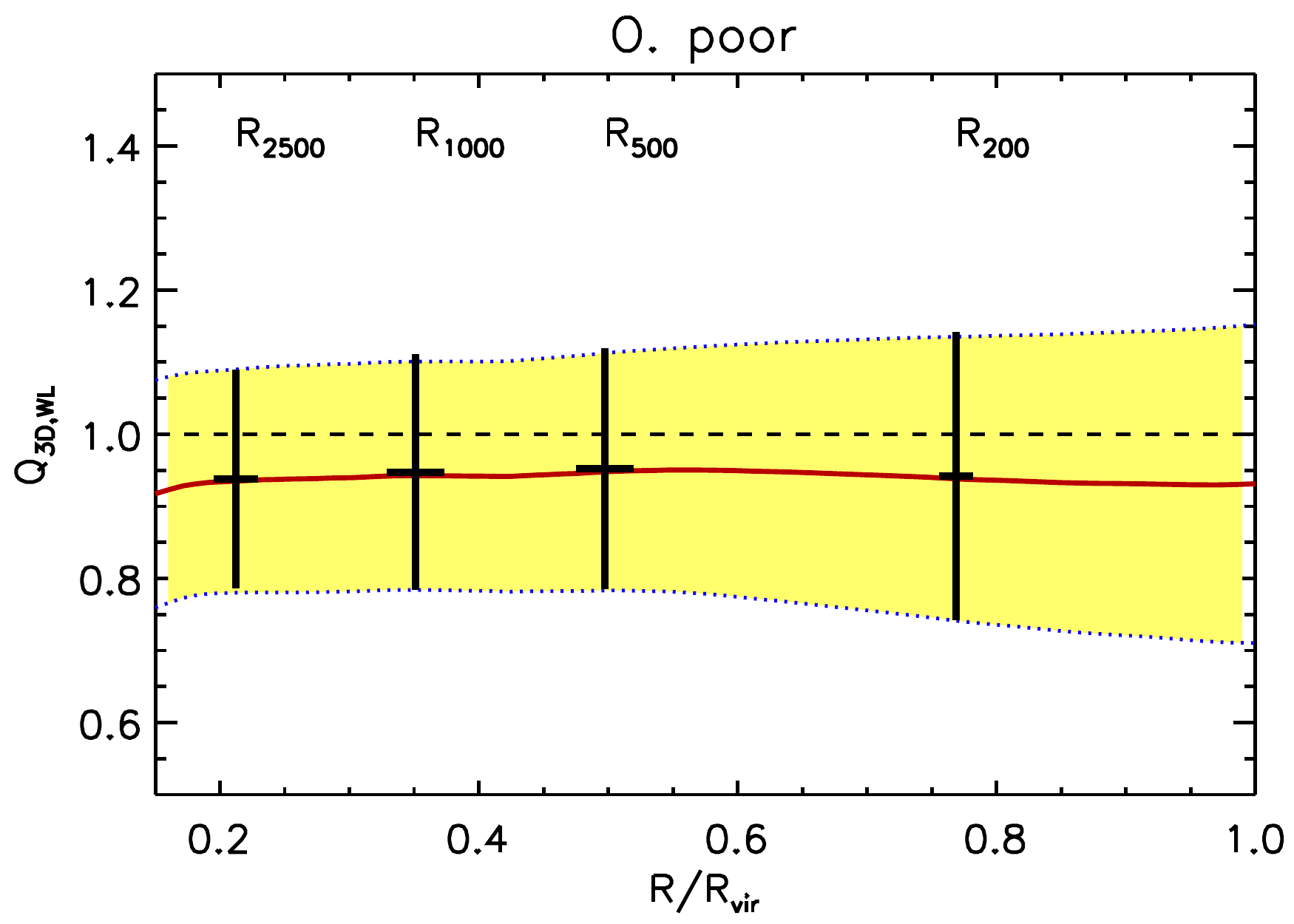}
\caption{Average ratios between 2D and 3D weak-lensing masses and true masses for different sub-samples: regular objects (on the top), cluster classified as lying in poor environment from the intrinsic maps (central panels), and from the optical maps (lower panels). The meaning of lines, crosses, and shaded regions is similar to Fig.~\ref{fig:wlall}. }
\label{fig:qclass}
\end{center}  
\end{figure}

We now proceed by checking whether the  lensing and true masses biases improve when selecting only the X-ray regular systems. Similarly to Fig.~\ref{fig:wlall}, we show in Fig.~\ref{fig:qclass}, $Q_{\rm WL}$ as a function of the distance from the cluster center for the sub-samples of regular systems. Quantitative results are listed in Table~\ref{tab:bias}  including those for $Q_X$.\\
 The scatter on lensing bias is reduced by 20\%-40\%. However, the
  bias itself worsen with respect to the whole sample. {\it Clearly, a
    selection based on the X-ray morphology is not  optimal for
    lensing purposes.} The reason of this behavior can be explained by
  comparing Table 3 and Table~B~6. Among the X-ray regular clusters,
  there are three images (projection 2 of CL1, CL9, and CL20) whose
  lensing measurements are severely under-estimated. All of them
  present in the outskirts of the optical images filaments or falling
  substructures, that do not have any obvious counterpart in the X-ray
  images or are lying outside the Chandra field of view.
      
\cite{jeltema.etal.08} and \cite{piffaretti&valdarnini} claimed to
find a significant trend of the X-ray masses bias on the morphological
estimators. We investigate this aspect further  by including an
  analysis of the biases in the weak-lensing mass reconstruction. To
  this purpose, we considered the absolute values of $|(1-Q_{3D,WL})|$
  to evaluate the dependence for any deviation.
A linear fit between the mass biases and the morphological estimators
has been computed accounting for measurement errors in both
  variables. The results are reported in Table~\ref{tab:fit}.  The
centroid shift performs better that the third order power ratio. The slopes
of the $Q-w$ relations are always significantly different from zero
with the correct sign (negative for the X-ray bias and positive for the
weak-lensing deviations).  The best fit values are similar to those
found by \cite{jeltema.etal.08} and \cite{piffaretti&valdarnini}.  We
further quantify the correlation between the mass biases and
$P_3/P_0$ or $w$ by means of the Pearson correlation coefficient.
 We find always a negative correlation, being the values between
-0.3 and -0.4 for $w$ and around -0.2 and -0.3 for $P_3/P_0$.

\begin{figure}
\begin{center}
\includegraphics[width=0.3\textwidth]{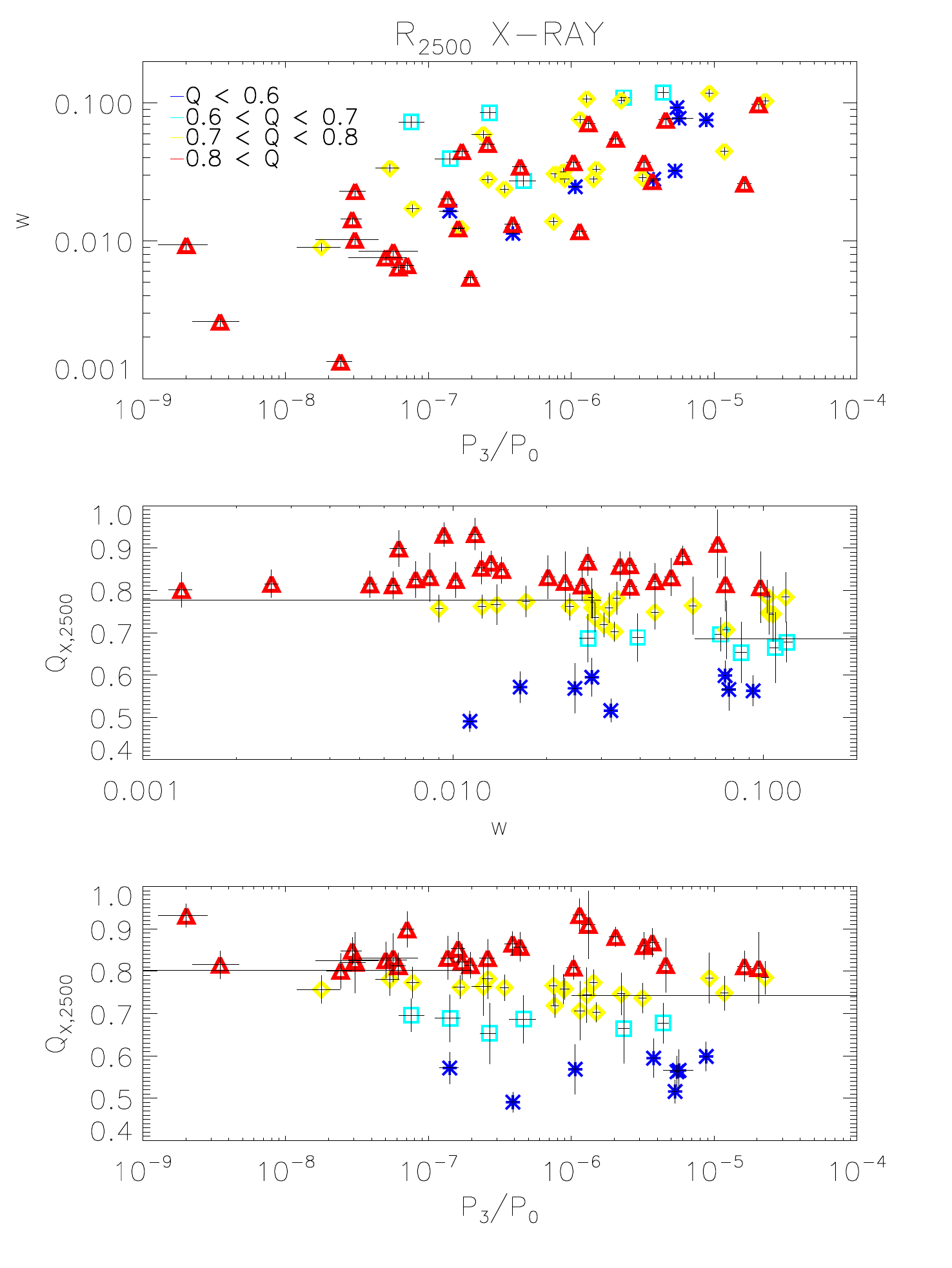}
\includegraphics[width=0.3\textwidth]{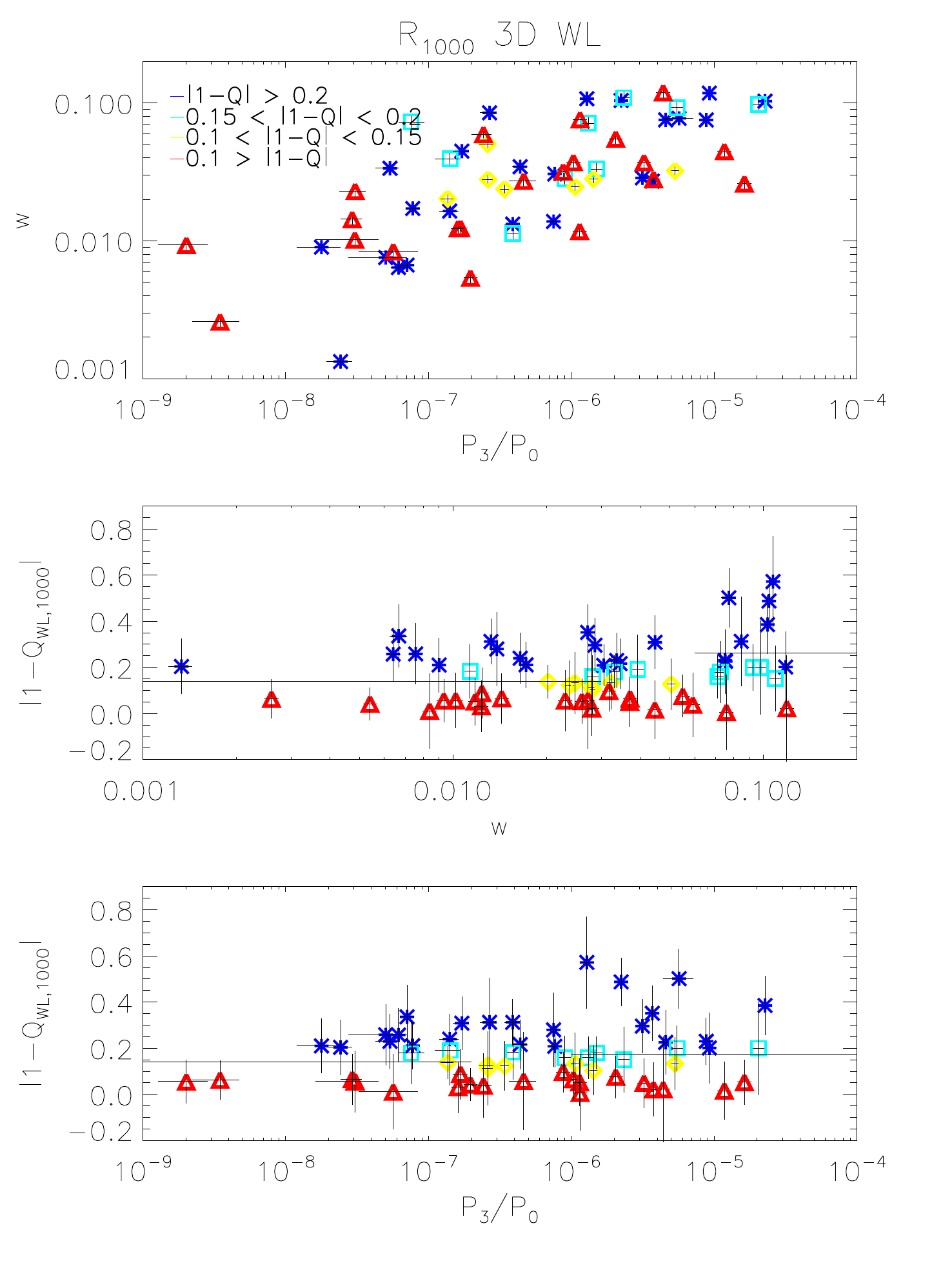}
\includegraphics[width=0.3\textwidth]{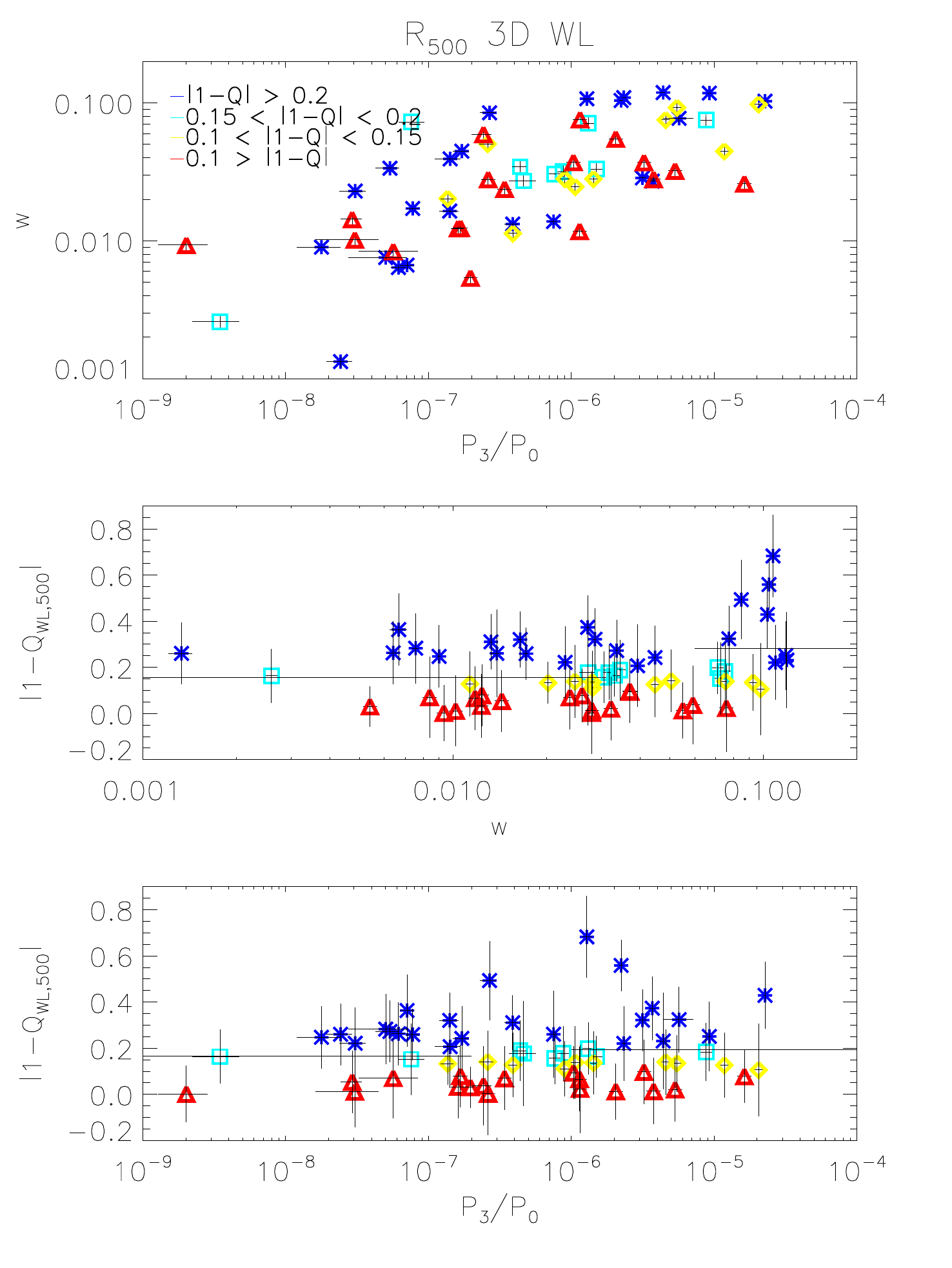}
\caption{Distribution of the clusters power ratio $P_3/P_0$ and centroid shift in dependance of the  mass biases. In the first column the bias shown refer to X-ray mass estimates computed within $R_{2500}$. In the second and third column, we plot the absolute deviation of the 3D weak-lensing mass bias computed at $R_{1000}$ and $R_{500}$, respectively. With red triangles and blue asterisks we show the weakest and strongest mass biases. The intermediate situations are shown with yellow rhombi and cyan squares.  In the central and lower panel, we separate the dependance of the bias by each morphological estimator. The horizontal lines represent the weighted average of the bias for particular values of the parameters ($w$ below 0.03 and above 0.06, $P_3/P_0$ below $2. \times10^{-7}$ and above $10^{-6}$.)} 
\label{fig:col}
\end{center}
\end{figure}

In  Fig.~\ref{fig:col} we present the best combinations: $Q_X-w$ for $R_{2500}$, and $|1-Q_{3D,WL}|-w$ for $ R_{1000}$ and $R_{500}$. The top panels, are similar to Fig.~\ref{fig:pw} where the different colors and symbols refer to different values of the mass bias. The red triangles refer to clusters whose X-ray mass biases, $Q_X$, are within 20\% or whose weak-lensing-masses deviations, $|(1-Q_{3D,WL})|$, are within 10\%. In all three top panels, we distinguish no segregation of colors. This implies that a better estimate of the total mass (red triangles) does not necessarily come from regular clusters defined on the basis of $P_3/P_0$ and $w$ values. However, for the centroid shift, even if this condition is not be necessary, it is sufficient at all radii: the weighted-average bias for clusters whose centroid shift is lower than 0.3 is 15-20\% lower than those with $w> 0.06$ (see difference on horizontal lines in the central panels). As confirmation, the third order power ratio weakly discriminate between good and bad estimates.

\paragraph{Other morphological indicators.}

 On top of the two estimators discussed, we tested other X-ray
  evidences used in literature to identify disturbed
  morphology. \cite{zhang.etal.10} and \cite{okabe.etal.10} (Locuss
  collaboration) considered the {\it asymmetry} and {\it fluctuation
    parameters} as introduced by \cite{conselice03}. Originally these
  parameters were created to quantitatively measure the distribution
  of stellar light in galaxies. The Locuss collaboration used them in
  12 clusters observed by XMM-Newton. The two parameters are defined
  as $A=\Sigma (|I-R|)/\Sigma I$ and $F=\Sigma (I-B)/ \Sigma I$, where
  $I$ is the [0.7-1.2] keV soft image, $R$ and $B$ are the same image
  rotated by 180 degrees, the first, and smoothed, the second. The
  smoothing kernel used by Locuss collaboration was equal to 2 arcmin
  or 400 kpc at redshift 
$z=0.2$
We choose three values for the
  FWHM of the smoothing Gaussian kernel: 320 kpc, 40 kpc, and 20 kpc. The
  first is similar to the one previous used in literature, the other
  two are smaller to take into account that our synthetic images,
  mimiking Chandra ACIS-S3 detector, have better spatial
  resolution.

  Subsequently, we tested two {\it hardness ratio
    indicators}. \cite{gitti.etal.11} built hardness ratio maps,
  obtained by dividing an hard-band image ([1.5-7.5 keV]) by a soft
  image ([0.3-1.5] keV), to identify presence of cold gas in Hydra A,
  a 3-4 keV cluster at redshift $z\sim0.05$. Similarly, we define two
  parameters $H1=\Sigma (H-S) / \Sigma S$ and $H2= \Sigma (H/S)$ where
  H and S are the hard and soft images smoothed with a gaussian of
  FWHM=320 kpc and 50 kpc.
 
 Finally, we consider the distance between the centers, $\Delta C$
 used in the our X-ray and weak-lensing analysis. In the former case,
 we considered the X-ray centroid, while in the latter we used the
 center of the BCG, which is also coincident with the minimum of the
 DM-potential well. A shift between the two centers might testimony a
 recent merger able to separate the two components, as for the Bullet
 cluster \citep{2004ApJ...606..819M}.

 All these parameters have been compared with the X-ray and weak
 lensing bias measures. For the weak lensing case we consider both the
 values of $Q_{3D,WL}$ and of $|Q_{3D,WL}-1|$ to evaluate general
 deviation from the true mass.  As done for the centroid-shift and the
 third order power ratio, we measured the correlation between all the
 parameters and the biases. We found that {\it none} of the new
 parameters is more strongly related to the bias than the {\it
   centroid shift} and the {\it third order power ratio}. On the opposite,
 their Pearson correlation coefficient is always smaller than 0.1 in
 absolute values.

\begin{table*}[htdp]
\caption{Best-fit relation of the form $Y=A+B \times X$ computed assuming errors in both $X,Y$. The variable $Y$ represents  the mass biases, while $X$ indicates the morphological estimators, in the first part of the table, and the temperature bias, in the second one. The signs + and - refer to significantly positive or negative slopes.}
\centering
\begin{tabular}{|c|c|c|c|c|c|c|c|c|c|}
\hline
&\multicolumn{3}{|c|}{$R_{2500}$}&\multicolumn{3}{|c|}{$R_{1000}$}&\multicolumn{3}{|c|}{$R_{500}$}\\
(Y,X) & A $\pm$ $\sigma_A$ & B   $\pm \sigma_B$& &A  $\pm \sigma_A$ & B  $\pm \sigma_B$&  & A $\pm \sigma_A$ & B $\pm \sigma_B$& \\
\hline

$Q_X,w$                                 &  0.79   $\pm$  0.01  & -0.88 $\pm$      0.18  &- &  0.73 $\pm$ 0.005 & -0.20 0.10      & - & 0.68 $\pm$ 0.0049& -0.27 $\pm$ 0.091 & - \\
$Q_X,P_3/P_0$                    &  0.51    $\pm$ 0.03  & -0.04 $\pm$     0.005 & - & 0.67 $\pm$ 0.02    &-0.007  0.003 & - &  0.71 $\pm$ 0.020 & 0.006 $\pm$ 0.003 & + \\

 $|1-Q_{3D,WL}|,w$              & 0.10    $\pm$   0.02  &  1.30  $\pm$  0.47 &  + &    0.11      $\pm$   0.02  &    1.54  $\pm$  0.50 & + &     0.12   $\pm$    0.03   &    1.68 $\pm$  0.56 & + \\
 $|1-Q_{3D,WL}|,P_3/P_0$ & 0.32    $\pm$   0.08   &  0.03 $\pm$  0.01 &   + &  0.32       $\pm$   0.10  &   0.02 $\pm$ 0.02 &   0  &    0.28    $\pm$    0.12   &    0.01 $\pm$ 0.02 & 0 \\
\hline
$Q_X,Q_T$ & -0.49 $\pm$ 0.15 & 1.48 $\pm$ 0.17 & +&-0.13 $\pm$ 0.10 & 1.04 $\pm$ 0.11 & +& -0.21 $\pm$ 0.10& 1.10 $\pm$ 0.12 & +\\
\hline

\end{tabular}
\label{tab:fit}
\end{table*}%


\subsection{Masses and cluster environment}

As shown in M10, the scatter in WL mass measurements is due to
substructures and triaxiality.  The combination of X-ray and lensing
data may help to further identify the most spherical systems or to
correct the mass estimates for triaxiality effects
\citep{2011arXiv1103.0202M,2010MNRAS.403.2077S}.  However, these
techniques are still model-dependent and subject to strong
assumptions. Furthermore, they require a certain amount of handling of
the data which cannot be applied to a large sample of objects.
Substructures, instead, may be more easily identified.

To classify clusters on the basis of the level of substructures in
their surroundings, we {\it visually} inspect the projected mass maps
of each cluster in our sample. We identify the objects whose
environment is poor of substructures within a region of $5
\;h^{-1}$Mpc around their centers. At first, we look directly at the
intrinsic density map from the simulations. We find 10 cluster
projections that match this criterium. We named this sample
``I. poor'' (Intrinsically
poor).  
 Often a regular cluster is not part of the poor environment class
  (see characterizations on Table~\ref{tab:class}). This is easily
  explained by presence of substructures outside the X-ray field of
  view which is limited to $\sim 2.5 h^{-1}$ Mpc. One of such clusters
  is the already-mentioned projection 2 of CL9. Despite being X-ray
  regular (Fig.~3) it shows evidence of filaments in its surroundings
  causing a strong underestimate of the weak-lensing masses in that
  projection (see Table~B~6). This is the main reason for which some X-ray
  regular objects show a large weak-lensing bias:
  they are lying in a rich environment that cannot be detected
  in the X-ray images.

The mass bias of the clusters classified `I. poor' is reported on Table~\ref{tab:bias} and showed in the middle panels of Fig.~4 for the weak-lensing masses.  The exciting result is that the projected true masses are  almost exactly recovered. For these systems $Q_{WL}$ deviates from unity by only a few percent for the 2D mass and by less than 5\% for the 3D mass. The scatter is  strongly reduced especially among  the two-dimensional masses, being only of order $\sim 5\%$ over a wide range of radii. {\it This is smaller than that found in M10 combining SL and WL non-parametrically}.
 For the three-dimensional masses, the scatter at the most external radii ($R_{500}$ and $R_{200}$) decreases by $\sim 50-60\%$ with respect to the whole sample and by 20-40\% with respect to the  systems with regular X-ray morphology.
 As shown in M10, this residual bias is caused by triaxiality.  It may be alleviated by means of introducing a parameter describing the elongation along the line of sight in the fitting model. This requires a combination of different probes, as proposed by \cite{2011arXiv1103.0202M} and \cite{2010MNRAS.403.2077S}, who combine lensing and X-ray data. However, a large uncertainty still remains, due to possible non-thermal pressure in the ICM, which is degenerate with the cluster triaxiality.

 As second step, we {\it visually} inspect the projected mass maps
 reconstructed from the synthetic weak lensing observations.  The method used for the reconstruction is described in \cite{2006A&A...458..349C} and in \cite{2009A&A...500..681M}, although we do not make use of the SL systems in this test. This
 approach, even if more ``observationally-oriented", is still
 subjective. Furthermore, the visual classification is more
 challenging because the resolution smears out possible features and
 the noise in the optical images reduces the detectability of
 clumps.  Clusters that appear isolated in the reconstructed mass maps are called ``O. poor''. This classification is also
   presented in Table~\ref{tab:class}. The results on the weak-lensing
   and X-ray mass biases are reported in Table~\ref{tab:bias} and
   shown in the bottom panels of Fig.~4.  The {\it net}
   improvement in terms of bias and scatter of both the 2D and 3D masses is now much less evident, but still masses are better recovered than in the sub-sample of X-ray regular clusters. In particular, the bias is reduced by about a factor two at $R_{500}$ and
   $R_{200}$, for both 2D and 3D masses.
 
Notice from Table 2 that the X-ray bias and scatter does not vary substantially in the three samples ( regular, I. poor, and O. poor).
This outcome is expected because, in general, X-ray masses have small scatter.  The intra-cluster medium is generally more spherical than the DM or the galaxy distribution, especially outside the core \citep{lau.etal.11}. As a consequence, the X-ray method is less prone to triaxiality. This implies that removing/adding a few objects, as long as they are not very disturbed, does not change significantly the result. 

\section{Discussion and conclusion} \label{sec:conclu}

This paper is an extension of the work of M10. We compared galaxy cluster masses derived from gravitational lensing and X-ray using 20 new massive halos simulated at high resolution including radiative gas physics. Each halo was observed along 3 different lines of sight and located at redshift 0.25. We used an optical and an X-ray simulators, namely {\tt Skylens} and {\tt X-MAS}, to build both optical and X-ray mock images mimicking, Subaru and Chandra observations, respectively. To perform the weak lensing  analysis, we measured the galaxy shapes using the KSB method and we derived the masses by fitting the tangential shear profiles using NFW functionals. For the X-ray, instead, we used the forward approach described in M10 and derived the mass under the hypothesis of hydrostatic equilibrium.  Then, we selected a subsample of regular clusters on the basis of the X-ray morphological estimators (the third moment of power ratios and the centroid shift). We further classified the objects in our sample based on the presence of substructures in the cluster environment. This classification was based on the {\it visual} inspection of both the true and the lensing-reconstructed  projected mass maps of the systems under investigation.

In the following, we discuss our main results:

\paragraph{$Q_{WL}$ and $Q_X$ for the whole sample.} The weak lensing mass bias is less than 10\% within $R_{500}$, and it grows to 13\% in the most external region.  The X-ray bias is around 25\% in the central region and increases to 33\% at $R_{500}$. The scatter of the bias is always higher by at least a factor of two for weak lensing than for X-ray mass measurements. The weak lensing bias and its large scatter is caused by presence of substructures in the cluster surroundings and by the triaxiality of the systems. The X-ray bias, instead, is mainly due to the lack of hydrostatic equilibrium, presence of clumps (in the external regions) and temperature dis-homogeneity (see further discussion).

\paragraph{$Q$ and morphological parameters.}
 We evaluate the effectiveness of some morphological estimators for
reducing the mass bias. We found that a selection based on centroid shift
($w<0.03$) and third order power ratio ($P_3/P_0 < 2 \times 10^{-7}$) reduces
the X-ray bias, especially in the central regions. This selection has
no effect on the weak lensing bias itself but it decreases the scatter by
20-40\% (see Table 2). Among the different morphological parameters
used to identify disturbed morphologies (including asymmetry and fluctuation parameters, two hardness ratios, and the optical-X-ray center offset), the only one that shows a mild
correlation with the bias is the centroid-shift. This is true also
for the weak-lensing masses bias. In terms of future X--ray
  missions, an optimal use the centroid-shift to identify ``ideal'' clusters to maximizing the efficiency of the mass measurements would require an imaging quality comparable to that of Chandra, but over a larger field of view so that the area probed by X-ray and optical observations become
comparable.

 \paragraph{$Q_{WL}$ substructures.} We established already that weak lensing methods based on single model fitting can severely fail to measure the mass of clusters in presence of massive substructures (M10). Working on single objects, the effect of substructures could be minimized by adopting multi-halo fitting techniques \citep{okabe.etal.11}, provided that substructures can be clearly identified as peaks in the weak lensing maps. Filtering techniques might also offer a possibility to mitigate the effect of structures perturbing the cluster shear profiles \citep[e.g.][]{2011MNRAS.416.1392G}.
We verified that removing from the sample those clusters which live in environments rich of substructures allows to minimize both the bias and the scatter in the weak lensing mass estimates. Unfortunately, the identification and the characterization of the substructures is a very difficult task. We will dedicate to another work the study of substructures detectability. Several galaxy surveys are planned for the next years which will scan large portions of the
sky \citep[see e.g.][]{des.05,lsst.09,2010arXiv1001.0061R}. The data
are expected to provide galaxy number densities in the range $15-40$
gals arcmin$^{-2}$, allowing to measure the shear signal of several
thousands of clusters. 
Having deep and sharp observations over a large field of view and
  with good spatial resolution would make it possible to detect
  substructure in a more efficient way than what presented in this
  analysis, thus enabling to virtually identify all the relevant
  substructures. Detection and mass measurements of sub-structures are
  already possible in the Coma cluster \citep{Okabe.etal.10a}. Methods based on higher order lensing distortion measurements (lensing flexion) also seem very promising \cite[e.g.][]{2007ApJ...660..995O, velander.etal.10}.

\paragraph{$Q_{WL}$ triaxiality.}
Triaxiality introduces a further scatter and bias in the three-dimensional lensing mass estimates. Even minimizing the impact of substructures, by restricting the analysis to the poor environment clusters, we notice a tendency to underestimate the total mass on
average. This is due to the fact that a large fraction of the systems in the sample is mostly elongated on the plane of the sky. Under this circumstance,  we expect to under-estimate
the mass in the de-projection phase \cite[see e.g.][]{feroz&hobson}. Conversely,  the mass is over-estimated in clusters seen along their major axis. Studies based on simulations showed that clusters forming in a CDM framework are generally prolate systems \citep{Shaw:2006id, allgood.etal.06, lau.etal.11}. For such mass distributions there is a larger chance to infer a smaller mass within a given radius from the projected density field
 and explains both the presence of the 2D bias and its increase after the
de-projection. This result depends on the selection of our sample
  which is mass limited. If our clusters were chosen for their lensing
  signals, and in particular for strong-gravitational lensing, we would have had 
  more objects strongly aligned  along the  line of sight \cite[e.g.][]{meneghetti.etal.10}. Then, the measured masses would have been  over-estimated on average. 

\paragraph{$Q_X$ and $Q_{WL}$ and dependence on cluster mass.} The mass range in our sample is too narrow to make a robust statistical analysis on the dependence of $Q_{X}$ and $Q_{WL}$ on the cluster masse. However, we can attempt to evaluate this effect by averaging these values on the three most (CL2, CL10, and CL11, all with $M_{500} > 7.5 \times 10^{14} h^{-1} M_{\odot}$) and least massive systems (CL 4, CL18, and CL19, all with $M_{500} < 3.5 \times 10^{14} h^{-1} M_{\odot}$ ).
The main result is that the bias $Q_{3D,WL}$ of the smallest clusters drops by almost a factor of 5  going from $R_{2500}$ ( $<Q_{3D,WL}>=0.94 \pm 0.20$) to $R_{200}$ ($<Q_{3D,WL}>=0.83 \pm 0.39$), while for massive clusters there is no change in the bias, always equal to 0.95. This indicates that at large distances from the cluster center, the mass estimates of the least massive clusters will be affected by a stronger bias compared to those of the most massive systems. This is not surprising, because M10 already showed that, when the tangential shear is used to measure the mass at a given radius, additional sources of shear, like massive substructures, located outside that radius lead to under-estimate the mass. The impact of such perturbers is obviously more significant in clusters of smaller mass.  Repeating the same test for the X-ray bias we do find the opposite trend. The bias of larger systems goes from $<Q_X>=0.77$ at $R_{2500}$ to  $<Q_X>=0.70 $ at $R_{500}$ (with error of order of 0.05), while for smaller objects the bias is constantly equal to 0.74. This can be ascribed to two main reasons: the massive objects are the most disturbed ones and they have a complex temperature structure \citep{ameglio.etal.09}. Its importance on the X-ray mass determination will be investigated more on the next paragraph.

\paragraph{$Q_X$ and temperature distribution.}
The values of $Q_X$ that we find in this work are consistent with what
previously found in R06. In M10 we find smaller deviations being the
average bias around 10\%.  There are some differences between this
paper and M10 which could affect the analysis, such as five times
smaller exposure time and a narrower field of view. However, we
believe that most of the difference is not due to the X-ray
preparation and analysis but to the physics adopted in the
simulation. 
The simulations analysed in M10 included a description of thermal
conduction, which is instead set to zero in the simulations presented
here and in those analysed by R06.
%

\begin{figure}
\begin{center}
\includegraphics[width=0.49\textwidth]{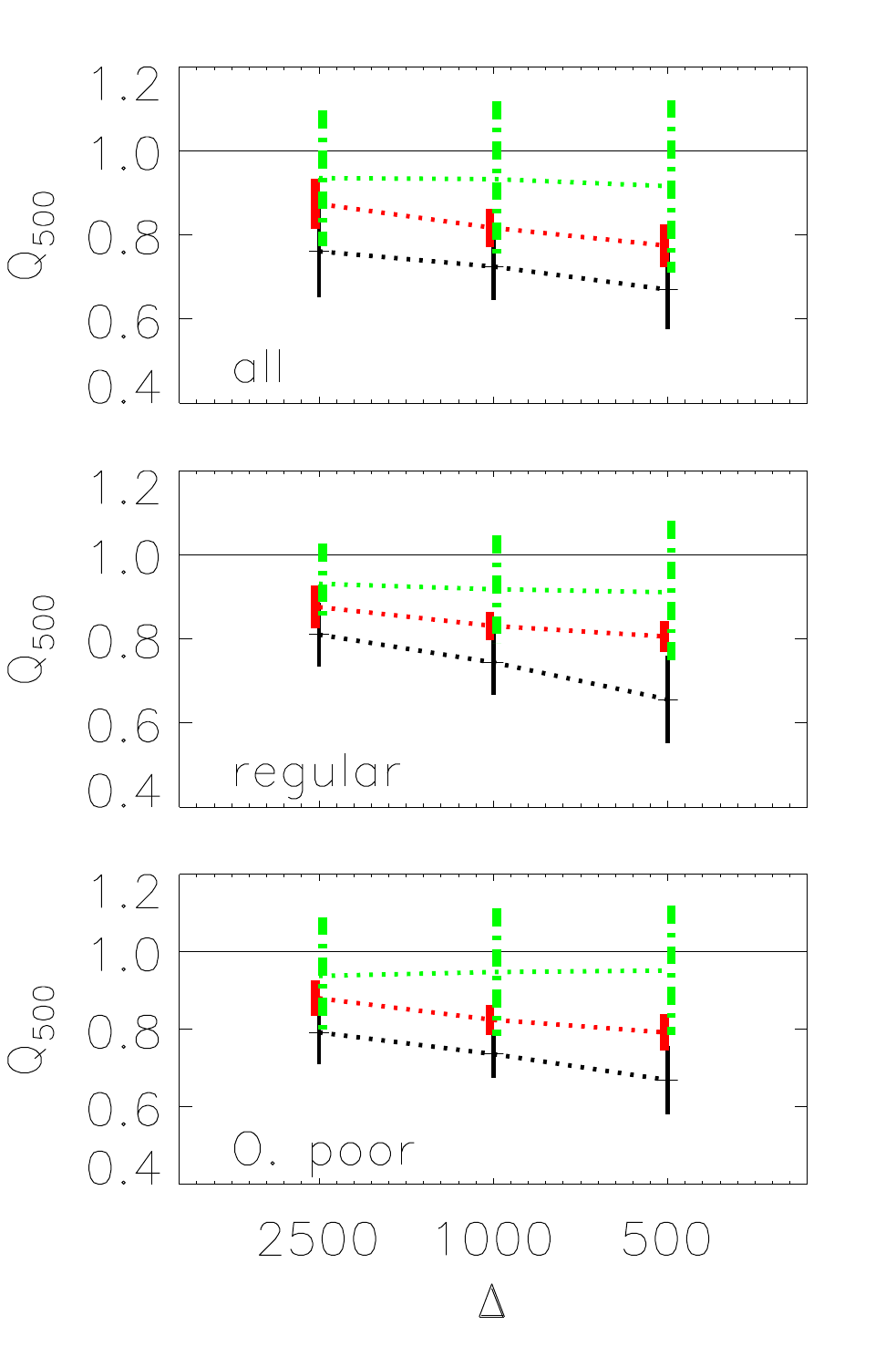}
\includegraphics[width=0.49\textwidth]{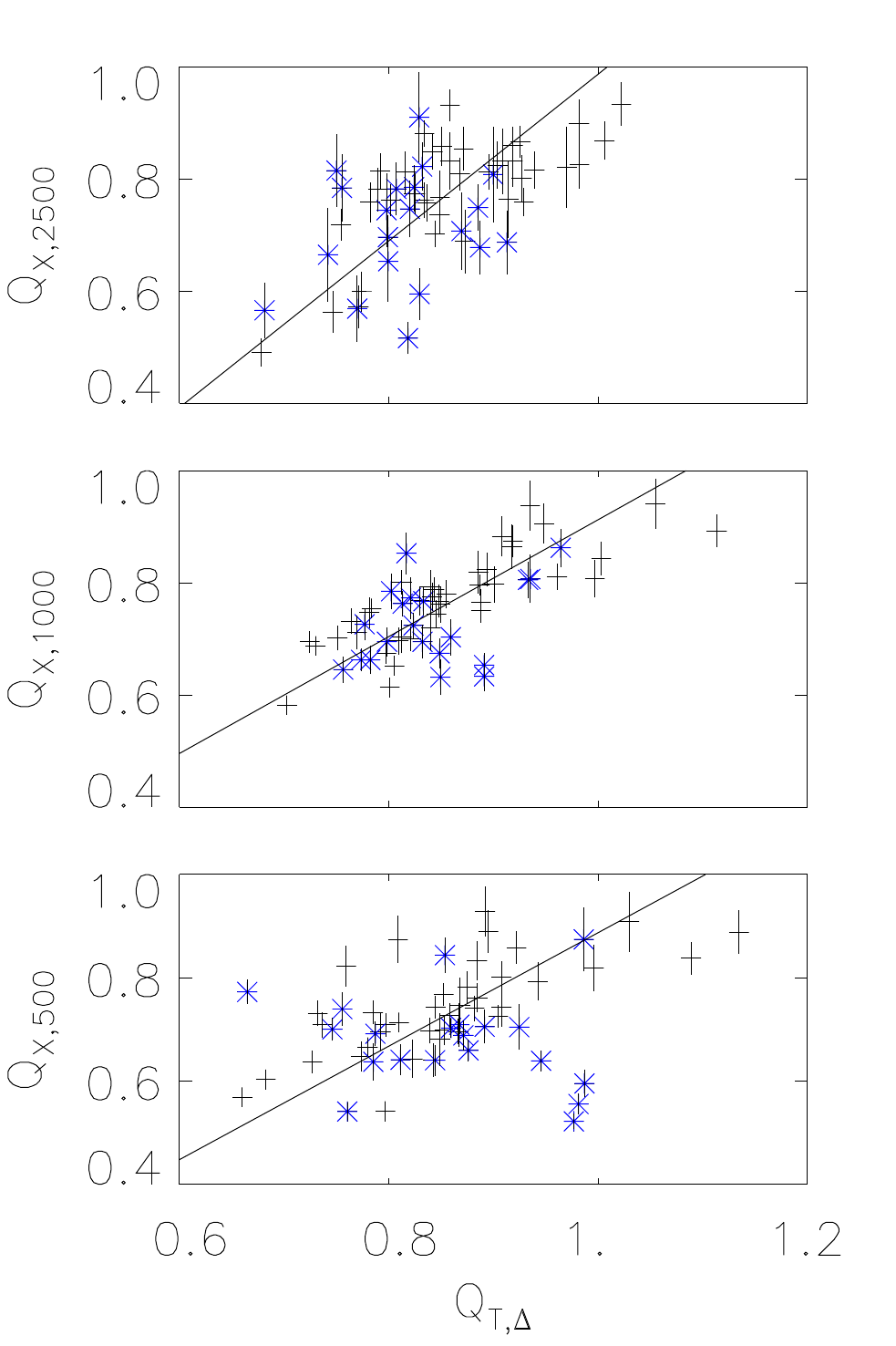}
\caption{Left panel: $Q_X$ (black), intrinsic $Q_X$ assuming
  mass-weighted temperature (red), and $Q_{3D,WL}$ (green) for all
  clusters, only regular, and only in poor environment. The values and scatter of $Q_X$ and $Q_{3D,WL}$ are reported in Table~2, while those of the intrinsic hydrostatic-mass bias are listed on the text (Section 8). Right panel:
  relation between $Q_X$ and $Q_T= T_X/T_{MW}$ at the three
  significant radii: $R_{2500}$ (top panel), $R_{1000}$ (central panel), and $R_{500}$ (bottom panels). The lines represent the best-fit in the form of
  $Q_X=A+B \times Q_T$ excluding the very disturbed clusters (blue asterisks).}
\label{fig:mhe}
\end{center}
\end{figure}

 Before elaborating more on the effect of changing the physical
description of the ICM, we remind that the X-ray mass is
derived from the hydrostatic equilibrium equation (Eq. ~\ref{eq:he})
where three terms are present: the derivative of the gas density, the
derivative of the temperature, and the temperature itself at the
radius considered. The over- or under-estimate of the temperature
leads to an over- or under- estimate on the X-ray mass of identical
amplitude.  If the temperature
structure is spherically homogenous, the measured temperature will be independent on the method used to derived
it. However, when the plasma presents   temperature structures in the annulus,
the X-ray measurements are biased low because the X-ray detectors of
Chandra and XMM-Newton have an higher efficiency on the soft band and,
thus, weight more colder gas \citep{mazzotta.etal.04}. In presence of
inhomogeneity, the X-ray temperature is, therefore, lower than the
mass-weighted one. As consequence, the hydrostatic masses computed
directly using the intrinsic gas density and mass-weighted
temperatures of the simulated clusters are higher than those obtained
following entirely the X-ray procedure. This is illustrated in
Fig.~\ref{fig:mhe}, where we plot the values of $Q_X$ reported in
Table.~\ref{tab:bias} in black and a similar ratio related to
intrinsic calculation in red. For all clusters, the {\it intrinsic bias} is
$12.6\% \pm 5.9\%$, $18.3\% \pm 4.5\%$, and $22.6\%\pm 5.1\%$ at the
three radii, respectively.  In the case of regular and poor systems,
these values decrease to $\sim 10\%, \sim 15\%$ and $\sim 17\%$ and
are in agreement with previous works based only on intrinsic
evaluation of the hydrostatic equilibrium mass using the mass-weighted
temperature \citep[e.g.][]{piffaretti&valdarnini,
  jeltema.etal.08,ameglio.etal.09,rtm, lau.etal.09}.

To stress more the dependence of the X-ray mass bias on the
temperature difference, we plot on the right panels of the
Fig.~\ref{fig:mhe} $Q_X$ versus $Q_T= T_X/T_{MW}$ for the three
X-ray-significant over-densities. The  uncertainty associated with the temperature bias are equal to 1 $\sigma$ error obtained from the spectroscopic analysis. The over-plotted line is the best
fit to the relation: $Q_X=A+B \times Q_T$ calculated excluding very disturbed objects (blue asterisks) and considering the errors in both coordinates. The values of B, for $R_{2500},
R_{1000}$ and $R_{500}$, are largely different from zero (see Table 4) and very close to 1.
The biases are strictly correlated one to the other showing
Pearson coefficients equal to $-0.7,-0.8,-0.7$.

 These results also suggest an explanation for the difference
  between our results and those \cite{nagai.etal.07}: their average
  bias is consistent with M10 and lower than the result of this
  paper. Their analysis and procedure are almost identical to ours,
  therefore, our results can be compared straightforwardly, despite
  the instrument-setting is slightly different (they consider
  reproduce synthetic observation of ACIS-I and not ACIS-S3). Indeed
  in both our works, the choice of the instrument is not as important
  as in real observation. Our mock images are, by construction, not
  affected by any calibration issues since we assume the same response
  files when generating and analyzing the images. In this condition of
  perfect calibration, we could have some differences only if the
  shape of the responses will be extremely different in shape (not in
  normalization), e.g. if an instruments weight a lot more plasma at 5
  keV in respect to the plasma at 8keV. However, as demonstrated by
  \cite{mazzotta.etal.04} in their Fig.~8 the detectors on board of
  Chandra and XMM-Newton give consistent answers in this
  respect. Furthermore, \cite{nevalainen.etal.10} in a study focused
  on real data showed that both the Coma Cluster and A1795 have broad
  band temperatures consistent within the 3\% statistical
  uncertainties and broad band fluxes may differ by up to 2-3\%. We,
therefore, exclude the possibility that the difference between our
results and Nagai et al. (2007) can be due to the instrument chosen.

As for the difference with respect to M10, as already mentioned
  the analysis presented in that paper was based on simulations which
  included the effect of thermal conduction, with conduction
  coefficient set to one-third of the Spitzer value. As discussed by
  \cite{dolag.etal.04}, thermal conduction in hot clusters is quite
  effective not only in removing cold blobs, but also in making the thermal structure
  of the ICM more homogeneous. 
  This leads  to an increase of the
  spectroscopic temperature and, therefore, of the hydrostatic mass.

  As for the comparison, with \cite{nagai.etal.07}, while their
  simulations do not include thermal conduction, they are based by an
  Eulerian Adaptive Mesh Refinement hydrodynamical scheme. As
  discussed by several authors
  \citep[e.g.][]{agertz.etal.07,mitchell.etal.09,vazza.etal.11},
  Eulerian hydrodynamics leads to a more efficient mixing of gas
  entropy. Therefore, one expects in Eulerian simulations that
  low--entropy gas residing in high--density clumps becomes more
  efficiently mixed, than in SPH simulations, to the high--entropy
  ICM. Again, this should result in a more homogenous temperature
  distribution, with a smaller bias  introduced in
  the estimate of X--ray temperature.

To summarize our X-ray mass bias derived intrinsically assuming
hydrostatic equilibrium and mass-weighted temperature are comparable
to previous results. Following the X-ray approach, instead, we confirm the
findings of R06 but we find a stronger bias in comparison to M10
and Nagai et al. 2007. The further $\sim$ 10-15\% is caused by
temperature inhomogeneity  \citep[see also][]{piffaretti&valdarnini,ameglio.etal.09}. This result highlights that, while
  hydrodynamical simulations are powerful tools to {\it understand biases}
  in observational mass estimates, a detailed assessment of this bias (e.g. its precise value)
  is still uncertain depending on the physical processes included in the simulations
  and on their numerical description. In this respect, we remind that none of the recent theoretical studies on X-ray mass biases includes the effect of feedback from AGNs.

\paragraph{Lensing and X-ray masses comparison.}
We compare the gravitational lensing masses with the X-ray masses following the same fitting procedure described in \cite{mahdavi.etal.08} and M10 to whom we address the curious reader for a detailed description. In brief, for each over-density $\Delta$, we define a parameter, $a_{\Delta}$,  as $M_{X}=a_{\Delta} \times M_{3D,WL}$. The error associated correspond to 68\% confidence level. In Table.~\ref{tab:compa}, we report our results for all  clusters and for the relaxed sample. We compared the lensing masses (green crosses in Fig. 6) to both the X-ray-strictly derived masses (black crosses in Fig. 6) and the intrinsic ones (red crosses in Fig. 6). For reference, we include the values found in M10 and in two observational papers: \cite{Zhang.etal.08} for the Locuss sample and \cite{mahdavi.etal.08} for the CCCP sample. 

Our intrinsic results are consistent within the errors with the observational data, especially for regular clusters.  Our ``observed'' X-ray to weak lensing mass ratios are, instead, consistent only with \cite{mahdavi.etal.08} for $R_{1000}$ and $R_{500}$. In all the other cases, our ratios are lower than the observed ones. This is due, once again, to the temperature in-homogeneity detected in our simulated clusters.

This last comparison has {\bf three} consequences. On one hand it could be that SPH codes generate more temperature structures, i.e. deviation from a spherically symmetric temperature distribution, than present in real clusters \citep{sijacki.etal.11}.  Unfortunately, current X-ray telescopes cannot provide detailed temperature maps for large sample of clusters with enough spatial resolution to confirm or dismiss this hypothesis.  Indeed, previous observational techniques to evaluate temperature structures \citep{bourdin.etal.08, zhang.etal.09} have been applied to a limited number of nearby objects. Increasing the size of the samples for which detailed observational studies are carried out would provide a unique opportunity to test the reliability of simulations in describing the complexity of the ICM thermal structure. Despite this precise match cannot be done at present,  some observations have already evidenced a significant level of gas clumpiness around the virial radius of some clusters \citep[][Vikhlinin et al. in prep.]{simionescu.etal.11}, leading to the conjecture that the plasma may be not completely thermalized at $R_{500}$. 
Another possible way to explain the mis-match between our results and the observations is that  weak-lensing masses of real clusters suffer of some additional bias. For example, as mentioned above, the contamination of shear catalogues by foreground galaxies may bias low the masses. The impact of this source of contamination and its possible correction using color-selection techniques is currently under investigation.  Finally, it is likely that the inclusion of AGN should significantly reduce the temperature in-homogeneity. In that case, the X-ray temperatures will be closer to the mass-weighted ones, leading to 5\%-10\% difference between X-ray and weak-lensing masses (see Table\ref{tab:compa}).

\begin{table*}[htdp]
\caption{$a_{\Delta}$ values for different overdensity and their uncertainty for our sample (first 4 rows), Meneghetti et al. 2010 ($5^{th}$ row), Zhang et al. 2008 ($6^{th}$ and $7^{th}$ rows), and Mahdavi et al. 2008 ($8^{th}$ row). For each of our sample, we report the values obtained from the X-ray analysis of the mock catalogue and those derived directly from the simulations, i.e. using the mass-weighted temperature in the hydrostatic equilibrium equation.}
\centering
\begin{tabular}{|l|c|c|c|}
\hline
 & $R_{2500}$ & $R_{1000}$& $R_{500}$\\
\hline
all                                        &  0.83 $\pm$ 0.02 & 0.80 $\pm$ 0.02 & 0.75 $\pm $0.02 \\
all-intrinsic                         &  0.94  $\pm$ 0.02 & 0.88 $\pm$ 0.02 & 0.83 $\pm$ 0.02 \\
regular                           &  0.87  $\pm $0.03 & 0.81 $\pm$ 0.04 & 0.75 $\pm$ 0.04 \\
regular-intrinsic                                           &  0.94  $\pm $0.03 & 0.91 $\pm$ 0.03 & 0.88 $\pm$ 0.03 \\
\hline
M10                                    &  0.90 $\pm $0.04 & 0.86 $\pm$ 0.02 &  0.88$ \pm $0.02 \\
\hline
Locuss all                &   1.00 $\pm$ 0.07  & 0.97  $\pm$ 0.05  & 0.99  $\pm$ 0.07 \\
Locuss relaxed       &   1.04 $\pm$ 0.08  & 0.96 $\pm$ 0.05   & 0.91 $\pm$ 0.06 \\
\hline
CCCP all                            & 1.03 $\pm$ 0.07 & 0.90 $\pm$ 0.09 & 0.78$\pm$ 0.09 \\

\hline
\end{tabular}
\label{tab:compa}
\end{table*}


\appendix

\section{A. Cold-particles-cut method}
A common characteristics of hydrodynamical simulations of galaxy
  clusters, is that gas associated to merging galaxies or small groups
  keep their identity for longer time within the hot ICM
  atmosphere. This is especially true for simulations including
  radiative cooling, without an efficient feedback mechanism, and for
  simulations based on SPH, which provide a rather inefficient mixing
  between high-- and low--entropy gas phases. These structures are
  revealed in the X-ray images as compact sources of dense gas, which
  are characterized by a strong emission. The majority of
  them 
  is located in the vicinity of the cluster center. A consequence of
  the overcooling problem is also that the central galaxy shows an
  extremely powerful X-ray peak. Since the presence of these features
  is mostly due to unaccounted physical processes, the standard
  procedure is to exclude them after their identification through a
  wavelet-detection algorithm \citep{vikh.etal.98} and to excise the
  central 15\% of $R_{500}$\citep[R06,][]{rasia.etal.08,nagai.etal.07}.
%

  This 
  procedure to remove high--density cold gas clumps is rather
  time-consuming, in particular for bright clusters, as those we are
  analyzing here, since they host a large number of sub-clumps. As a
  more efficient approach, we explore a method to exclude {\it a
    priori} particles which have short cooling time. These particles,
  which have high density and low temperature, can be
  identified in a well--determined region of the phase diagram of
  temperature, $T_p$, and gas densitie, $\rho_p$.
  Empirically, we found that {\it all} clusters in our sample have a
  separated phase of cooling particles which satisfy the following
  condition:
 \begin{equation}
 T_{\rm p}< N \times \rho_{\rm gas, p}^{0.25}.
\label{eq:tp}
 \end{equation} 
 
 The normalization factor, $N$, depends on the mass of the cluster
 being higher for the more massive systems. In our sample it does not
 vary substantially since the mass range considered is relatively
 small. Therefore, we consider a fixed value of $N$ equal to $3 \times
 10^6$ with density expressed in units of (g cm$^{-3}$) and
 temperature in keV. This normalization is conservative, meaning that
 {\it all} the particles belonging to a genuine hot phase of {\it all}
 our cluster lye above the relation set by this limit. The exponential
 factor, 0.25, depends on the polytropic index. If we assume that the
 pressure, $P \equiv {\rm constant} \times T \times \rho_{\rm gas}$,
 is related to the gas density through the polytropic equation: $P
 \propto \rho_{\rm gas}^{\gamma}$, we obtain $T \propto
 \rho_{\rm gas}^{\gamma -1}$ . The polytropic index, $\gamma$,
 physically can vary between $\gamma =1$ (isothermal plasma) and
 $\gamma=5/3$ (adiabatic gas) \cite[see][for a
 review]{sarazin}. For simulated clusters, the polytropic index lies
 between 1.15 and 1.25 \citep{ascasibar.etal.03,rtm,
   ostriker.etal.05,bode.etal.09}, constraining the exponential factor
 in the equation \ref{eq:tp} between 0.15 and 0.25.

\begin{figure}
\begin{center}
\includegraphics[width=0.43\textwidth]{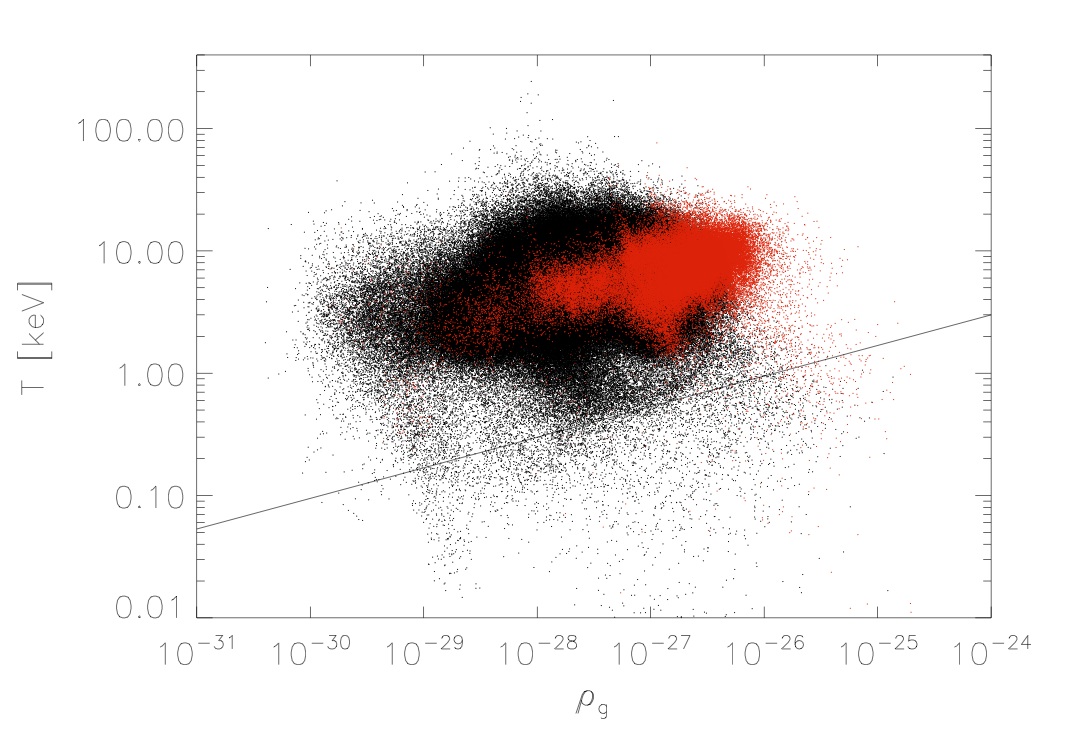}
\includegraphics[width=0.56\textwidth]{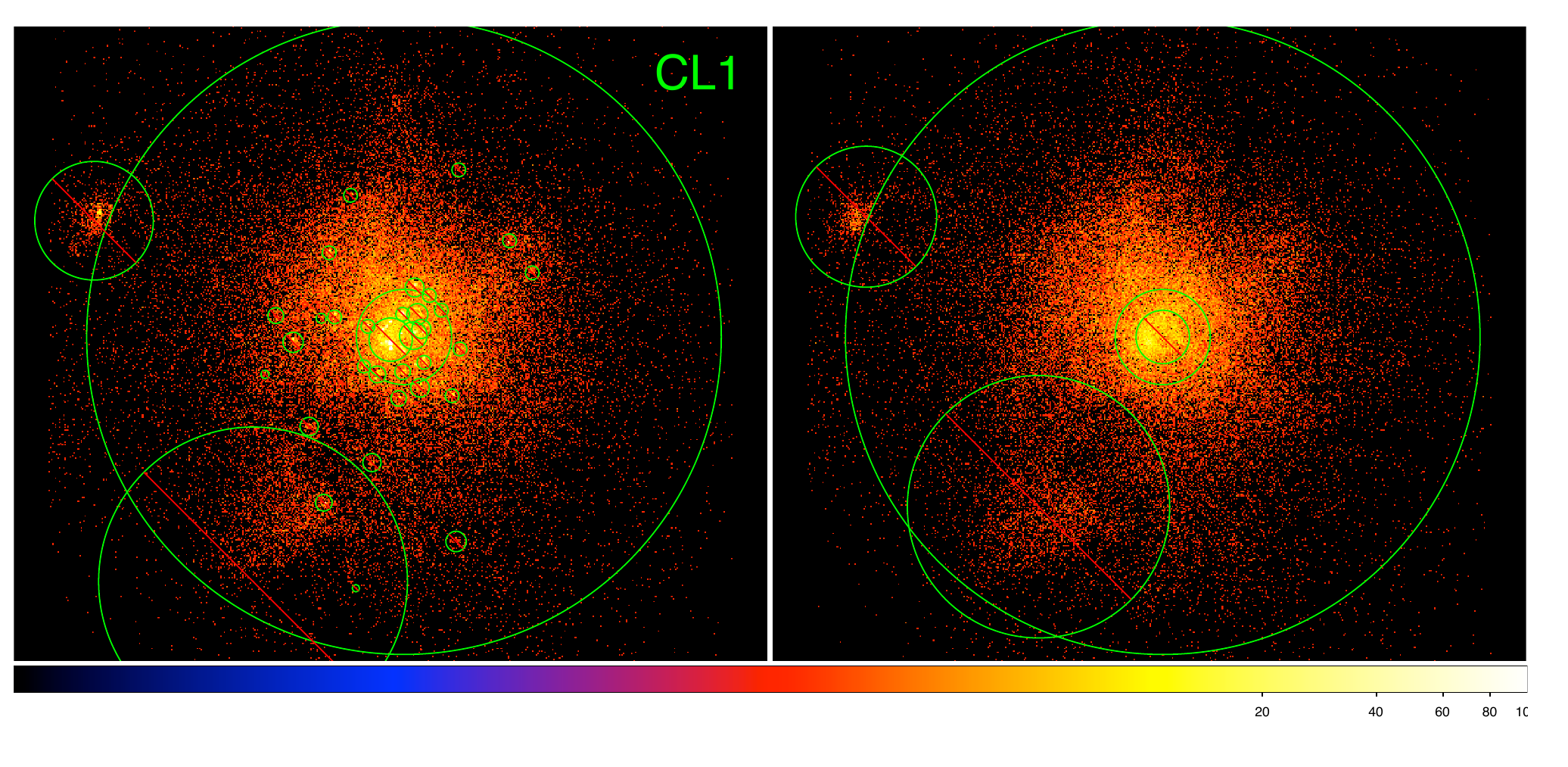}
\includegraphics[width=0.46\textwidth]{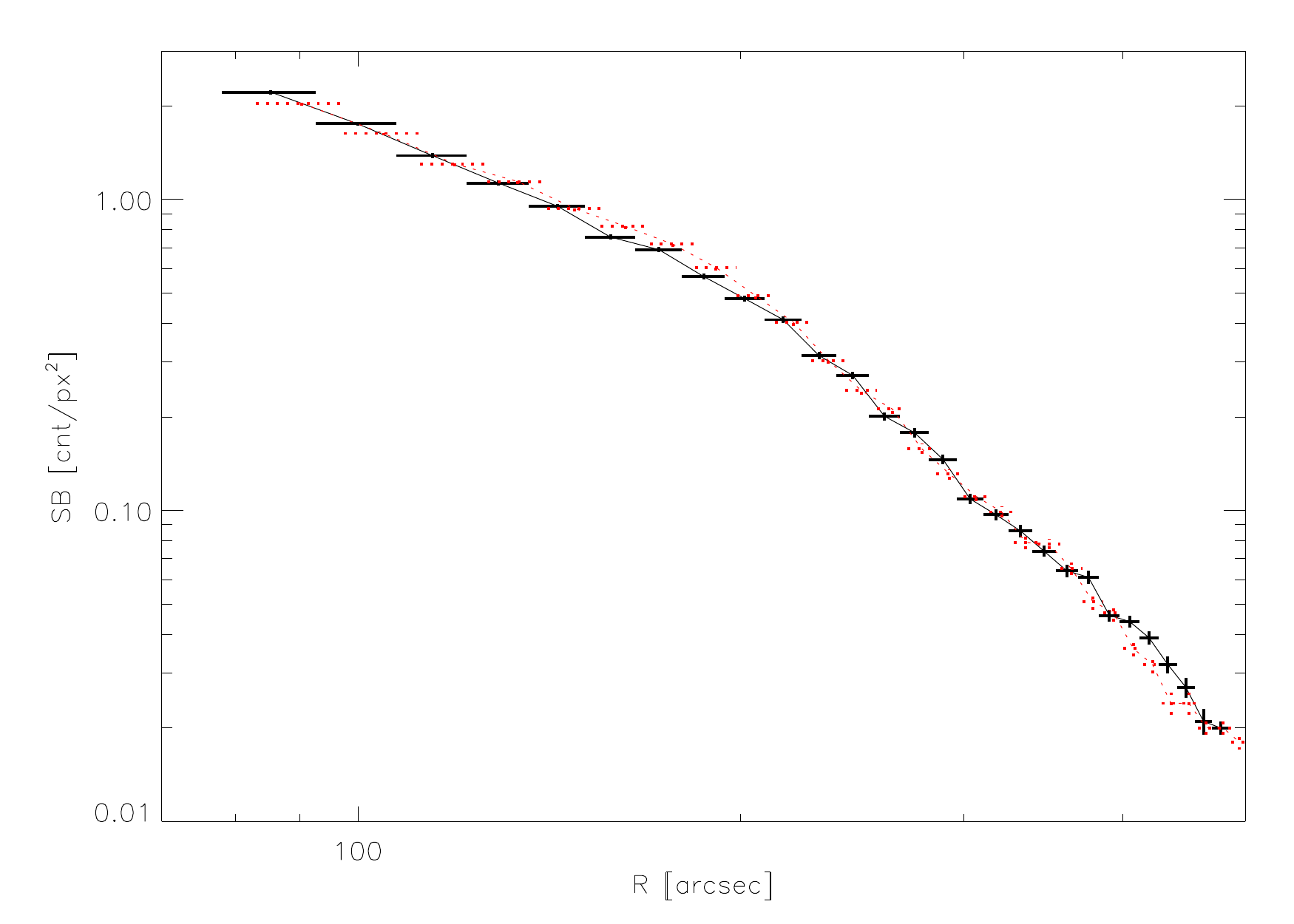}
\includegraphics[width=0.46\textwidth]{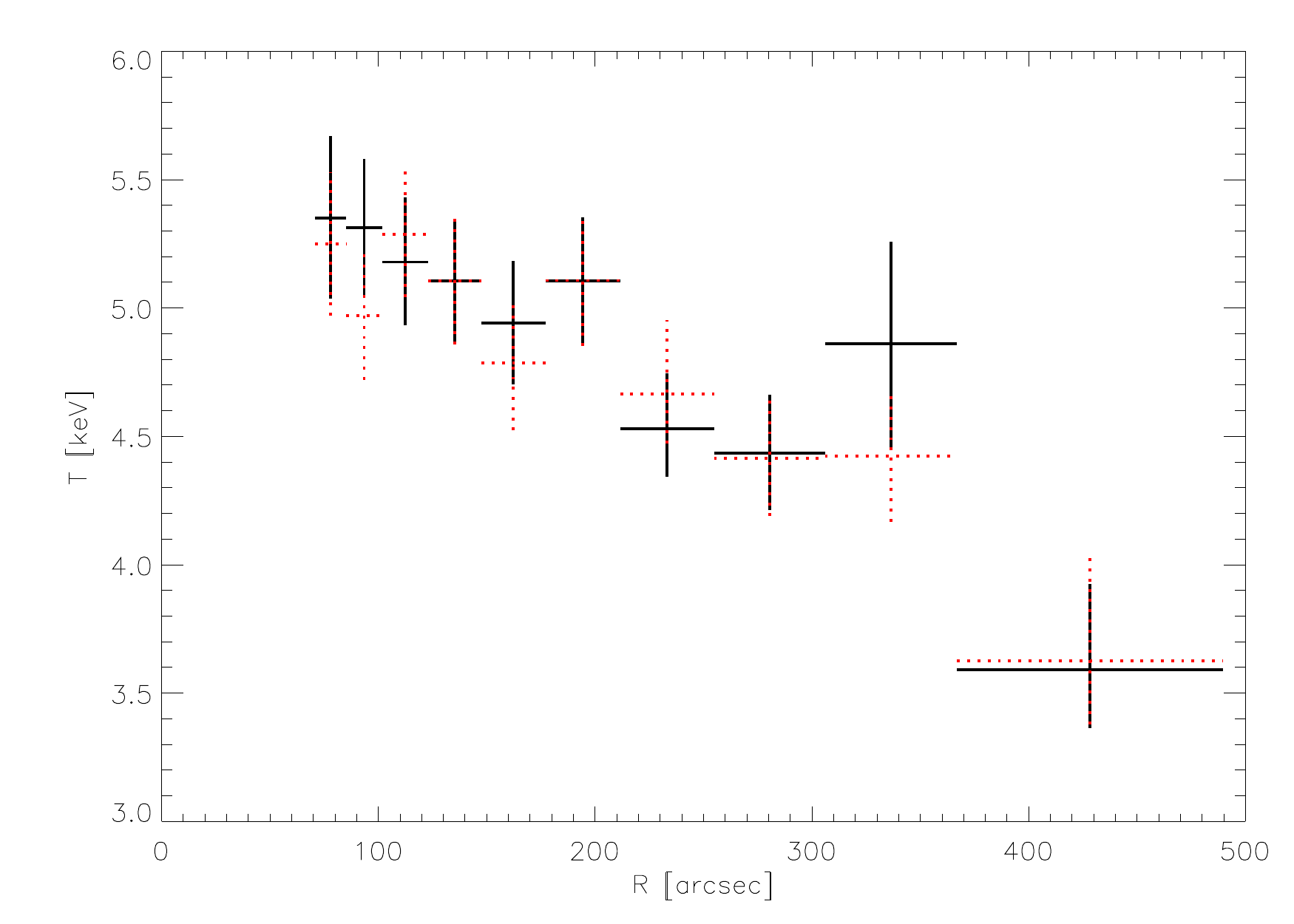}
\caption{In the top-right panels we show the soft X-ray images
  ([0.7-2] keV) of CL1 with the over-cooled particles included and
  excluded in the creation of the synthetic image. The top right panel
  shows the space-density plane for the same cluster where each dot is
  a particles. Red dots are the particles inside 10\% of
  $R_{500}$. The condition expressed in eq.\ref{eq:tp} is represented
  by the black line. In the bottom panels we compare the surface
  brightness profiles of the two soft X-ray images once excluded the
  regions marked (on the left) and the temperature profiles on the
  right.}
\label{fig:cut}
\end{center}
\end{figure}

This prescription to remove gas particles belonging to a cold phase is
visually illustrated in the top panels of Fig.~\ref{fig:cut}. The
right panel shows all the particles centered on CL1, in a field of
view of 16 $\times$ 16 arcmin$^2$ and within 10 Mpc $h^{-1}$ along the
line of sight (projection 1). The corresponding X-ray synthetic soft
energy image is in the central panel. The black line in the right
panel corresponds to the cut applied in our sample (see
Eq.~\ref{eq:tp}). Only the particles above that line contribute to the
X--ray emission shown in second soft map (right panel), which is the
image used for the analysis presented in this work. The small green
circles are the regions identified and removed with the wavelet
algorithm. 
The annuli in the two images have radii equal to 0.15$\times
R_{500}$ and $R_{500}$. The difference between the two emission maps
is evident. Our cutting technique allow us to clean the image from
about 30 small blobs while does not affect the emission of the
clusters or of the large sub-clumps. To test this last statement,
we analyzed both synthetic images (considering and excluding the
overcooled particles) for three clusters in our sample. In the six
resulting images, we run the wavelet algorithm to identify the
X-ray peaks and exclude them. Subsequently, we extract both surface
brightness and temperature profiles and confirm that the cutting
technique does not introducing any bias. In the bottom panels of
Fig.\ref{fig:cut} we plot the comparison for CL1, projection 0. Both
the surface brightness profiles and the temperature profiles are
consistent with each other.

\section{B. Measured Masses} 

 In Tables~\ref{tab:qwl} and
  ~\ref{tab:qx} we report the values of $Q_{3D,WL}$ and $Q_X$
 computed at different radii: $R_{\rm    2500}, R_{\rm 1000}, R_{\rm 500},$ and $R_{200}$.  Uncertainties
 are computed
  following the procedure described in Sections~6.1 and 5.2, respectively.  

\begin{table*}[htdp]\small
\caption{$Q_{\rm 3D,WL} $ and uncertainties at different overdensity ($\Delta= 2500,1000, 500, 200$).}
\centering
\begin{tabular}{lccccccccccccc}
\hline
\hline
cluster & proj&$Q_{\rm WL,2500}$ &  $eM_{\rm WL,2500}$ & $Q_{\rm WL,1000}$ & $eM_{\rm WL,1000}$ & $Q_{\rm WL,500}$  & $eM_{\rm WL,500}$ &$Q_{\rm WL,200}$ & $eM_{\rm WL,200}$\\
\hline

  CL1 &     1 & 0.6833& 0.1086& 0.7700& 0.1013& 0.8168& 0.1260 &0.9045& 0.1980 \\
  CL1  &     2 & 0.8517& 0.0939& 0.7605& 0.1095& 0.6793& 0.1218 &0.6209& 0.1368 \\
CL1  &     3 & 1.2358& 0.0894& 1.1995& 0.1003& 1.1348& 0.1229 &1.1018& 0.1563 \\
  CL2  &     1 & 0.8879& 0.0551& 0.8618& 0.0720& 0.8673& 0.0902 &0.9271& 0.1190 \\
  CL2  &     2 & 1.1038& 0.0531& 1.0424& 0.0714& 1.0306& 0.0884 &1.0820& 0.1139 \\
  CL2  &     3 & 0.8285& 0.0554& 0.8202& 0.0712& 0.8358& 0.0910 &0.9061& 0.1233 \\
   CL3  &     1 & 0.8288& 0.1303& 0.8092& 0.1520& 0.7933& 0.1793 &0.8226& 0.2255 \\
   CL3  &     2 & 0.8594& 0.1388& 1.0555& 0.1329& 1.2216& 0.1563 &1.5281& 0.2558 \\
   CL3  &     3 & 0.9560& 0.1207& 0.9623& 0.1399& 0.9627& 0.1715 &1.0194& 0.2262 \\
  CL4  &     1& 0.9539& 0.1526& 1.0058& 0.1631& 0.9762& 0.1907 &0.9058& 0.2299 \\
  CL4  &     2 & 1.1896& 0.1530& 1.2796& 0.1609& 1.2606& 0.1894 &1.1909& 0.2347 \\
  CL4  &     3 & 0.8191& 0.1763& 0.9889& 0.1633& 1.0707& 0.1754 &1.1380& 0.2526 \\
  CL5  &     1 & 1.0187& 0.0822& 1.0885& 0.1103& 1.0790& 0.1352 &0.9535& 0.1524 \\
  CL5  &     2 & 0.9999& 0.0833& 1.0656& 0.1098& 1.0541& 0.1350 &0.9290& 0.1532 \\
  CL5  &     3 & 0.7505& 0.0919& 0.8772& 0.1075& 0.9301& 0.1383 &0.8959& 0.1817 \\
   CL6  &     1 & 1.0115& 0.0771& 0.9470& 0.0979& 0.9218& 0.1156 &0.8726& 0.1330 \\
   CL6  &     2 & 0.8944& 0.0838& 0.7891& 0.1009& 0.7407& 0.1127 &0.6739& 0.1215 \\
   CL6  &     3 & 1.0089& 0.0712& 1.0969& 0.0876& 1.1777& 0.1182 &1.2504& 0.1662 \\
  CL7  &     1 & 1.2435& 0.0805& 1.2160& 0.1077& 1.1889& 0.1318 &1.1044& 0.1539 \\
  CL7  &     2 & 0.8814& 0.0987& 0.7962& 0.1198& 0.7397& 0.1338 &0.6483& 0.1411 \\
  CL7 &     3 & 0.8838& 0.0924& 0.8727& 0.1116& 0.8589& 0.1354 &0.8038& 0.1601 \\
   CL8  &     1 & 0.8335& 0.0970& 0.7703& 0.1192& 0.7275& 0.1341 &0.5987& 0.1333 \\
   CL8  &     2 & 0.9627& 0.0855& 0.9692& 0.1115& 0.9671& 0.1371 &0.8510& 0.1531 \\
   CL8  &     3 & 0.8433& 0.0961& 0.7901& 0.1184& 0.7527& 0.1359 &0.6256& 0.1386 \\
   CL9  &     1 & 0.9297& 0.1026& 0.9437& 0.1206& 0.9889& 0.1532 &1.0755& 0.2074 \\
   CL9  &     2 & 0.7573& 0.1195& 0.6639& 0.1383& 0.6366& 0.1563 &0.6290& 0.1807 \\
   CL9  &     3 & 0.8340& 0.1133& 0.7418& 0.1325& 0.7171& 0.1524 &0.7151& 0.1799 \\
   CL10  &     1 & 1.2901& 0.0940& 1.2083& 0.0902& 1.1570& 0.1125 &1.1863& 0.1541 \\
   CL10  &     2 & 0.6441& 0.1026& 0.5124& 0.1044& 0.4414& 0.1115 &0.4103& 0.1257 \\
   CL10  &     3 & 0.9038& 0.0976& 0.8409& 0.0931& 0.8014& 0.1136 &0.8178& 0.1527 \\
   CL11  &     1 & 0.8060& 0.0715& 0.8404& 0.0934& 0.8888& 0.1207 &0.9449& 0.1599 \\
   CL11  &     2 & 0.9677& 0.0661& 1.0629& 0.0851& 1.1634& 0.1170 &1.2880& 0.1692 \\
   CL11  &     3 & 1.0703& 0.0662& 1.0653& 0.0912& 1.0938& 0.1146 &1.1242& 0.1442 \\
   CL12  &     1 & 0.9912& 0.1403& 0.8486& 0.1421& 0.7792& 0.1618 &0.7000& 0.1830 \\
   CL12  &     2 & 0.8985& 0.1681& 0.8670& 0.1346& 0.8613& 0.1578 &0.8458& 0.2117 \\
   CL12  &     3 & 0.8050& 0.1632& 0.8206& 0.1338& 0.8481& 0.1560 &0.8731& 0.2204 \\
  CL13  &     1 & 0.7265& 0.1903& 0.7989& 0.1543& 0.7491& 0.1496 &0.7926& 0.2234 \\
  CL13  &     2 & 1.1241& 0.1710& 1.2254& 0.1398& 1.1396& 0.1427 &1.1953& 0.2151 \\
  CL13  &     3 & 1.4571& 0.1467& 1.5012& 0.1290& 1.3240& 0.1432 &1.3141& 0.1979 \\
   CL14  &     1 & 0.7052& 0.0775& 0.6879& 0.0998& 0.6892& 0.1189 &0.6833& 0.1406 \\
   CL14 &     2 & 0.8859& 0.0718& 0.9445& 0.0938& 1.0020& 0.1223 &1.0601& 0.1631 \\
   CL14  &     3 & 0.8582& 0.0695& 0.9242& 0.0925& 0.9869& 0.1212 &1.0523& 0.1628 \\
   CL15 &     1 & 0.5843& 0.1164& 0.6148& 0.1276& 0.5712& 0.1451 &0.5154& 0.1696 \\
   CL15  &     2 & 0.9911& 0.1000& 0.9838& 0.1259& 0.8737& 0.1410 &0.7492& 0.1523 \\
   CL15 &     3 & 1.2096& 0.1026& 1.3084& 0.1158& 1.2424& 0.1398 &1.1479& 0.1740 \\
   CL16  &     1 & 1.0100& 0.0889& 1.0497& 0.1075& 1.0976& 0.1388 &1.1350& 0.1837 \\
   CL16  &     2 & 0.8213& 0.0964& 0.9480& 0.1037& 1.0663& 0.1388 &1.2055& 0.2124 \\
   CL16  &     3 & 0.7075& 0.1008& 0.6486& 0.1204& 0.6265& 0.1381 &0.5920& 0.1557 \\
   CL17  &     1 & 0.7084& 0.1252& 0.8665& 0.1157& 0.9788& 0.1385 &1.0575& 0.2264 \\
   CL17  &     2 & 1.2043& 0.0962& 1.1826& 0.1174& 1.1280& 0.1408 &0.9769& 0.1573 \\
   CL17  &     3 & 0.9235& 0.1151& 0.9787& 0.1174& 0.9862& 0.1440 &0.9129& 0.1813 \\
   CL18  &     1 & 1.0900& 0.1865& 0.9420& 0.2121& 0.8231& 0.2282 &0.6701& 0.2286 \\
   CL18  &     2 & 0.6677& 0.2273& 0.8004& 0.2031& 0.8942& 0.2001 &0.9848& 0.2770 \\
   CL18  &     3 & 0.7849& 0.2333& 1.0221& 0.2300& 1.2313& 0.2086 &1.5211& 0.2555 \\
   CL19  &     1 & 1.0025& 0.1936& 0.6878& 0.1931& 0.5067& 0.1724 &0.3975& 0.1580 \\
   CL19  &     2 & 1.1994& 0.1679& 1.1137& 0.1603& 0.9976& 0.1787 &0.9378& 0.2160 \\
   CL19  &     3 & 0.6173& 0.2038& 0.4284& 0.1990& 0.3175& 0.1778 &0.2566& 0.1679 \\
   CL20  &     1 & 1.0924& 0.0816& 1.1033& 0.1075& 1.1363& 0.1362 &1.1429& 0.1710 \\
   CL20  &     2 & 0.7797& 0.0930& 0.7427& 0.1157& 0.7371& 0.1374 &0.7103& 0.1597 \\
   CL20  &     3 & 0.7763& 0.0983& 0.7041& 0.1177& 0.6783& 0.1345 &0.6322& 0.1492 \\
\hline
\hline
\end{tabular}
\label{tab:qwl}
\end{table*}%

\begin{table*}[htdp] \small
\caption{$Q_{\rm X} $ and uncertainties at different overdensity ($\Delta= 2500,1000, 500, 200, vir$).} 
\centering
\begin{tabular}{lccccccccccccc}
\hline
\hline
cluster & proj&$Q_{\rm X,2500}$ &  $eM_{\rm X,2500}$ &$Q_{\rm X,1000}$ & $eM_{\rm X,1000}$ &$Q_{\rm X,500}$  & $eM_{\rm X,500}$ &$Q_{\rm X,200}$ & $eM_{\rm X,200}$\\
\hline
CL1   &   1 & 0.5987 & 0.0356 & 0.6742 & 0.0182 & 0.6956 & 0.0251 & 0.7059 & 0.0266\\
CL1   &   2 & 0.5715 & 0.0378 & 0.7003 & 0.0220 & 0.7613 & 0.0281 & 0.7834 & 0.0338\\
CL1   &   3 & 0.5626 & 0.0371 & 0.6521 & 0.0176 & 0.7251 & 0.0222 & 0.8064 & 0.0268\\
 CL2  &    1 &  0.8315 & 0.0515 & 0.8194 & 0.0383 & 0.7456 & 0.0555 & 0.6555 & 0.0366\\
 CL2  &    2 &  0.8144 & 0.0319 & 0.6965 & 0.0248 & 0.6401 & 0.0322 & 0.5914 & 0.0256\\
 CL2  &    3 &  0.7025 & 0.0236 & 0.8240 & 0.0296 & 0.8581 & 0.0314 & 0.7965 & 0.0506\\
CL3    &   1 & 0.6887 & 0.0567 & 0.8736 & 0.0287 & 0.8902 & 0.0417 & 0.8032 & 0.0485\\
CL3    &   2 & 0.8202 & 0.0721 & 0.9378 & 0.0447 & 0.8330 & 0.0389 & 0.7032 & 0.0352\\
CL3    &   3 & 0.7640 & 0.0686 & 0.8644 & 0.0403 & 0.9281 & 0.0484 & 0.8785 & 0.0463\\
CL4    &   1 &  0.7069 & 0.0698 & 0.8624 & 0.0340 & 0.8742 & 0.0620 & 0.7890 & 0.0555\\
CL4    &   2 &  0.7663 & 0.0486 & 0.7678 & 0.0254 & 0.7130 & 0.0186 & 0.6313 & 0.0327 \\
CL4    &   3 & 0.8315  & 0.0580 & 0.7934 & 0.0303 & 0.7331 & 0.0225 & 0.6350 & 0.0196\\
CL5   &    1 & 0.7620 & 0.0282& 0.7040&  0.0298& 0.6425&  0.0368 & 0.5195 &  0.0373\\
CL5   &    2 & 0.8483 & 0.0319& 0.6946& 0.0198 & 0.5683&  0.0162 & 0.4005 & 0.0199\\
CL5   &    3 & 0.7613 & 0.0316& 0.7014& 0.0221& 0.6028 & 0.0182 & 0.4334 & 0.0246\\
CL6   &   1 & 0.8118& 0.0368 &0.7468 &0.0281 &0.6951& 0.0359 &0.6008 &0.0594\\
CL6   &   2 &0.7735 &0.0371 &0.7198 &0.0406 &0.6805 &0.0515 &0.5928& 0.0470\\
CL6   &   3 & 0.7582 &0.0357 &0.6871 &0.0113 &0.6370& 0.0217& 0.5637 &0.0373\\
CL7   &  1 & 0.8576 & 0.0350 & 0.7882 & 0.0209 & 0.7269 & 0.0234 & 0.6348 & 0.0303\\
CL7   &  2 & 0.8009 & 0.0421 & 0.7499 & 0.0208 & 0.7406 & 0.0244 & 0.6904 & 0.0321\\
CL7   &  3 & 0.8312 & 0.0450 & 0.7615 & 0.0324 & 0.6952 & 0.0270 & 0.5708 & 0.0221\\
CL8   &  1 & 0.7812 &0.0386 &0.7116 &0.0301 &0.6472 &0.0292 &0.4972 &0.0489 \\
CL8   &  2 & 0.8534 &0.0388 &0.7670 &0.0292 &0.6985 &0.0296 &0.5493 &0.0307 \\
CL8   &  3 & 0.7571 &0.0330 &0.6133 &0.0162 &0.5413 &0.0099 &0.4218 &0.0114 \\
CL9    &  1 & 0.8245 &0.0424& 0.7967& 0.0186& 0.8011& 0.0302 &0.8192& 0.0539\\
CL9    &  2 & 0.8990 &0.0432& 0.8827& 0.0355& 0.8227& 0.0401 &0.6596 &0.0313\\
CL9    &  3 & 0.8261 &0.0433& 0.9058& 0.0353& 0.8744 &0.0457& 0.6831 &0.0260\\
CL10   &   1 &  0.7187 & 0.0286 & 0.7751 & 0.0247 & 0.7814  &0.0315  & 0.7281 & 0.0636 \\
CL10   &    2 & 0.7466 & 0.0510 & 0.7628 & 0.0226 & 0.7024  & 0.0189 & 0.6600 & 0.0395 \\
CL10   &    3 &  0.9098 & 0.0809 & 0.6453  & 0.0234 & 0.5407 &0.0168  & 0.4864 & 0.0285 \\
CL11   &      1 & 0.7586 &0.0256 &0.8068 &0.0334 &0.8190 &0.0447 &0.7335 &0.0419\\
CL11   &      2 & 0.8156 &0.0333 &0.7969 &0.0327 &0.7432 &0.0221 &0.6238 &0.0267\\
CL11   &      3 & 0.8087 &0.0289 &0.7443 &0.0257 &0.6972 &0.0251 &0.6203 &0.0236\\
CL12    &  1 & 0.6647 & 0.0834 & 0.6625 &0.0257 &0.6402 &0.0299& 0.6069& 0.0597\\
CL12    &  2 & 0.5686 & 0.0597 & 0.6319 &0.0323 &0.7051 &0.0329 &0.7190 &0.0489\\
CL12    &  3 & 0.6961 & 0.0397& 0.6948& 0.0299 &0.6915 &0.0239 &0.5577& 0.0353\\
CL13  &   1 & 0.7840 & 0.0598 & 0.7031 & 0.0322 & 0.5553 & 0.0204 & 0.5161 & 0.0257\\
CL13  &   2 & 0.8149 & 0.0647  & 0.6742 & 0.0270 & 0.5217  & 0.0151  & 0.4848  & 0.0226\\
CL13  &   3 & 0.5658 & 0.0499 & 0.6951  & 0.0259 & 0.6389 & 0.0198 & 0.6488  & 0.0334\\
CL14    &  1 & 0.8654 & 0.0285 & 0.8114 & 0.0237 & 0.7431 & 0.0380 & 0.5746 & 0.0299\\
CL14    &  2 & 0.9314 & 0.0285 & 0.8010 & 0.0223 & 0.7091 & 0.0230 & 0.5769 & 0.0261\\
CL14    &  3 & 0.8812 & 0.0239 & 0.7542 & 0.0167 & 0.6650 & 0.0155 & 0.5473 & 0.0251\\
CL15   &   1 & 0.7851 & 0.0393 & 0.7670 & 0.0248 & 0.6591 & 0.0214 & 0.5374 & 0.0213\\
CL15   &   2 & 0.7488 & 0.0411 & 0.8052 & 0.0323 & 0.7042 & 0.0441 & 0.5387 & 0.0442\\
CL15   &   3 & 0.8219 & 0.0430 & 0.7730 & 0.0374 & 0.6408 & 0.0298 & 0.4896 & 0.0245\\
CL16   &     1 & 0.8596 & 0.0329& 0.8435 &0.0303& 0.8379 &0.0317 &0.8107 &0.0435\\
CL16   &     2 & 0.9331 & 0.0387 &0.9406 &0.0445 &0.9082 &0.0580 &0.7638 &0.0492\\
CL16   &     3 & 0.8681 & 0.0341 &0.8923& 0.0291 &0.8883 &0.0424 &0.7895& 0.0448\\
CL17   &   1 & 0.5160 & 0.0286 & 0.6332 & 0.0265 & 0.6884 & 0.0297 & 0.5743 & 0.0313 \\
CL17   &   2 & 0.4908 & 0.0248 & 0.5815 & 0.0120 & 0.7308 & 0.0249 & 0.8929 & 0.0582\\
CL17   &   3 & 0.5947 & 0.0462 & 0.6626 & 0.0201 & 0.7002 & 0.0227 & 0.6082 &  0.0428\\
CL18     & 1 & 0.6868 & 0.0571 & 0.8068 & 0.0429 & 0.8437 & 0.0348 & 0.6430 & 0.0663\\
CL18     & 2 & 0.8076 & 0.0845 & 0.7850 & 0.0322 & 0.7393 & 0.0318 & 0.6284 & 0.0580\\
CL18     & 3 & 0.6776 & 0.0481 & 0.8526 & 0.0370 & 0.7726 & 0.0236 & 0.4716 & 0.0443\\
   CL19    & 1 & 0.6531 & 0.0722 & 0.7239 & 0.0227 & 0.7090& 0.0427 & 0.7159 & 0.0834\\
   CL19    & 2 & 0.7823 & 0.0473 & 0.6532 & 0.0221 & 0.5951& 0.0265 & 0.6032 & 0.0423\\
   CL19    & 3 & 0.7441 & 0.0654 & 0.7260 & 0.0291 & 0.6371& 0.0357 & 0.5712 & 0.0379\\
CL20     &  1 & 0.7731 & 0.0307 & 0.7315 & 0.0229 & 0.7085 & 0.0243 & 0.6335 & 0.0269\\
CL20     &  2 & 0.8122 & 0.0322 & 0.7795 & 0.0251 & 0.7680 & 0.0225 & 0.7132 & 0.0298\\
CL20     &  3 & 0.7365 & 0.0356 & 0.7646 & 0.0240 & 0.7930 & 0.0370 & 0.7807 & 0.0320 \\
\hline
\hline
\end{tabular}
\label{tab:qx}
\end{table*}%

\section{acknowledgements} 
We acknowledge financial contribution from contracts ASI-INAF
I/023/05/0, ASI-INAF I/088/06/0, PRIN-INAF-2009 Grant "Towards an
Italian Network for Computational Cosmology" and INFN PD51.  The work
has been performed under the HPC-EUROPA2 project (project number:
228398) with the support of the European Commission - Capacities Area
- Research Infrastructures. ER acknowledges the Michigan Society of
Fellow. MM, ER, PM, SB and SE thank the organizers of the workshop
"Galaxy cluster at the crossroads between Astrophysics and Cosmology"
and the KITP for hospitality and financial support by the National
Science Foundation under Grant No. PHY05-51164. SB acknowledges
partial support by the European Commissions FP7 Marie Curie Initial
Training Network CosmoComp (PITN-GA-2009-238356). DF acknowledges support by the European Union and Ministry of Higher Education, Science and Technology of Slovenia. This research was
supported in part by the Michigan Center for Theoretical
Physics. Simulations have been carried out at CINECA (Bologna, Italy),
with CPU time allocated through an Italian SuperComputing Resource
Allocation (ISCRA) project.

\bibliographystyle{apj}
\bibliography{test_ele}

\begin{thebibliography}{134}
\expandafter\ifx\csname natexlab\endcsname\relax\def\natexlab#1{#1}\fi

\bibitem[{{Agertz} {et~al.}(2007){Agertz}, {Moore}, {Stadel}, {Potter},
  {Miniati}, {Read}, {Mayer}, {Gawryszczak}, {Kravtsov}, {Nordlund}, {Pearce},
  {Quilis}, {Rudd}, {Springel}, {Stone}, {Tasker}, {Teyssier}, {Wadsley}, \&
  {Walder}}]{agertz.etal.07}
{Agertz}, O., {Moore}, B., {Stadel}, J., {Potter}, D., {Miniati}, F., {Read},
  J., {Mayer}, L., {Gawryszczak}, A., {Kravtsov}, A., {Nordlund}, {\AA}.,
  {Pearce}, F., {Quilis}, V., {Rudd}, D., {Springel}, V., {Stone}, J.,
  {Tasker}, E., {Teyssier}, R., {Wadsley}, J., \& {Walder}, R. 2007, \mnras,
  380, 963

\bibitem[{{Allgood} {et~al.}(2006){Allgood}, {Flores}, {Primack}, {Kravtsov},
  {Wechsler}, {Faltenbacher}, \& {Bullock}}]{allgood.etal.06}
{Allgood}, B., {Flores}, R.~A., {Primack}, J.~R., {Kravtsov}, A.~V.,
  {Wechsler}, R.~H., {Faltenbacher}, A., \& {Bullock}, J.~S. 2006, \mnras, 367,
  1781

\bibitem[{{Ameglio} {et~al.}(2009){Ameglio}, {Borgani}, {Pierpaoli}, {Dolag},
  {Ettori}, \& {Morandi}}]{ameglio.etal.09}
{Ameglio}, S., {Borgani}, S., {Pierpaoli}, E., {Dolag}, K., {Ettori}, S., \&
  {Morandi}, A. 2009, \mnras, 394, 479

\bibitem[{Anders \& Grevesse(1989)}]{anders&grevesse}
Anders, E. \& Grevesse, N. 1989, \gca, 53, 197

\bibitem[{Arnaud(1996)}]{arnaud96}
Arnaud, K.~A. 1996, in ASPC, Vol. 110, 17

\bibitem[{Ascasibar {et~al.}(2003)Ascasibar, Yepes, M{\"u}ller, \&
  Gottl{\"o}ber}]{ascasibar.etal.03}
Ascasibar, Y., Yepes, G., M{\"u}ller, V., \& Gottl{\"o}ber, S. 2003, \mnras,
  346, 731

\bibitem[{Bardeau {et~al.}(2007)Bardeau, Soucail, Kneib, Czoske, Ebeling,
  Hudelot, Smail, \& Smith}]{Bardeau:2007jj}
Bardeau, S., Soucail, G., Kneib, J.-P., Czoske, O., Ebeling, H., Hudelot, P.,
  Smail, I., \& Smith, G.~P. 2007, A\&A, 470, 449

\bibitem[{Bartelmann(1996)}]{BA96.1}
Bartelmann, M. 1996, A\&A, 313, 697

\bibitem[{Bartelmann \& Schneider(2001)}]{bart&schne}
Bartelmann, M. \& Schneider, P. 2001, Physics Reports, 340, 291

\bibitem[{{Becker} \& {Kravtsov}(2011)}]{becker&kravtsov}
{Becker}, M.~R. \& {Kravtsov}, A.~V. 2011, \apj, 740, 25

\bibitem[{Beckwith {et~al.}(2006)Beckwith, Stiavelli, Koekemoer, Caldwell,
  Ferguson, Hook, Lucas, Bergeron, Corbin, Jogee, Panagia, Robberto, Royle,
  Somerville, \& Sosey}]{BECK06.1}
Beckwith, S.~V.~W., Stiavelli, M., Koekemoer, A.~M., Caldwell, J.~A.~R.,
  Ferguson, H.~C., Hook, R., Lucas, R.~A., Bergeron, L.~E., Corbin, M., Jogee,
  S., Panagia, N., Robberto, M., Royle, P., Somerville, R., \& Sosey, M. 2006,
  \aj, 132, 1729

\bibitem[{Benitez(2000)}]{2000ApJ...536..571B}
Benitez, N. 2000, \apj, 536, 571

\bibitem[{Bertin \& Arnouts(1996)}]{BE96.1}
Bertin, E. \& Arnouts, S. 1996, A\&AS, 117, 393

\bibitem[{{Bode} {et~al.}(2009){Bode}, {Ostriker}, \&
  {Vikhlinin}}]{bode.etal.09}
{Bode}, P., {Ostriker}, J.~P., \& {Vikhlinin}, A. 2009, \apj, 700, 989

\bibitem[{B{\"o}hringer {et~al.}(2010)B{\"o}hringer, Pratt, Arnaud, Borgani,
  Croston, Ponman, Ameglio, Temple, \& Dolag}]{boehringer.etal.10}
B{\"o}hringer, H., Pratt, G.~W., Arnaud, M., Borgani, S., Croston, J.~H.,
  Ponman, T.~J., Ameglio, S., Temple, R.~F., \& Dolag, K. 2010, A\&A, 514, 32

\bibitem[{{Bonafede} {et~al.}(2011){Bonafede}, {Dolag}, {Stasyszyn}, {Murante},
  \& {Borgani}}]{bonafede.etal.11}
{Bonafede}, A., {Dolag}, K., {Stasyszyn}, F., {Murante}, G., \& {Borgani}, S.
  2011, \mnras, 418, 2234

\bibitem[{Borgani \& Kravtsov(2009)}]{borgani&kravtsov}
Borgani, S. \& Kravtsov, A. 2009, arXiv, 0906.4370

\bibitem[{{Bourdin} \& {Mazzotta}(2008)}]{bourdin.etal.08}
{Bourdin}, H. \& {Mazzotta}, P. 2008, \aap, 479, 307

\bibitem[{Brada{\v c} {et~al.}(2005)Brada{\v c}, Schneider, Lombardi, \&
  Erben}]{Bradac:PbBNSt9K}
Brada{\v c}, M., Schneider, P., Lombardi, M., \& Erben, T. 2005, \aap, 437, 39

\bibitem[{{Buote} \& {Tsai}(1995)}]{buote&tsai}
{Buote}, D.~A. \& {Tsai}, J.~C. 1995, \apj, 452, 522

\bibitem[{{Cacciato} {et~al.}(2006){Cacciato}, {Bartelmann}, {Meneghetti}, \&
  {Moscardini}}]{2006A&A...458..349C}
{Cacciato}, M., {Bartelmann}, M., {Meneghetti}, M., \& {Moscardini}, L. 2006,
  \aap, 458, 349

\bibitem[{{Cassano} {et~al.}(2010){Cassano}, {Ettori}, {Giacintucci},
  {Brunetti}, {Markevitch}, {Venturi}, \& {Gitti}}]{cassano.etal.10}
{Cassano}, R., {Ettori}, S., {Giacintucci}, S., {Brunetti}, G., {Markevitch},
  M., {Venturi}, T., \& {Gitti}, M. 2010, \apj Letters, 721, L82

\bibitem[{Cen(1997)}]{1997ApJ...485...39C}
Cen, R. 1997, \apj, 485, 39

\bibitem[{Clowe {et~al.}(2004)Clowe, De~Lucia, \& King}]{Clowe:2004cb}
Clowe, D., De~Lucia, G., \& King, L. 2004, \mnras, 350, 1038

\bibitem[{Clowe \& Schneider(2002)}]{CL02.1}
Clowe, D. \& Schneider, P. 2002, A\&A, 395, 385

\bibitem[{Coe {et~al.}(2006)Coe, Ben~i tez, S{\'a}nchez, Jee, Bouwens, \&
  Ford}]{2006AJ....132..926C}
Coe, D., Ben~i tez, N., S{\'a}nchez, S.~F., Jee, M., Bouwens, R., \& Ford, H.
  2006, \aj, 132, 926

\bibitem[{Comerford {et~al.}(2006)Comerford, Meneghetti, Bartelmann, \&
  Schirmer}]{2006ApJ...642...39C}
Comerford, J.~M., Meneghetti, M., Bartelmann, M., \& Schirmer, M. 2006, \apj,
  642, 39

\bibitem[{{Conselice}(2003)}]{conselice03}
{Conselice}, C.~J. 2003, \apjs, 147, 1

\bibitem[{Dahle(2006)}]{2006ApJ...653..954D}
Dahle, H. 2006, \apj, 653, 954

\bibitem[{Diego {et~al.}(2007)Diego, Tegmark, Protopapas, \& Sandvik}]{DI07.1}
Diego, J.~M., Tegmark, M., Protopapas, P., \& Sandvik, H.~B. 2007, \mnras, 375,
  958

\bibitem[{{Dolag} {et~al.}(2004){Dolag}, {Jubelgas}, {Springel}, {Borgani}, \&
  {Rasia}}]{dolag.etal.04}
{Dolag}, K., {Jubelgas}, M., {Springel}, V., {Borgani}, S., \& {Rasia}, E.
  2004, \apj Letters, 606, L97

\bibitem[{Donnarumma {et~al.}(2011)Donnarumma, Ettori, Meneghetti, Gavazzi,
  Fort, Moscardini, Romano, Fu, Giordano, Radovich, Maoli, Scaramella, \&
  Richard}]{2011A&A...528A..73D}
Donnarumma, A., Ettori, S., Meneghetti, M., Gavazzi, R., Fort, B., Moscardini,
  L., Romano, A., Fu, L., Giordano, F., Radovich, M., Maoli, R., Scaramella,
  R., \& Richard, J. 2011, A\&A, 528, 73

\bibitem[{Donnarumma {et~al.}(2009)Donnarumma, Ettori, Meneghetti, \&
  Moscardini}]{2009MNRAS.398..438D}
Donnarumma, A., Ettori, S., Meneghetti, M., \& Moscardini, L. 2009, \mnras,
  398, 438

\bibitem[{{Ettori} \& {Balestra}(2009)}]{ettori&balestra}
{Ettori}, S. \& {Balestra}, I. 2009, \aap, 496, 343

\bibitem[{Ettori {et~al.}(2002)Ettori, de~Grandi, \& Molendi}]{ettori.etal.02}
Ettori, S., de~Grandi, S., \& Molendi, S. 2002, \aap, 391, 841

\bibitem[{{Ettori} {et~al.}(2010){Ettori}, {Gastaldello}, {Leccardi},
  {Molendi}, {Rossetti}, {Buote}, \& {Meneghetti}}]{ettori.etal.10}
{Ettori}, S., {Gastaldello}, F., {Leccardi}, A., {Molendi}, S., {Rossetti}, M.,
  {Buote}, D., \& {Meneghetti}, M. 2010, \aap, 524, 68

\bibitem[{{Fabjan} {et~al.}(2011){Fabjan}, {Borgani}, {Rasia}, {Bonafede},
  {Dolag}, {Murante}, \& {Tornatore}}]{fabjan.etal.11}
{Fabjan}, D., {Borgani}, S., {Rasia}, E., {Bonafede}, A., {Dolag}, K.,
  {Murante}, G., \& {Tornatore}, L. 2011, \mnras, 416, 801

\bibitem[{{Feroz} \& {Hobson}(2011)}]{feroz&hobson}
{Feroz}, F. \& {Hobson}, M.~P. 2011, ArXiv e-prints

\bibitem[{Fruscione {et~al.}(2006)Fruscione, McDowell, Allen, Brickhouse,
  Burke, Davis, Durham, Elvis, Galle, Harris, Huenemoerder, Houck, Ishibashi,
  Karovska, Nicastro, Noble, Nowak, Primini, Siemiginowska, Smith, \&
  Wise}]{fruscione.etal.06}
Fruscione, A., McDowell, J.~C., Allen, G.~E., Brickhouse, N.~S., Burke, D.~J.,
  Davis, J.~E., Durham, N., Elvis, M., Galle, E.~C., Harris, D.~E.,
  Huenemoerder, D.~P., Houck, J.~C., Ishibashi, B., Karovska, M., Nicastro, F.,
  Noble, M.~S., Nowak, M.~A., Primini, F.~A., Siemiginowska, A., Smith, R.~K.,
  \& Wise, M. 2006, SPIE, 6270

\bibitem[{{Gardini} {et~al.}(2004){Gardini}, {Rasia}, {Mazzotta}, {Tormen}, {De
  Grandi}, \& {Moscardini}}]{xmas}
{Gardini}, A., {Rasia}, E., {Mazzotta}, P., {Tormen}, G., {De Grandi}, S., \&
  {Moscardini}, L. 2004, \mnras, 351, 505

\bibitem[{Giavalisco {et~al.}(2004)Giavalisco, Ferguson, Koekemoer, Dickinson,
  Alexander, Bauer, Bergeron, Biagetti, Brandt, Casertano, Cesarsky,
  Chatzichristou, Conselice, Cristiani, Da~Costa, Dahlen, de~Mello, Eisenhardt,
  Erben, Fall, Fassnacht, Fosbury, Fruchter, Gardner, Grogin, Hook,
  Hornschemeier, Idzi, Jogee, Kretchmer, Laidler, Lee, Livio, Lucas, Madau,
  Mobasher, Moustakas, Nonino, Padovani, Papovich, Park, Ravindranath, Renzini,
  Richardson, Riess, Rosati, Schirmer, Schreier, Somerville, Spinrad, Stern,
  Stiavelli, Strolger, Urry, Vandame, Williams, \& Wolf}]{GIA04.1}
Giavalisco, M., Ferguson, H.~C., Koekemoer, A.~M., Dickinson, M., Alexander,
  D.~M., Bauer, F.~E., Bergeron, J., Biagetti, C., Brandt, W.~N., Casertano,
  S., Cesarsky, C., Chatzichristou, E., Conselice, C., Cristiani, S., Da~Costa,
  L., Dahlen, T., de~Mello, D., Eisenhardt, P., Erben, T., Fall, S.~M.,
  Fassnacht, C., Fosbury, R., Fruchter, A., Gardner, J.~P., Grogin, N., Hook,
  R.~N., Hornschemeier, A.~E., Idzi, R., Jogee, S., Kretchmer, C., Laidler, V.,
  Lee, K.~S., Livio, M., Lucas, R., Madau, P., Mobasher, B., Moustakas, L.~A.,
  Nonino, M., Padovani, P., Papovich, C., Park, Y., Ravindranath, S., Renzini,
  A., Richardson, M., Riess, A., Rosati, P., Schirmer, M., Schreier, E.,
  Somerville, R., Spinrad, H., Stern, D., Stiavelli, M., Strolger, L., Urry,
  C.~M., Vandame, B., Williams, R., \& Wolf, C. 2004, \apj Letters, 600, L93

\bibitem[{{Giocoli} {et~al.}(2011){Giocoli}, {Meneghetti}, {Bartelmann},
  {Moscardini}, \& {Boldrin}}]{giocoli.etal.11}
{Giocoli}, C., {Meneghetti}, M., {Bartelmann}, M., {Moscardini}, L., \&
  {Boldrin}, M. 2011, ArXiv, 1109.0285

\bibitem[{{Gitti} {et~al.}(2011){Gitti}, {Nulsen}, {David}, {McNamara}, \&
  {Wise}}]{gitti.etal.11}
{Gitti}, M., {Nulsen}, P.~E.~J., {David}, L.~P., {McNamara}, B.~R., \& {Wise},
  M.~W. 2011, \apj, 732, 13

\bibitem[{{Gruen} {et~al.}(2011){Gruen}, {Bernstein}, {Lam}, \&
  {Seitz}}]{2011MNRAS.416.1392G}
{Gruen}, D., {Bernstein}, G.~M., {Lam}, T.~Y., \& {Seitz}, S. 2011, \mnras,
  416, 1392

\bibitem[{{Heinz} \& {Br{\"u}ggen}(2009)}]{heinz&brueggen}
{Heinz}, S. \& {Br{\"u}ggen}, M. 2009, ArXiv, 0903.0043

\bibitem[{Hennawi {et~al.}(2007)Hennawi, Dalal, Bode, \&
  Ostriker}]{Hennawi:2007fj}
Hennawi, J.~F., Dalal, N., Bode, P., \& Ostriker, J.~P. 2007, \apj, 654, 714

\bibitem[{Henriksen \& Mushotzky(1986)}]{henriksen&mushotzky}
Henriksen, M.~J. \& Mushotzky, R.~F. 1986, \apj, 302, 287

\bibitem[{Hoekstra(2001)}]{2001A&A...370..743H}
Hoekstra, H. 2001, \aap, 370, 743

\bibitem[{Hoekstra(2003)}]{2003MNRAS.339.1155H}
---. 2003, \mnras, 339, 1155

\bibitem[{Hoekstra {et~al.}(2000)Hoekstra, Franx, \& Kuijken}]{Hoekstra:2000cq}
Hoekstra, H., Franx, M., \& Kuijken, K. 2000, \apj, 532, 88

\bibitem[{Hoekstra {et~al.}(1998)Hoekstra, Franx, Kuijken, \&
  Squires}]{1998ApJ...504..636H}
Hoekstra, H., Franx, M., Kuijken, K., \& Squires, G. 1998, \apj, 504, 636

\bibitem[{Hoekstra {et~al.}(2011)Hoekstra, Hartlap, Hilbert, \& van
  Uitert}]{2011MNRAS.tmp...72H}
Hoekstra, H., Hartlap, J., Hilbert, S., \& van Uitert, E. 2011, \mnras, 72

\bibitem[{Jee {et~al.}(2005)Jee, White, Benitez, Ford, Blakeslee, Rosati,
  Demarco, \& Illingworth}]{2005ApJ...618...46J}
Jee, M.~J., White, R.~L., Benitez, N., Ford, H.~C., Blakeslee, J.~P., Rosati,
  P., Demarco, R., \& Illingworth, G.~D. 2005, \apj, 618, 46

\bibitem[{Jeltema {et~al.}(2008)Jeltema, Hallman, Burns, \&
  Motl}]{jeltema.etal.08}
Jeltema, T.~E., Hallman, E.~J., Burns, J.~O., \& Motl, P.~M. 2008, \apj, 681,
  167

\bibitem[{Jenkins {et~al.}(2001)Jenkins, Frenk, White, Colberg, Cole, Evrard,
  Couchman, \& Yoshida}]{JE01.1}
Jenkins, A., Frenk, C.~S., White, S. D.~M., Colberg, J., Cole, S., Evrard,
  A.~E., Couchman, H., \& Yoshida, N. 2001, \mnras, 321, 372

\bibitem[{Jullo {et~al.}(2007)Jullo, Kneib, Limousin, El~i asd{\'o}ttir,
  Marshall, \& Verdugo}]{2007NJPh....9..447J}
Jullo, E., Kneib, J.-P., Limousin, M., El~i asd{\'o}ttir, {\'A}., Marshall,
  P.~J., \& Verdugo, T. 2007, New Journal of Physics, 9, 447

\bibitem[{Kaiser {et~al.}(1995)Kaiser, Squires, \&
  Broadhurst}]{1995ApJ...449..460K}
Kaiser, N., Squires, G., \& Broadhurst, T. 1995, \apj, 449, 460

\bibitem[{Komatsu {et~al.}(2011)Komatsu, Smith, Dunkley, Bennett, Gold,
  Hinshaw, Jarosik, Larson, Nolta, Page, Spergel, Halpern, Hill, Kogut, Limon,
  Meyer, Odegard, Tucker, Weiland, Wollack, \& Wright}]{komatsu.etal.11}
Komatsu, E., Smith, K.~M., Dunkley, J., Bennett, C.~L., Gold, B., Hinshaw, G.,
  Jarosik, N., Larson, D., Nolta, M.~R., Page, L., Spergel, D., Halpern, M.,
  Hill, R.~S., Kogut, A., Limon, M., Meyer, S.~S., Odegard, N., Tucker, G.~S.,
  Weiland, J.~L., Wollack, E., \& Wright, E.~L. 2011, arXiv, 192, 18

\bibitem[{Kubo {et~al.}(2007)Kubo, Stebbins, Annis, Dell'Antonio, Lin,
  Khiabanian, \& Frieman}]{2007ApJ...671.1466K}
Kubo, J.~M., Stebbins, A., Annis, J., Dell'Antonio, I.~P., Lin, H., Khiabanian,
  H., \& Frieman, J.~A. 2007, \apj, 671, 1466

\bibitem[{Lau {et~al.}(2009)Lau, Kravtsov, \& Nagai}]{lau.etal.09}
Lau, E.~T., Kravtsov, A.~V., \& Nagai, D. 2009, \apj, 705, 1129

\bibitem[{{Lau} {et~al.}(2011){Lau}, {Nagai}, {Kravtsov}, \&
  {Zentner}}]{lau.etal.11}
{Lau}, E.~T., {Nagai}, D., {Kravtsov}, A.~V., \& {Zentner}, A.~R. 2011, \apj,
  734, 93

\bibitem[{{Leccardi} \& {Molendi}(2008)}]{leccardi&molendi}
{Leccardi}, A. \& {Molendi}, S. 2008, \aap, 486, 359

\bibitem[{{Lee} {et~al.}(2011){Lee}, {Kashyap}, {van Dyk}, {Connors}, {Drake},
  {Izem}, {Meng}, {Min}, {Park}, {Ratzlaff}, {Siemiginowska}, \&
  {Zezas}}]{lee.etal.11}
{Lee}, H., {Kashyap}, V.~L., {van Dyk}, D.~A., {Connors}, A., {Drake}, J.~J.,
  {Izem}, R., {Meng}, X.-L., {Min}, S., {Park}, T., {Ratzlaff}, P.,
  {Siemiginowska}, A., \& {Zezas}, A. 2011, \apj, 731, 126

\bibitem[{{LSST Science Collaborations} {et~al.}(2009){LSST Science
  Collaborations}, {Abell}, {Allison}, {Anderson}, {Andrew}, {Angel}, {Armus},
  {Arnett}, {Asztalos}, {Axelrod}, \& et~al.}]{lsst.09}
{LSST Science Collaborations}, {Abell}, P.~A., {Allison}, J., {Anderson},
  S.~F., {Andrew}, J.~R., {Angel}, J.~R.~P., {Armus}, L., {Arnett}, D.,
  {Asztalos}, S.~J., {Axelrod}, T.~S., \& et~al. 2009, ArXiv e-prints

\bibitem[{Luppino \& Kaiser(1997)}]{1997ApJ...475...20L}
Luppino, G.~A. \& Kaiser, N. 1997, \apj, 475, 20

\bibitem[{Mahdavi {et~al.}(2008)Mahdavi, Hoekstra, Babul, \&
  Henry}]{mahdavi.etal.08}
Mahdavi, A., Hoekstra, H., Babul, A., \& Henry, J.~P. 2008, \mnras, 384, 1567

\bibitem[{Markevitch {et~al.}(2004)Markevitch, Gonzalez, Clowe, Vikhlinin,
  Forman, Jones, Murray, \& Tucker}]{2004ApJ...606..819M}
Markevitch, M., Gonzalez, A.~H., Clowe, D., Vikhlinin, A., Forman, W., Jones,
  C., Murray, S., \& Tucker, W. 2004, \apj, 606, 819

\bibitem[{{Maughan} {et~al.}(2008){Maughan}, {Jones}, {Forman}, \& {Van
  Speybroeck}}]{maughan.etal.08}
{Maughan}, B.~J., {Jones}, C., {Forman}, W., \& {Van Speybroeck}, L. 2008,
  \apjs, 174, 117

\bibitem[{Mazzotta {et~al.}(2004)Mazzotta, Rasia, Moscardini, \&
  Tormen}]{mazzotta.etal.04}
Mazzotta, P., Rasia, E., Moscardini, L., \& Tormen, G. 2004, \mnras, 320

\bibitem[{{Medezinski} {et~al.}(2007){Medezinski}, {Broadhurst}, {Umetsu},
  {Coe}, {Ben{\'{\i}}tez}, {Ford}, {Rephaeli}, {Arimoto}, \&
  {Kong}}]{medezinski.etal.07}
{Medezinski}, E., {Broadhurst}, T., {Umetsu}, K., {Coe}, D., {Ben{\'{\i}}tez},
  N., {Ford}, H., {Rephaeli}, Y., {Arimoto}, N., \& {Kong}, X. 2007, \apj, 663,
  717

\bibitem[{{Medezinski} {et~al.}(2010){Medezinski}, {Broadhurst}, {Umetsu},
  {Oguri}, {Rephaeli}, \& {Ben{\'{\i}}tez}}]{medezinski.etal.10}
{Medezinski}, E., {Broadhurst}, T., {Umetsu}, K., {Oguri}, M., {Rephaeli}, Y.,
  \& {Ben{\'{\i}}tez}, N. 2010, \mnras, 405, 257

\bibitem[{Meneghetti {et~al.}(2003)Meneghetti, Bartelmann, \&
  Moscardini}]{ME03.1}
Meneghetti, M., Bartelmann, M., \& Moscardini, L. 2003, \mnras, 340, 105

\bibitem[{{Meneghetti} {et~al.}(2010){Meneghetti}, {Fedeli}, {Pace},
  {Gottl{\"o}ber}, \& {Yepes}}]{meneghetti.etal.10}
{Meneghetti}, M., {Fedeli}, C., {Pace}, F., {Gottl{\"o}ber}, S., \& {Yepes}, G.
  2010, \aap, 519, 90

\bibitem[{{Meneghetti} {et~al.}(2011){Meneghetti}, {Fedeli}, {Zitrin},
  {Bartelmann}, {Broadhurst}, {Gottl{\"o}ber}, {Moscardini}, \&
  {Yepes}}]{meneghetti.etal.11}
{Meneghetti}, M., {Fedeli}, C., {Zitrin}, A., {Bartelmann}, M., {Broadhurst},
  T., {Gottl{\"o}ber}, S., {Moscardini}, L., \& {Yepes}, G. 2011, \aap, 530, 17

\bibitem[{Meneghetti {et~al.}(2008)Meneghetti, Melchior, Grazian, De~Lucia,
  Dolag, Bartelmann, Heymans, Moscardini, \& Radovich}]{2008A&A...482..403M}
Meneghetti, M., Melchior, P., Grazian, A., De~Lucia, G., Dolag, K., Bartelmann,
  M., Heymans, C., Moscardini, L., \& Radovich, M. 2008, arXiv, 482, 403

\bibitem[{Meneghetti {et~al.}(2010)Meneghetti, Rasia, Merten, Bellagamba,
  Ettori, Mazzotta, Dolag, \& Marri}]{2010A&A...514A..93M}
Meneghetti, M., Rasia, E., Merten, J., Bellagamba, F., Ettori, S., Mazzotta,
  P., Dolag, K., \& Marri, S. 2010, Astronomy and Astrophysics, 514, 93

\bibitem[{Merten {et~al.}(2009)Merten, Cacciato, Meneghetti, Mignone, \&
  Bartelmann}]{2009A&A...500..681M}
Merten, J., Cacciato, M., Meneghetti, M., Mignone, C., \& Bartelmann, M. 2009,
  Astronomy and Astrophysics, 500, 681

\bibitem[{Merten {et~al.}(2011)Merten, Coe, Dupke, Massey, Zitrin, Cypriano,
  Okabe, Frye, Braglia, Jimenez-Teja, Benitez, Broadhurst, Rhodes, Meneghetti,
  Moustakas, Sodre, Krick, \& Bregman}]{2011arXiv1103.2772M}
Merten, J., Coe, D., Dupke, R., Massey, R., Zitrin, A., Cypriano, E.~S., Okabe,
  N., Frye, B., Braglia, F., Jimenez-Teja, Y., Benitez, N., Broadhurst, T.,
  Rhodes, J., Meneghetti, M., Moustakas, L.~A., Sodre, L.~J., Krick, J., \&
  Bregman, J.~N. 2011, \mnras, 1103, 2772

\bibitem[{Metzler {et~al.}(2001)Metzler, White, \& Loken}]{Metzler:2001gg}
Metzler, C.~A., White, M., \& Loken, C. 2001, \apj, 547, 560

\bibitem[{Metzler {et~al.}(1999)Metzler, White, Norman, \&
  Loken}]{1999ApJ...520L...9M}
Metzler, C.~A., White, M., Norman, M., \& Loken, C. 1999, \apj Letters, 520, L9

\bibitem[{{Mitchell} {et~al.}(2009){Mitchell}, {McCarthy}, {Bower}, {Theuns},
  \& {Crain}}]{mitchell.etal.09}
{Mitchell}, N.~L., {McCarthy}, I.~G., {Bower}, R.~G., {Theuns}, T., \& {Crain},
  R.~A. 2009, \mnras, 395, 180

\bibitem[{Morandi {et~al.}(2007)Morandi, Ettori, \&
  Moscardini}]{2007MNRAS.379..518M}
Morandi, A., Ettori, S., \& Moscardini, L. 2007, \mnras, 379, 518

\bibitem[{Morandi {et~al.}(2011)Morandi, Limousin, Rephaeli, Umetsu, Barkana,
  Broadhurst, \& Dahle}]{2011arXiv1103.0202M}
Morandi, A., Limousin, M., Rephaeli, Y., Umetsu, K., Barkana, R., Broadhurst,
  T., \& Dahle, H. 2011, \mnras, 416, 2567

\bibitem[{{Nagai} \& {Lau}(2011)}]{nagai.etal.11}
{Nagai}, D. \& {Lau}, E.~T. 2011, \apj Letters, 731, 10

\bibitem[{Nagai {et~al.}(2007)Nagai, Vikhlinin, \& Kravtsov}]{nagai.etal.07}
Nagai, D., Vikhlinin, A., \& Kravtsov, A.~V. 2007, \apj, 655, 98

\bibitem[{Navarro {et~al.}(1997)Navarro, Frenk, \& White}]{nfw}
Navarro, J.~F., Frenk, C.~S., \& White, S. D.~M. 1997, Astrophysical Journal
  v.490, 490, 493

\bibitem[{Neumann(2005)}]{neumann05}
Neumann, D.~M. 2005, preprint, astro-ph/0505049

\bibitem[{{Nevalainen} {et~al.}(2010){Nevalainen}, {David}, \&
  {Guainazzi}}]{nevalainen.etal.10}
{Nevalainen}, J., {David}, L., \& {Guainazzi}, M. 2010, \aap, 523, A22+

\bibitem[{Oguri {et~al.}(2009)Oguri, Hennawi, Gladders, Dahle, Natarajan,
  Dalal, Koester, Sharon, \& Bayliss}]{2009ApJ...699.1038O}
Oguri, M., Hennawi, J.~F., Gladders, M.~D., Dahle, H., Natarajan, P., Dalal,
  N., Koester, B.~P., Sharon, K., \& Bayliss, M. 2009, \apj, 699, 1038

\bibitem[{{Okabe} {et~al.}(2011){Okabe}, {Bourdin}, {Mazzotta}, \&
  {Maurogordato}}]{okabe.etal.11}
{Okabe}, N., {Bourdin}, H., {Mazzotta}, P., \& {Maurogordato}, S. 2011, \apj,
  741, 1160

\bibitem[{{Okabe} {et~al.}(2010){Okabe}, {Okura}, \&
  {Futamase}}]{Okabe.etal.10a}
{Okabe}, N., {Okura}, Y., \& {Futamase}, T. 2010, \apj, 713, 291

\bibitem[{Okabe {et~al.}(2010)Okabe, Takada, Umetsu, Futamase, \&
  Smith}]{2010PASJ...62..811O}
Okabe, N., Takada, M., Umetsu, K., Futamase, T., \& Smith, G.~P. 2010, \\pasp,
  62, 811

\bibitem[{{Okabe} {et~al.}(2010){Okabe}, {Zhang}, {Finoguenov}, {Takada},
  {Smith}, {Umetsu}, \& {Futamase}}]{okabe.etal.10}
{Okabe}, N., {Zhang}, Y.-Y., {Finoguenov}, A., {Takada}, M., {Smith}, G.~P.,
  {Umetsu}, K., \& {Futamase}, T. 2010, \apj, 721, 875

\bibitem[{{Okura} {et~al.}(2007){Okura}, {Umetsu}, \&
  {Futamase}}]{2007ApJ...660..995O}
{Okura}, Y., {Umetsu}, K., \& {Futamase}, T. 2007, \apj, 660, 995

\bibitem[{{Ostriker} {et~al.}(2005){Ostriker}, {Bode}, \&
  {Babul}}]{ostriker.etal.05}
{Ostriker}, J.~P., {Bode}, P., \& {Babul}, A. 2005, \apj, 634, 964

\bibitem[{Paulin-Henriksson {et~al.}(2007)Paulin-Henriksson, Antonuccio-Delogu,
  Haines, Radovich, Mercurio, \& Becciani}]{PaulinHenriksson:2007ex}
Paulin-Henriksson, S., Antonuccio-Delogu, V., Haines, C.~P., Radovich, M.,
  Mercurio, A., \& Becciani, U. 2007, Astronomy and Astrophysics, 467, 427

\bibitem[{Piffaretti {et~al.}(2003)Piffaretti, Jetzer, \&
  Schindler}]{Piffaretti:2003kn}
Piffaretti, R., Jetzer, P., \& Schindler, S. 2003, Astronomy and Astrophysics,
  398, 41

\bibitem[{Piffaretti \& Valdarnini(2008)}]{piffaretti&valdarnini}
Piffaretti, R. \& Valdarnini, R. 2008, A\&A, 491, 71

\bibitem[{{Poole} {et~al.}(2006){Poole}, {Fardal}, {Babul}, {McCarthy},
  {Quinn}, \& {Wadsley}}]{poole.etal.06}
{Poole}, G.~B., {Fardal}, M.~A., {Babul}, A., {McCarthy}, I.~G., {Quinn}, T.,
  \& {Wadsley}, J. 2006, \mnras, 373, 881

\bibitem[{Press \& Schechter(1974)}]{PR74.1}
Press, W. \& Schechter, P. 1974, \apj, 187, 425

\bibitem[{Rasia {et~al.}(2006)Rasia, Ettori, Moscardini, Mazzotta, Borgani,
  Dolag, Tormen, Cheng, \& Diaferio}]{rasia.etal.06}
Rasia, E., Ettori, S., Moscardini, L., Mazzotta, P., Borgani, S., Dolag, K.,
  Tormen, G., Cheng, L.~M., \& Diaferio, A. 2006, \mnras, 369, 2013

\bibitem[{Rasia {et~al.}(2005)Rasia, Mazzotta, Borgani, Moscardini, Dolag,
  Tormen, Diaferio, \& Murante}]{rasia.etal.05}
Rasia, E., Mazzotta, P., Borgani, S., Moscardini, L., Dolag, K., Tormen, G.,
  Diaferio, A., \& Murante, G. 2005, \apj Letters, 618, L1

\bibitem[{Rasia {et~al.}(2008)Rasia, Mazzotta, Bourdin, Borgani, Tornatore,
  Ettori, Dolag, \& Moscardini}]{rasia.etal.08}
Rasia, E., Mazzotta, P., Bourdin, H., Borgani, S., Tornatore, L., Ettori, S.,
  Dolag, K., \& Moscardini, L. 2008, \apj, 674, 728

\bibitem[{Rasia {et~al.}(2011)Rasia, Mazzotta, Evrard, Markevitch, Dolag, \&
  Meneghetti}]{rasia.etal.11}
Rasia, E., Mazzotta, P., Evrard, A., Markevitch, M., Dolag, K., \& Meneghetti,
  M. 2011, \apj, 729, 45

\bibitem[{Rasia {et~al.}(2004)Rasia, Tormen, \& Moscardini}]{rtm}
Rasia, E., Tormen, G., \& Moscardini, L. 2004, \mnras, 351, 237

\bibitem[{Refregier(2003)}]{RE03.1}
Refregier, A. 2003, \mnras, 338, 35

\bibitem[{Refregier {et~al.}(2010)Refregier, Amara, Kitching, Rassat,
  Scaramella, Weller, \& Euclid Imaging~Consortium}]{2010arXiv1001.0061R}
Refregier, A., Amara, A., Kitching, T.~D., Rassat, A., Scaramella, R., Weller,
  J., \& Euclid Imaging~Consortium, f.~t. 2010, arXiv.org, 1001, 61

\bibitem[{Romano {et~al.}(2010)Romano, Fu, Giordano, Maoli, Martini, Radovich,
  Scaramella, Antonuccio-Delogu, Donnarumma, Ettori, Kuijken, Meneghetti,
  Moscardini, Paulin-Henriksson, Giallongo, Ragazzoni, Baruffolo, Dipaola,
  Diolaiti, Farinato, Fontana, Gallozzi, Grazian, Hill, Pedichini, Speziali,
  Smareglia, \& Testa}]{2010A&A...514A..88R}
Romano, A., Fu, L., Giordano, F., Maoli, R., Martini, P., Radovich, M.,
  Scaramella, R., Antonuccio-Delogu, V., Donnarumma, A., Ettori, S., Kuijken,
  K., Meneghetti, M., Moscardini, L., Paulin-Henriksson, S., Giallongo, E.,
  Ragazzoni, R., Baruffolo, A., Dipaola, A., Diolaiti, E., Farinato, J.,
  Fontana, A., Gallozzi, S., Grazian, A., Hill, J., Pedichini, F., Speziali,
  R., Smareglia, R., \& Testa, V. 2010, Astronomy and Astrophysics, 514, 88

\bibitem[{Sarazin(1988)}]{sarazin88}
Sarazin, C.~L. 1988, X-Ray Emission from Clusters of Galaxies (Cambridge:
  Cambridge Univ. Press)

\bibitem[{{Sarazin}(1988)}]{sarazin}
{Sarazin}, C.~L. 1988, {X-ray emission from clusters of galaxies}, ed.
  {Sarazin, C.~L.}

\bibitem[{Sereno {et~al.}(2010)Sereno, Jetzer, \& Lubini}]{2010MNRAS.403.2077S}
Sereno, M., Jetzer, P., \& Lubini, M. 2010, MNRAS, 403, 2077

\bibitem[{Shaw {et~al.}(2006)Shaw, Weller, Ostriker, \& Bode}]{Shaw:2006id}
Shaw, L.~D., Weller, J., Ostriker, J.~P., \& Bode, P. 2006, \apj, 646, 815

\bibitem[{Sheth \& Tormen(2002)}]{SH02.1}
Sheth, R. \& Tormen, G. 2002, \mnras, 329, 61

\bibitem[{{Sijacki} {et~al.}(2011){Sijacki}, {Vogelsberger}, {Keres},
  {Springel}, \& {Hernquist}}]{sijacki.etal.11}
{Sijacki}, D., {Vogelsberger}, M., {Keres}, D., {Springel}, V., \& {Hernquist},
  L. 2011, ArXiv e-prints

\bibitem[{{Simionescu} {et~al.}(2011){Simionescu}, {Allen}, {Mantz}, {Werner},
  {Takei}, {Morris}, {Fabian}, {Sanders}, {Nulsen}, {George}, \&
  {Taylor}}]{simionescu.etal.11}
{Simionescu}, A., {Allen}, S.~W., {Mantz}, A., {Werner}, N., {Takei}, Y.,
  {Morris}, R.~G., {Fabian}, A.~C., {Sanders}, J.~S., {Nulsen}, P.~E.~J.,
  {George}, M.~R., \& {Taylor}, G.~B. 2011, Science, 331, 1576

\bibitem[{Springel(2005)}]{sp05.1}
Springel, V. 2005, \mnras, 364, 1105

\bibitem[{Springel \& Hernquist(2003)}]{springel&hernquist03}
Springel, V. \& Hernquist, L. 2003, \mnras, 339, 289

\bibitem[{{The Dark Energy Survey Collaboration}(2005)}]{des.05}
{The Dark Energy Survey Collaboration}. 2005, ArXiv Astrophysics e-prints

\bibitem[{Tormen {et~al.}(1997)Tormen, Bouchet, \& White}]{tormen.etal.97}
Tormen, G., Bouchet, F., \& White, S. D.~M. 1997, \mnras, 286, 865

\bibitem[{Tornatore {et~al.}(2007)Tornatore, Borgani, Dolag, \&
  Matteucci}]{tornatore.etal.07}
Tornatore, L., Borgani, S., Dolag, K., \& Matteucci, F. 2007, arXiv, 382, 1050

\bibitem[{Umetsu {et~al.}(2011)Umetsu, Broadhurst, Zitrin, Medezinski, \&
  Hsu}]{2011ApJ...729..127U}
Umetsu, K., Broadhurst, T., Zitrin, A., Medezinski, E., \& Hsu, L.-Y. 2011,
  \apj, 729, 127

\bibitem[{{Vazza} {et~al.}(2011){Vazza}, {Dolag}, {Ryu}, {Brunetti}, {Gheller},
  {Kang}, \& {Pfrommer}}]{vazza.etal.11}
{Vazza}, F., {Dolag}, K., {Ryu}, D., {Brunetti}, G., {Gheller}, C., {Kang}, H.,
  \& {Pfrommer}, C. 2011, ArXiv e-prints

\bibitem[{{Velander} {et~al.}(2011){Velander}, {Kuijken}, \&
  {Schrabback}}]{velander.etal.10}
{Velander}, M., {Kuijken}, K., \& {Schrabback}, T. 2011, \mnras, 412, 2665

\bibitem[{Vikhlinin {et~al.}(2009)Vikhlinin, Burenin, Ebeling, Forman,
  Hornstrup, Jones, Kravtsov, Murray, Nagai, Quintana, \&
  Voevodkin}]{2009ApJ...692.1033V}
Vikhlinin, A., Burenin, R.~A., Ebeling, H., Forman, W.~R., Hornstrup, A.,
  Jones, C., Kravtsov, A.~V., Murray, S.~S., Nagai, D., Quintana, H., \&
  Voevodkin, A. 2009, \apj, 692, 1033

\bibitem[{Vikhlinin {et~al.}(2006)Vikhlinin, Kravtsov, Forman, Jones,
  Markevitch, Murray, \& Van~Speybroeck}]{vikh.etal.06}
Vikhlinin, A., Kravtsov, A., Forman, W., Jones, C., Markevitch, M., Murray,
  S.~S., \& Van~Speybroeck, L. 2006, \apj, 640, 691

\bibitem[{Vikhlinin {et~al.}(1998)Vikhlinin, McNamara, Forman, Jones, Quintana,
  \& Hornstrup}]{vikh.etal.98}
Vikhlinin, A., McNamara, B.~R., Forman, W., Jones, C., Quintana, H., \&
  Hornstrup, A. 1998, \apj, 502, 558

\bibitem[{Warren {et~al.}(2006)Warren, Abazajian, Holz, \&
  Teodoro}]{2006ApJ...646..881W}
Warren, M.~S., Abazajian, K., Holz, D.~E., \& Teodoro, L. 2006, \apj, 646, 881

\bibitem[{White \& Vale(2004)}]{2004APh....22...19W}
White, M. \& Vale, C. 2004, Astroparticle Physics, 22, 19

\bibitem[{{Wiersma} {et~al.}(2009){Wiersma}, {Schaye}, \&
  {Smith}}]{wiersma.etal.09}
{Wiersma}, R.~P.~C., {Schaye}, J., \& {Smith}, B.~D. 2009, \mnras, 393, 99

\bibitem[{Wright \& Brainerd(2000)}]{WR00.1}
Wright, C. \& Brainerd, T.~G. 2000, \apj, 534, 34

\bibitem[{Zhang {et~al.}(2008)Zhang, Finoguenov, B{\"o}hringer, Kneib, Smith,
  Kneissl, Okabe, \& Dahle}]{Zhang.etal.08}
Zhang, Y.~Y., Finoguenov, A., B{\"o}hringer, H., Kneib, J.-P., Smith, G.~P.,
  Kneissl, R., Okabe, N., \& Dahle, H. 2008, Astronomy and Astrophysics, 482,
  451

\bibitem[{{Zhang} {et~al.}(2010){Zhang}, {Okabe}, {Finoguenov}, {Smith},
  {Piffaretti}, {Valdarnini}, {Babul}, {Evrard}, {Mazzotta}, {Sanderson}, \&
  {Marrone}}]{zhang.etal.10}
{Zhang}, Y.-Y., {Okabe}, N., {Finoguenov}, A., {Smith}, G.~P., {Piffaretti},
  R., {Valdarnini}, R., {Babul}, A., {Evrard}, A.~E., {Mazzotta}, P.,
  {Sanderson}, A.~J.~R., \& {Marrone}, D.~P. 2010, \apj, 711, 1033

\bibitem[{{Zhang} {et~al.}(2009){Zhang}, {Reiprich}, {Finoguenov}, {Hudson}, \&
  {Sarazin}}]{zhang.etal.09}
{Zhang}, Y.-Y., {Reiprich}, T.~H., {Finoguenov}, A., {Hudson}, D.~S., \&
  {Sarazin}, C.~L. 2009, \apj, 699, 1178

\bibitem[{Zitrin {et~al.}(2011)Zitrin, Broadhurst, Coe, Umetsu, Postman,
  Meneghetti, Medezinski, Jouvel, Bradley, Koekemoer, Zheng, Ford, Merten,
  Kelson, Lahav, Lemze, Molino, Nonino, Donahue, Rosati, Van~der Wel,
  Bartelmann, Bouwens, Graur, Graves, Host, Infante, Jha, Jimenez-Teja, Lazkoz,
  Maoz, McCully, Melchior, Moustakas, Ogaz, Patel, Regoes, Riess, Rodney, \&
  Seitz}]{Zitrin:2011ut}
Zitrin, A., Broadhurst, T., Coe, D., Umetsu, K., Postman, M., Meneghetti, M.,
  Medezinski, E., Jouvel, S., Bradley, L., Koekemoer, A., Zheng, W., Ford, H.,
  Merten, J., Kelson, D., Lahav, O., Lemze, D., Molino, A., Nonino, M.,
  Donahue, M., Rosati, P., Van~der Wel, A., Bartelmann, M., Bouwens, R., Graur,
  O., Graves, G., Host, O., Infante, L., Jha, S., Jimenez-Teja, Y., Lazkoz, R.,
  Maoz, D., McCully, C., Melchior, P., Moustakas, L.~A., Ogaz, S., Patel, B.,
  Regoes, E., Riess, A., Rodney, S., \& Seitz, S. 2011, \apj, 742, 117

\end{thebibliography}

\end{document}